\documentclass{article}
\usepackage{ijcai24}

\usepackage{times}
\usepackage{soul}
\usepackage{url}
\usepackage[hidelinks]{hyperref}
\usepackage[utf8]{inputenc}
\usepackage[small]{caption}
\usepackage{graphicx}
\usepackage{amsmath}
\usepackage{amsthm}
\usepackage{amssymb}
\usepackage{booktabs}
\usepackage{algorithmic}
\usepackage[linesnumbered,ruled,vlined]{algorithm2e}
\usepackage[switch]{lineno}
\usepackage{natbib}
\usepackage{subcaption}
\usepackage{multicol}
\usepackage{multirow}
\usepackage{xspace}
\usepackage{mathtools}

\usepackage{cleveref} %

\newtheorem{definition}{Definition}

\DeclareMathOperator*{\argmax}{arg\,max}

\DeclarePairedDelimiter{\ceil}{\lceil}{\rceil}

\newcommand{\warehouseSmallR}{\textit{warehouse-20-17}\xspace}
\newcommand{\warehouselargeW}{\textit{warehouse-33-36}\xspace}
\newcommand{\randomSmall}{\textit{random-32-32-20}\xspace}
\newcommand{\roomLarge}{\textit{room-64-64-8}\xspace}
\newcommand{\mazeSmall}{\textit{maze-32-32-4}\xspace}
\newcommand{\emptyMid}{\textit{empty-48-48}\xspace}
\newcommand{\randomLarge}{\textit{random-64-64-20}\xspace}
\newcommand{\denSmall}{\textit{den312d}\xspace}

\usepackage{xcolor}

\newcommand{\xxnote}[3]{}
\ifx\hidenotes\undefined
  \renewcommand{\xxnote}[3]{\color{#2}{(#1: #3)}}
\fi

\title{Guidance Graph Optimization for Lifelong Multi-Agent Path Finding}

\author{
Yulun Zhang$^1$
\and
He Jiang$^1$\and
Varun Bhatt$^2$\and
Stefanos Nikolaidis$^2$\And
Jiaoyang Li$^1$
\affiliations
$^1$Robotics Institute, Carnegie Mellon University\\
$^2$Thomas Lord Department of Computer Science, University of Southern California
\emails
\{yulunzhang,hejiangrivers\}@cmu.edu,
\{vsbhatt,nikolaid\}@usc.edu,
jiaoyangli@cmu.edu
}

\begin{document}

\maketitle

\begin{abstract}
We study how to use guidance to improve the throughput of lifelong Multi-Agent Path Finding (MAPF). Previous studies have demonstrated that, while incorporating guidance, such as highways, can accelerate MAPF algorithms, this often results in a trade-off with solution quality. In addition, how to generate good guidance automatically remains largely unexplored, with current methods falling short of surpassing manually designed ones. In this work, we introduce the guidance graph as a versatile representation of guidance for lifelong MAPF, framing Guidance Graph Optimization as the task of optimizing its edge weights. We present two GGO algorithms to automatically generate guidance for arbitrary lifelong MAPF algorithms and maps. The first method directly optimizes edge weights, while the second method optimizes an update model capable of generating edge weights.  
Empirically, we show that (1) our guidance graphs improve the throughput of three representative lifelong MAPF algorithms in eight benchmark maps, and (2) our update model can generate guidance graphs for as large as $93 \times 91$ maps and as many as 3,000 agents. We include the source code at: \url{https://github.com/lunjohnzhang/ggo_public}. All optimized guidance graphs are available online at: \url{https://yulunzhang.net/publication/zhang2024ggo}.

\end{abstract}

\section{Introduction}

We study the problem of leveraging a guidance graph with optimized edge weights to guide agent movement, thereby improving the throughput of lifelong Multi-Agent Path Finding (MAPF). MAPF~\citep{SternSoCS19} aims to plan collision-free paths for a set of agents from their start to goal locations on a given map, depicted as a graph $G$. 
Lifelong MAPF~\citep{Li2020LifelongMP} extends MAPF by assigning new goals to agents as soon as they reach their current ones. Example applications include character control in video games~\citep{MaAIIDE17,Jansen2008DirectionMF} and automated warehouses in which hundreds of robots are continually assigned new tasks to transport inventory pods~\citep{VaramballySoCS22}.
Driven by these real-world demands, numerous studies have focused on improving the throughput, namely the average number of reached goals per timestep, by developing better lifelong MAPF algorithms~\citep{MaAAMAS17,LiAAMAS20a,KouAAAI20,DamaniRAL21} or optimizing map layouts (i.e., map structures)~\citep{ZhangNCA2023,zhangLayout23}. 

\begin{figure}[!t]
    \centering
    \begin{subfigure}[b]{0.15\textwidth}
      \centering
      \includegraphics[width=1\textwidth]{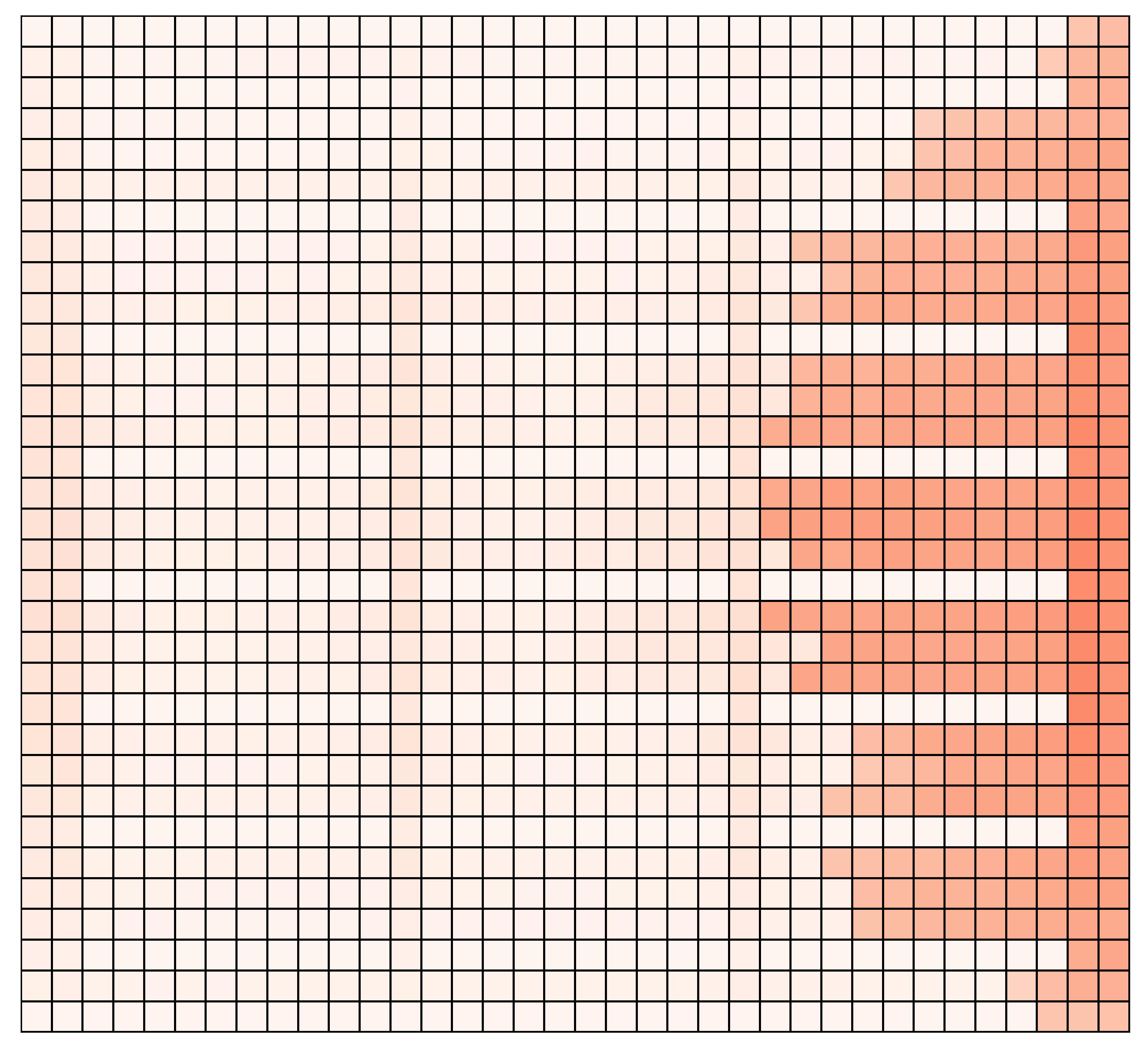}
      \caption{No guidance}
      \label{fig:front-fig:unweighted}
    \end{subfigure}%
    \begin{subfigure}[b]{0.15\textwidth}
      \centering
      \includegraphics[width=1\textwidth]{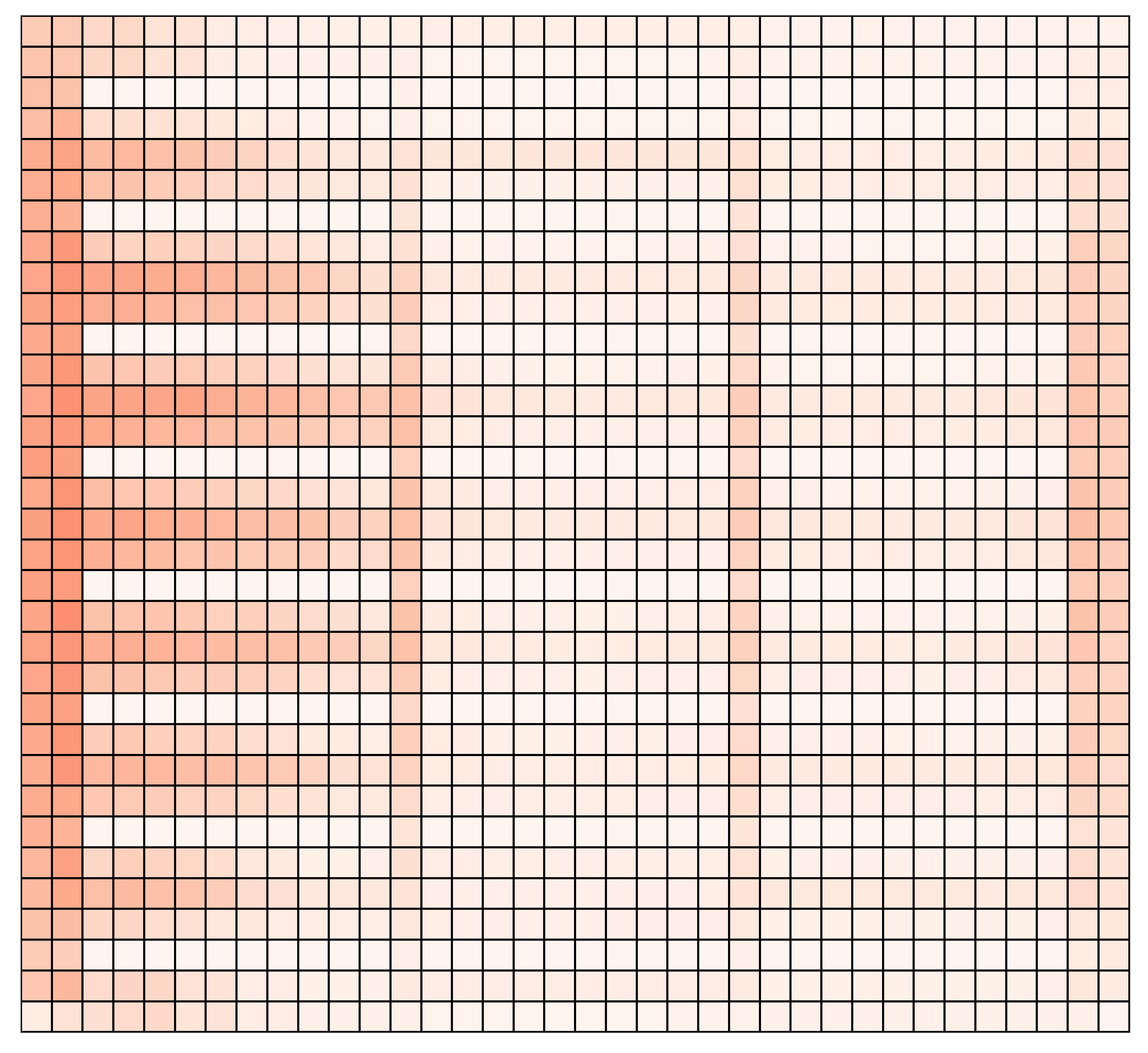}
      \caption{Crisscross}
      \label{fig:front-fig:criss-cross}
    \end{subfigure}%
    \begin{subfigure}[b]{0.15\textwidth}
      \centering
      \includegraphics[width=1\textwidth]{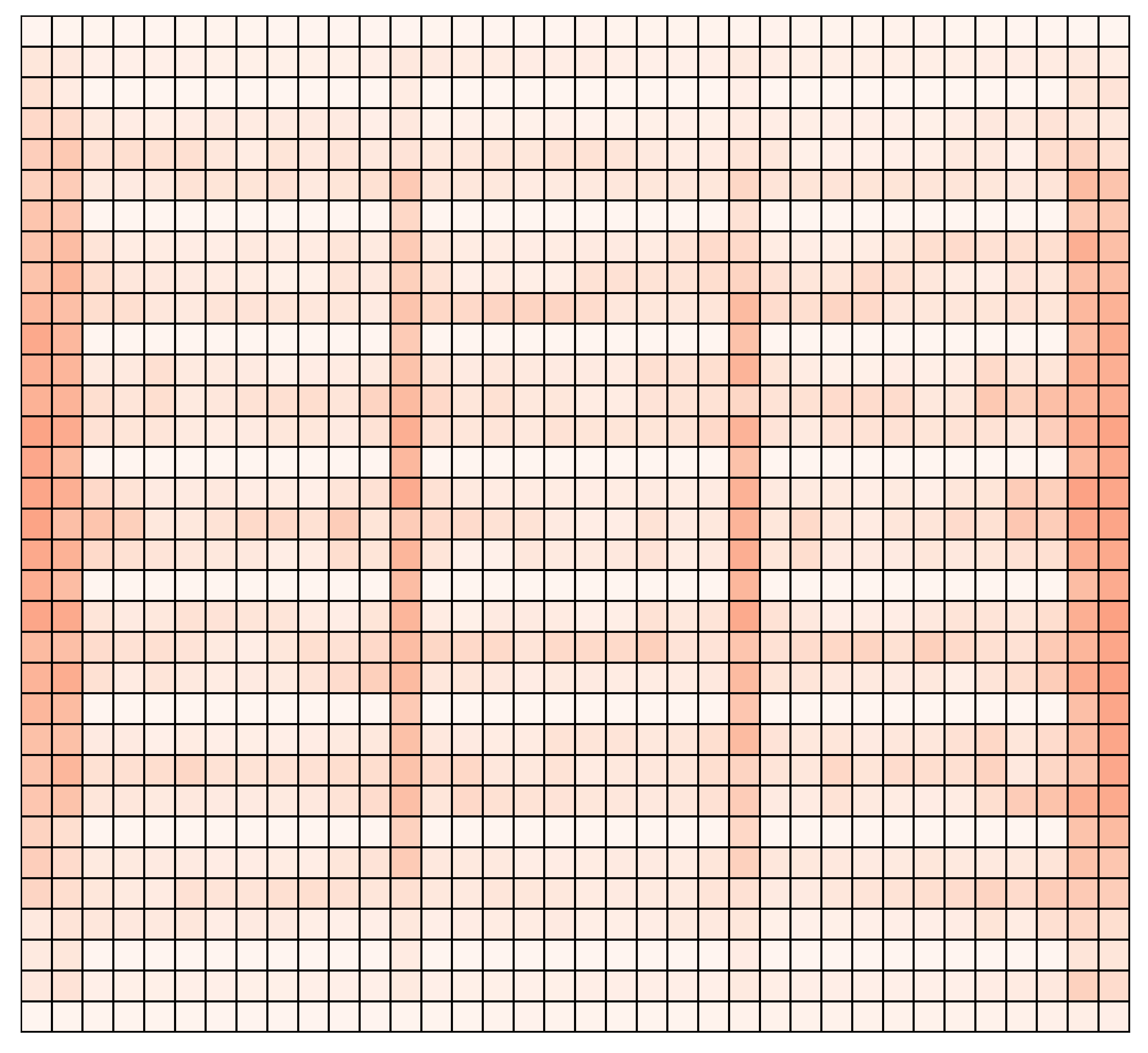}
      \caption{Our guidance}
      \label{fig:front-fig:cma-es-opt}
    \end{subfigure}%
    \begin{subfigure}[b]{0.035\textwidth}
      \centering
      \includegraphics[width=1\textwidth]{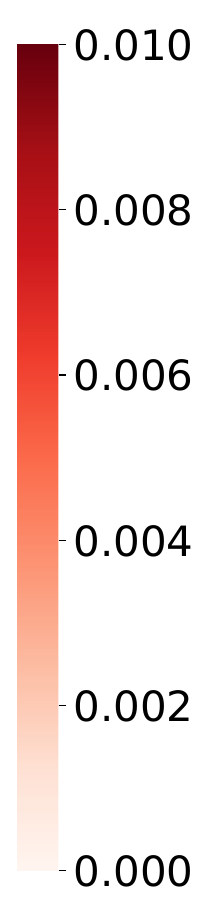}
      \label{fig:front-fig:cbar}
    \end{subfigure}%
    \caption{
    Comparison of no guidance, human-designed crisscross guidance, and our guidance with a simulation of 240 agents in a 33 $\times$ 36 warehouse map, shown in \Cref{fig:warehouse-large-w}. The heatmaps show the tile-usage (the frequency that each tile is occupied). Our guidance results in the most balanced traffic with the highest throughput.
    }
    \label{fig:front-fig}
\end{figure}

Given that lifelong MAPF requires online computation of new paths as agents are continuously assigned to new goal locations, lifelong MAPF algorithms always decompose the problem into a series of (one-shot) MAPF instances and solve them sequentially. However, such methods are myopic because each MAPF instance involves only the current goal locations.
Achieving (near-)optimal solutions for individual MAPF instances does not necessarily result in the best throughput. In this work, we propose to foster implicit cooperation among agents over the long term by introducing global guidance for agent movement.
Our guidance takes the form of a directed weighted graph that alters the costs of agents moving along each edge and waiting at each vertex of graph $G$. Intuitively, such a guidance graph serves two purposes. First, by amplifying the cost difference of traveling through an edge in opposite directions, we encourage agents to move in the same direction, reducing the number of \textit{head-on} collisions, which happen when two agents try to traverse through the same edge in opposite directions. Second, by increasing the cost of moving in areas prone to congestion, we motivate agents to navigate through less congested areas, ultimately reducing traffic congestion. 
\Cref{fig:front-fig} illustrates the traffic resulting from different guidance strategies.

One closely related work to our guidance graph is highways~\citep{Cohen2015FeasibilitySU,li2023study}, which are a subset of edges selected from graph $G$ with assigned directions and a lower traversal cost. This strategy incentivizes agents to move along the highways, reducing the number of collisions to be resolved by MAPF algorithms. However, the question of how to select these edges and determine their directions and costs remains largely unexplored. The \emph{crisscross} approach~\citep{lironPhDthesis}, where edge directions alternate in even and odd rows and columns, is common but not optimal. \citet{li2023study} show that, while crisscross highways speed up lifelong MAPF algorithms, they do not always improve throughput. Additionally, \citet{liron_highway16} proposed two methods to select edges and directions for highways, but neither of them outperforms crisscross highways.

Therefore, we introduce the Guidance Graph Optimization (GGO) problem to improve throughput by optimizing the edge weights of a guidance graph. We present two automatic GGO methods. The first applies Covariance Matrix Adaptation Evolutionary Strategy (CMA-ES)~\citep{hansen2016cmaes}, a state-of-the-art black-box optimizer, to solve GGO, but its solution is map-specific. 
The second method, Parameterized Iterative Update (PIU), uses CMA-ES to optimize an update model. This update model, represented by a neural network,  starts with an unweighted guidance graph and iteratively updates it with traffic information obtained from a lifelong MAPF simulator. It is capable of optimizing guidance graphs for different maps with similar layouts.

We make the following contributions:
(1) introducing the guidance graph, a versatile representation of guidance for lifelong MAPF, and guidance graph optimization (GGO) to improve its throughput, 
(2) conducting an in-depth study of various existing guidance works in MAPF, %
and (3) proposing two automatic GGO methods, CMA-ES and PIU,  showcasing their superior performance over unweighted graphs and previous guidance methods, along with the transferability of PIU to larger maps with similar layouts.

\section{Problem Definition and Preliminaries}

\subsection{Lifelong Multi-Agent Path Finding}

\begin{definition}[MAPF]
    The (one-shot) MAPF problem takes as inputs a graph $G(V,E)$
    and $k$ agents %
    with their start and goal locations.
    At each timestep, an agent can move to an adjacent vertex or stay at its current vertex. Two agents collide when they arrive at the same vertex or swap locations at the same timestep. The MAPF problem searches for collision-free paths that move all agents from their start to goal locations
    with minimum sum-of-cost, defined as the total number of move and wait actions that the agents need to take.
\end{definition}

\begin{definition}[Lifelong MAPF]
    Lifelong MAPF extends one-shot MAPF by constantly assigning new goals to agents when they reach their current ones. Lifelong MAPF searches for collision-free paths that maximize throughput, namely the average number of reached goals per timestep.
\end{definition}

\subsubsection{Lifelong MAPF Algorithms}
Solving MAPF optimally is known to be NP-hard~\citep{YuAAAI13}. Lifelong MAPF poses an even greater challenge as agents consistently receive new goals, requiring the continuous computation of new paths. Consequently, state-of-the-art algorithms approach lifelong MAPF by decomposing it into a series of (modified) one-shot MAPF instances, usually one at each timestep, assuming that minimizing their sum-of-costs enhances lifelong MAPF throughput. They can be divided into three categories. To show the generality of our GGO algorithms, we select a leading algorithm from each category to conduct our experiments.

\noindent \textbf{Replan All.} We replan all agents at every timestep (or every few timesteps)~\citep{WanICARCV18,Li2020LifelongMP}. In each replanning cycle, we solve a MAPF instance with the start locations being the current locations of all agents and the goals being their current goals.
We select RHCR~\citep{Li2020LifelongMP} as a representative algorithm from this category.

\noindent \textbf {Replan New.} This category is similar to the previous one except that, at every timestep, we replan only agents that have just reached their current goals and have been assigned new goals~\citep{CapICAPS15,MaAAMAS17,GrenouilleauICAPS19,LiuAAMAS19}. Since agents being replanned must avoid collisions with agents not being replanned, methods in this category need to impose constraints on the map structure and goal locations, often denoted as well-formed maps, to ensure the existence of collision-free paths.  
We select Dummy Path Planning (DPP)~\citep{LiuAAMAS19,Li2020LifelongMP} as a representative algorithm from this category.

\noindent \textbf{Reactive.} In contrast to the previous two categories, reactive methods plan paths for each agent without considering collisions with other agents (resulting in paths with no wait actions) and then resolve collisions reactively through pre-defined rules~\citep{WangICAPS08,okumura2019priority,Yu2023}, such as inserting wait actions or taking short detours. 
We select PIBT~\citep{okumura2019priority} as a representative algorithm from this category. It is complete on biconnected graphs and runs significantly faster than algorithms in other categories.

\begin{table*}[!t]
    \centering
    \resizebox{0.95\linewidth}{!}{
    \begin{tabular}{l|l|ccc|ccc|c}
        \toprule
                       & & \multicolumn{3}{c|}{Representation} & \multicolumn{3}{c|}{Generation} & Usage\\ 
          &                                    & Edge direction & Move cost & Wait cost  & Design & MAPF & Online Update & Method\\
        \midrule
        1 & \citet{Jansen2008DirectionMF}      & soft           & $\mathbb{R^+}$ & N/A            & handcrafted procedure & lifelong & Yes & reactive %
        \\
        2 & \citet{WangICAPS08}                & strict         & 1              & N/A              & crisscross            & one-shot & No  & reactive %
        \\
        3 & \citet{Cohen2015FeasibilitySU}     & soft           & 1 or $c$     & 1              & crisscross            & one-shot & No  & ECBS %
        \\
        4 & \citet{liron_highway16}            & soft           & 1 or $c$     & 1              & handcrafted procedure & one-shot & No  & ECBS %
        \\
        5 & \citet{han2022spaceutil}           & soft           & $\mathbb{R^+}$ & N/A & handcrafted procedure             & both & Yes  & reactive \& RHCR \\
        6 & \citet{li2023study}                & soft           & 1 or $c$     & 1              & crisscross            & lifelong & No  & RHCR %
        \\
        7 & \citet{Yu2023}                     & soft           & $\mathbb{R^+}$ & N/A            & handcrafted procedure & lifelong & Yes & reactive %
        \\
        8 & \citet{ChenAAAI24}                 & soft           & $\mathbb{R^+}$ & N/A            & handcrafted procedure & both     & Yes & PIBT \\
        9 & GGO (ours)                         & soft           & $\mathbb{R^+}$ & $\mathbb{R^+}$ & automatic             & lifelong & No  & many \\
        \bottomrule
    \end{tabular}
    }
    \caption{Overview of previous works on representation, generation, and usage of guidance in MAPF. For edge direction, strict means movement is unidirectional along each edge, and soft means bidirectional. Move cost refers to the cost of moving to an adjacent vertex and wait cost refers to the cost of waiting at the current vertex. For move cost, ``1 or $c$'' means that the value considers 1 and a scalar $c > 1$ only.
    For design, handcrafted procedure refers to using a manually designed procedure to generate guidance, while crisscross refers to the popular human-designed guidance~\citep{lironPhDthesis}.
    }
    \label{tab:guidance_mapf}
\end{table*}

\subsection{Guidance Graph Optimization}

To maintain generality, we consider graph $G(V,E)$ to be either directed, undirected, or mixed. We use $\{u,v\}$ and $(u,v)$  to denote an undirected edge and a directed edge, respectively. We use $E_{und}$ and $E_{dir}$ to denote the subsets of edges in $E$ that are undirected and directed, respectively. 

\begin{definition}[Guidance Graph]
    Given a graph $G(V,E)$ for lifelong MAPF,
    we define a guidance graph as a directed weighted graph $G_g(V_g,E_g,\boldsymbol{\omega})$ with the same vertex set $V_g = V$. Each edge in $E_g$ corresponds to an action that an agent can take at each vertex, with the edge weight indicating the action cost. Formally, we define $E_g = E_{wait} \cup E_{move}$ with
    \begin{align}
        E_{wait} &= \bigcup_{v \in V}\{(v,v)\} \\
        E_{move} &= 
                \{\bigcup_{\{u,v\} \in E_{und}}\{(u,v),(v,u)\}\} \cup E_{dir}.
    \end{align}%
    All edges weights are collectively represented as a vector $\boldsymbol{\omega} \in \mathbb{R}^{|E_g|}_{>0}$. %
\end{definition}

\noindent\textbf{Planning with guidance graphs.} To utilize guidance graphs in lifelong MAPF, we redefine the sum-of-costs of the underlying (one-shot) MAPF instances as the sum of the action costs across all paths for all agents (instead of the total number of actions). This leads to a minor modification to existing lifelong MAPF algorithms. Specifically, when planning paths for each agent, instead of seeking the shortest path on $G$, we aim to find a cost-minimal path on $G_g$. This modification alters the MAPF objective without compromising feasibility.

\begin{definition}[Guidance Graph Optimization (GGO)]
    Given a graph $G(V,E)$, an objective function $f: \mathbb{R}^{|E_g|} \rightarrow \mathbb{R}$, as well as predefined lower and upper bounds $\boldsymbol{\omega_{lb}}$ and $\boldsymbol{\omega_{ub}}$ ($0 < \boldsymbol{\omega_{lb}} \le \boldsymbol{\omega_{ub}}$) for edge weights, the GGO problem searches for the optimal guidance graph $G_g^*(V_g,E_g,\boldsymbol{\omega^*})$ with
    \begin{equation}
        \boldsymbol{\omega^*} = \argmax_{\boldsymbol{\omega_{lb}} \leq \boldsymbol{\omega} \leq \boldsymbol{\omega_{ub}}}
        f(\boldsymbol{\omega}).
    \end{equation}
\end{definition}

In this paper, our objective function $f$ is a simulator that runs a given lifelong MAPF algorithm on a given guidance graph and returns the throughput.

\section{Guidance in MAPF} \label{sec:guidance-review}

While the term ``guidance'' has not been explicitly proposed in the MAPF literature, the concept of enforcing global guidance and rules to enhance MAPF has been employed by numerous works in various ways. 
In this section, we present a summary of these works and provide a comprehensive review of how they represent, generate, and utilize guidance. %
\Cref{tab:guidance_mapf} shows a comparison between them and our GGO. We refer to them by their indices in the table for the rest of this section.

\noindent \textbf{Representing Guidance.} Previous works primarily represent guidance through modified edge directions or movement costs. They are all particular cases within the definition of our guidance graph, and none of them consider varying wait costs. We roughly divide them into 4 categories.
(1) Inspired by potential-field and flow-field methods used in swarm robotics, Work 1 represents guidance through a direction map. This map assigns a direction vector to every vertex of $G$ and sets the movement cost along an edge to be inversely related to the dot product of its direction vector and the vector of the edge, encouraging agents to move along the direction vectors. 
Consequently, a direction map can be transformed into a guidance graph with edge weights defined as dot products mentioned above.
(2) Work 2 turns some undirected edges into unidirectional, strictly prohibiting agents from moving against the assigned edge directions. This can be seen as a special guidance graph with infinite weights for constrained edges and 1 for others.
(3) Works 3, 4, and 6 use the highway idea that converts undirected edges into directed edges in both directions and then selects a subset of directed edges to be highways. They assign a weight of 1 for all highway edges and a weight of a predefined constant $c > 1$ for other edges, encouraging agents to move along the highway edges. Highways are special guidance graphs with restrictions on the values of movement and wait costs. 
(4) Work 5 uses a temporal heuristic function to estimate the movement cost between adjacent vertices at each timestep. This temporal heuristic function can be viewed as a time-extended guidance graph where edge weights can be different at different timesteps. However, the time-extended guidance graph is only applicable while an online update mechanism is incorporated (see below for details).
(5) Works 7 and 8 represent guidance similarly to our guidance graphs, with the distinction that they do not allow self-edges and thus cannot represent wait costs.

\noindent \textbf{Generating Guidance.} 
An important distinction between our work and previous works is that we are the first to propose an automated method for generating guidance. All previous works either use handcrafted guidance, such as crisscross highways (Works 2, 3, and 6), or use handcrafted procedures to generate guidance from a heatmap or a similar data structure that predicts traffic flows (Works 1, 4, 5, 7, and 8). 
More specifically, Work 1 computes direction vectors (of the direction map) from past traffic flows and then uses a handcrafted equation to convert them into movement costs. 
Work 4 introduces two methods, GM and HM, for generating highways. GM uses a graphic model~\citep{koller2009PGM} with a number of handcrafted features obtained from the estimated traffic flow. HM converts the estimated traffic flow into a score for each edge using a handcrafted score function and selects edges based on a predefined score threshold. 
Works 5 and 8 collect the planned paths of all agents and convert them into movement costs through handcrafted equations. 
Last, Work 7 uses a data-driven model to predict traffic flow, or more specifically, the delays that the agents will encounter (due to collision avoidance etc.) and directly uses the predicted delays as movement costs.

Therefore, we select 4 baseline methods to generate guidance graphs in our experiments: (1) \emph{Unweighted}, where no guidance is used, (2) \emph{Crisscross}, (3) \emph{HM Cost}, adapted from HM in Work 4, and (4) \emph{Traffic Flow}, adapted from Work 8. We did not compare Work 1, as HM from Work 4 is inspired by it. We did not compare GM from Work 4, as its performance is similar to HM. We did not compare Work 5 because the non-temporal version of their proposed guidance is similar to Works 4 and 8. We did not compare Work 7, because, while it obtains predicted traffic flow differently from Work 4, the procedure of converting predicted traffic flow to guidance is similar. 

Please note that the HM Cost and Traffic Flow used in our experiments do not use the original traffic flow models in their papers. This is because Work 4 tackles one-shot MAPF and predicts traffic flows by planning shortest paths between the start and goal locations of the agents, 
which is not realistic in lifelong MAPF as goal locations are unknown in advance. Work 8 tackles lifelong MAPF but assumes the guidance graph can be updated on the fly using real-time traffic information, %
while we assume that our guidance graph is optimized offline, and thus we do not have access to real-time traffic information. 
We focus on optimizing the guidance graphs offline because adding online adaptation not only requires additional computation to update the guidance graph but can also dramatically slow down path planning. This is because we need to either update the heuristic for the single-agent A$^*$ search when the guidance graph is updated or use a less informed heuristic without update. %
For example, Work 8 incorporates an online adaptation mechanism in PIBT, but it slows down the algorithm by 2-10 times.
Thus, in this paper, we use the same traffic flow model, namely the tile-usage map obtained from simulation, for both HM Cost and Traffic Flow methods. 
More details can be found in \Cref{appen:baseline}.

\begin{figure}[!t]
    \centering
    \includegraphics[width=0.48\textwidth]{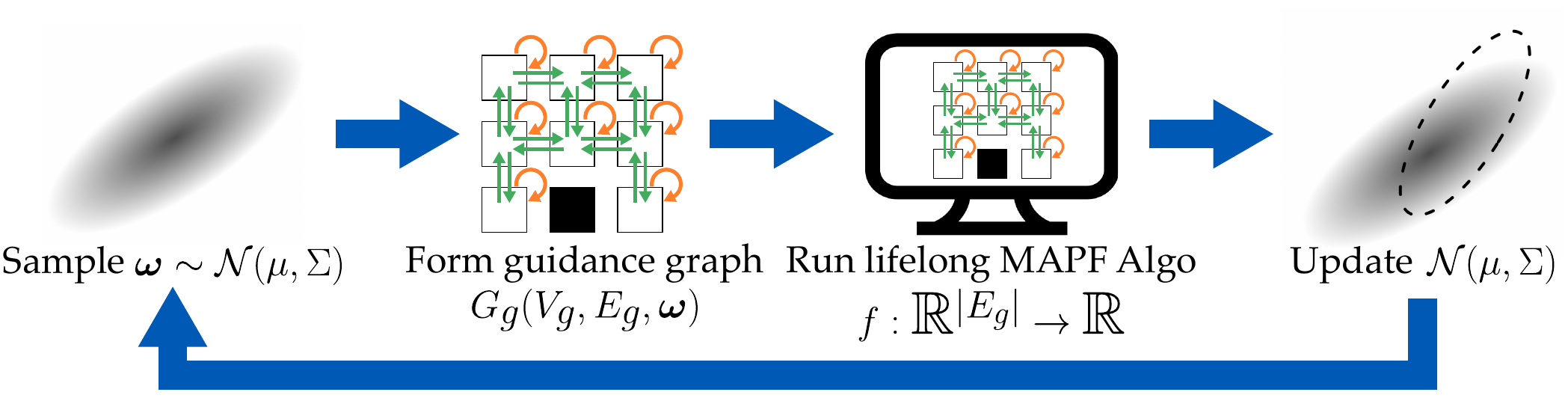}
    \caption{CMA-ES for GGO. The edge weights are iteratively sampled from a Gaussian distribution and then evaluated by a lifelong MAPF simulator. The simulated results are used to update the Gaussian distribution towards high-throughput regions.}
    \label{fig:cma-es_ggo}
\end{figure}

\noindent \textbf{Using Guidance.} All previous works study their guidance methods with a specific MAPF algorithm, such as ECBS~\citep{BarrerSoCS14} and RHCR~\citep{Li2020LifelongMP}, so it remains unclear whether and how well their methods can generalize to other MAPF algorithms. For example, an evident limitation of methods designed for reactive (lifelong) MAPF algorithms (namely Works 1, 2, 5, 7, and 8) is that, since paths planned by reactive methods do not include wait actions, these guidance methods, by design, do not define wait costs, making it non-trivial to extend them to (lifelong) MAPF algorithms in other categories. In contrast, we assess our GGO methods with three leading lifelong MAPF algorithms from different categories, thereby demonstrating their generality.

\section{Approach} \label{sec:approach}

We first introduce CMA-ES to solve GGO directly. Then we introduce Parameterized Iterative Update (PIU), which uses CMA-ES to optimize an update model that iteratively generates a guidance graph based on simulated traffic information.

\subsection{CMA-ES} 

CMA-ES~\citep{hansen2016cmaes} is a derivative-free, black-box optimization algorithm based on covariance matrix adaptation. 
\Cref{fig:cma-es_ggo} gives an overview of using CMA-ES to solve GGO. Specifically, we model the edge weight vector $\boldsymbol{\omega}$ as a 
multivariate Gaussian distribution. We then iteratively sample from the distribution for a new batch of  $b$ edge weight vectors, forming $b$ guidance graphs. We normalize each $\boldsymbol{\omega}$ to meet the bound constraint $\boldsymbol{\omega_{lb}} \le \boldsymbol{\omega} \le \boldsymbol{\omega_{ub}}$. We then evaluate each guidance graph by running $N_{e\_cma}$ simulations in a given lifelong MAPF simulator and computing the average throughput. The evaluated guidance graphs are ranked based on their throughput, and the top $N_{best}$ of them are used to update the mean and covariance of the Gaussian distribution. We run CMA-ES for $I$ iterations and return the guidance graph with the highest throughput as the solution.

\noindent \textbf{Handling Bounds through Normalization.} %
We use min-max normalization to enforce the bound constraint because it does not affect path-planning solutions. To prove it, consider two guidance graphs with edge weights $\boldsymbol{\omega_1}$ and $\boldsymbol{\omega_2}=C \cdot \boldsymbol{\omega_1}$, where $C \in \mathbb{R^+}$. Since the weight of every edge is scaled by the same scaler $C$, the paths returned by the lifelong MAPF algorithms with low-level single agent solvers minimizing the sum of edge weights do not change. We show additional experiments in \Cref{appen:bounds-handle} that min-max normalization yields better solutions than representative bounds handling methods introduced in a prior study~\citet{BIEDRZYCKI_cma-es_bounds2020}.

\subsection{Parameterized Iterative Update}

\begin{figure}[!t]
    \centering
    \includegraphics[width=0.45\textwidth]{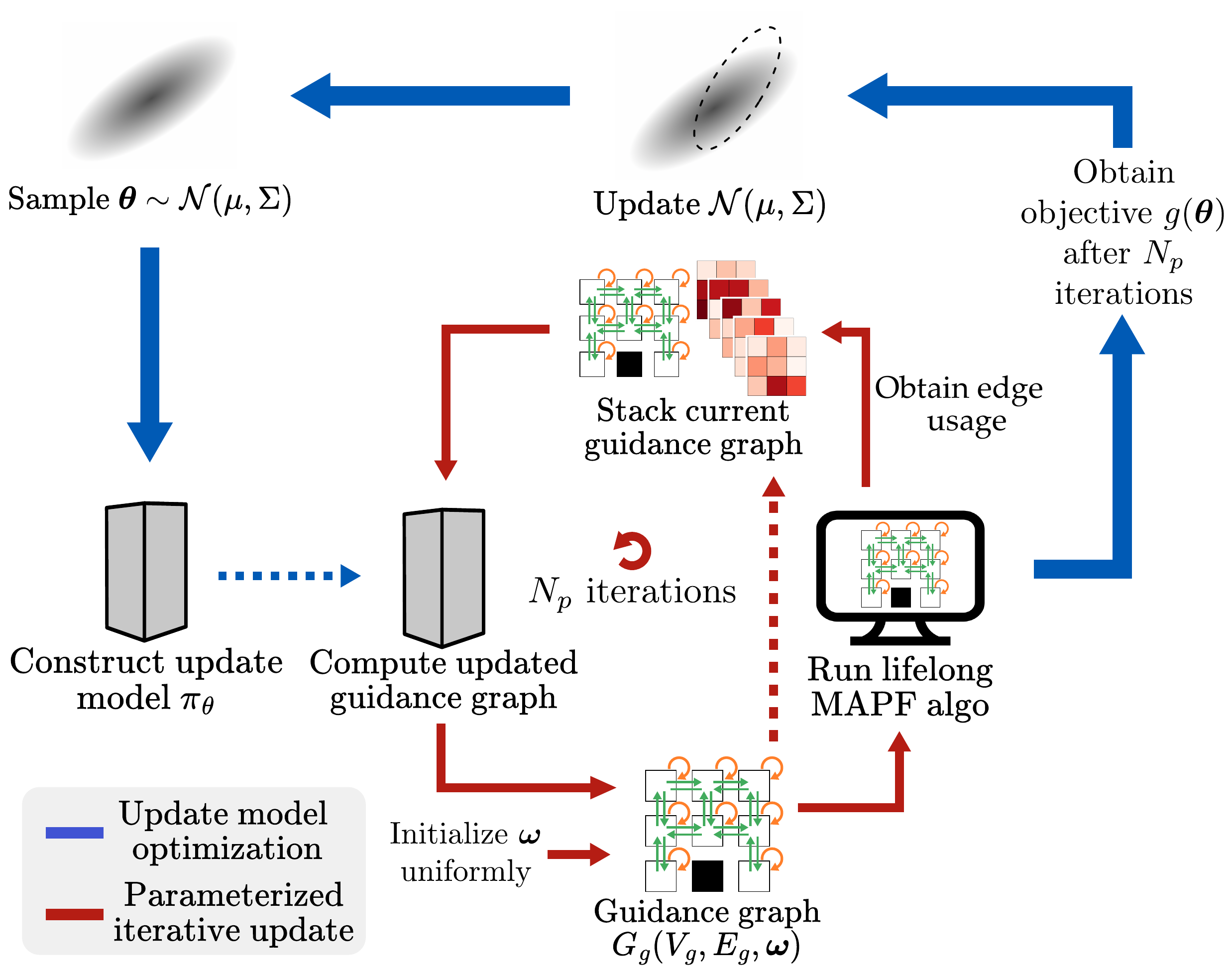}
    \caption{PIU for GGO. Starting with a guidance graph with uniform edge weights, we run MAPF simulations to get the edge usage. We then use an update model $\pi_{\boldsymbol{\theta}}$ to update the edge weights. We run this process iteratively for $N_p$ iterations.
    The update model $\pi_{\boldsymbol{\theta}}$ is optimized using CMA-ES.}
    \label{fig:piu}
\end{figure}

CMA-ES is known to scale poorly to high dimensional search spaces, making it challenging to optimize guidance graph for large maps.
Therefore, we propose Parameterized Iterative Update (PIU). \Cref{fig:piu} gives an overview of PIU, and \Cref{alg:cma-es-piu} provides the pseudocode. On a high level, PIU leverages a parameterized \textit{update model} to iteratively update the edge weights of the guidance graph using traffic information obtained from lifelong MAPF simulations. PIU can work with a wide variety of optimization methods. In this work, we choose to use CMA-ES to optimize the update model.

\begin{definition}[Update model]
    Given a guidance graph $G_g(V_g, E_g, \boldsymbol{\omega})$,
    an update model is a function $\pi_{\boldsymbol{\theta}}: \mathbb{R}^{|E_g|} \times\mathbb{R}^{|E_g|} \rightarrow \mathbb{R}^{|E_g|}$ that computes the updated edge weights $\boldsymbol{\omega'} \in \mathbb{R}^{|E_g|}_{>0}$ given the current edge weights $\boldsymbol{\omega} \in \mathbb{R}^{|E_g|}_{>0}$ and edge usage $U_{E_g} \in \mathbb{R}^{|E_g|}_{\geq0}$. 
    The edge usage is the frequency with which each edge is used by the agents in the lifelong MAPF simulation.
    The model $\pi_{\boldsymbol{\theta}}$ is parameterized by a vector $\boldsymbol{\theta} \in \boldsymbol{\Theta}$, where $\boldsymbol{\Theta}$ is the space of all parameters.
\end{definition}

\noindent \textbf{PIU.} The red loop in \Cref{fig:cma-es_ggo} gives an overview of PIU, and \Cref{alg:piu_f:1,alg:piu_f:2,alg:piu_f:3,alg:piu_f:4,alg:piu_f:avg1,alg:piu_f:5,alg:piu_f:avg2,alg:piu_f:avg4,alg:piu_f:6,alg:piu_f:7,alg:piu_f:8,alg:piu_f:init1,alg:piu_f:init2} of \Cref{alg:cma-es-piu} provides the pseudocode. 
We first form an update model $\pi_{\boldsymbol{\theta}}$ parameterized by a given parameter vector $\boldsymbol{\theta}$ (\Cref{alg:piu_f:2}).
We then 
start an iterative update procedure (\Cref{alg:piu_f:4,alg:piu_f:avg1,alg:piu_f:5,alg:piu_f:avg2,alg:piu_f:avg4,alg:piu_f:6,alg:piu_f:7,alg:piu_f:init1,alg:piu_f:init2,alg:piu_f:3}).
In each iteration, we either initialize the edge weights to $\mathbf{1}$ (\Cref{alg:piu_f:3}) in the first iteration or use the update model $\pi_{\boldsymbol{\theta}}$ to update the edge weights (\Cref{alg:piu_f:7}) in the following iterations.
We then construct the current guidance graph using $\boldsymbol{\omega}$ (\Cref{alg:piu_f:avg1}). 
We then run $N_{e\_piu}$ lifelong MAPF simulations (\Cref{alg:piu_f:5,alg:piu_f:avg2}), computing the average throughput $f$ and edge usage $U_{E_g}$ (\Cref{alg:piu_f:avg4,alg:piu_f:6}).
We run PIU for $N_p$ iterations. Finally, we return the throughput $f$ of the last iteration (\Cref{alg:piu_f:8}).

\begin{algorithm}[!t] 

\caption{Update model optimization}\label{alg:cma-es-piu}
\LinesNumbered
\SetKwInput{KwInput}{Input}             %
\SetKwInput{KwOutput}{Output}           %

\SetKwFunction{PIU}{PIU} \DontPrintSemicolon
  \KwInput
  {
  $\mu_0, \Sigma_0$: initial mean and covariance matrix of the multivariate Gaussian distribution.\newline
  $Np$: number of iterations to run in \PIU.
  $update\_gaussian$: function to update $\mu$ and $\Sigma$ according to evaluated parameter vectors.\newline
  $simulate$: function to run lifelong MAPF simulation and return edge usage $U_{E_g}$ and throughput $f$.
  }

\label{alg:piu_f:ini_umo}
Initialize $\mu \gets \mu_0, \Sigma \gets \Sigma_0, \boldsymbol{\theta^*} \gets$ NULL $,g^* \gets -\infty$\\
\label{alg:piu_f:10}
\For{$i \gets 1$ \KwTo $I$}
{
    \label{alg:piu_f:for_umo}
    Sample $b$ parameter
vectors $\boldsymbol{\theta}_1,...,\boldsymbol{\theta}_b \sim \mathcal{N}(\mu, \Sigma)$ \\
    \label{alg:piu_f:sample}
    \For{$k \gets 1$ \KwTo $b$}
    {
        \label{alg:piu_f:run_piu1}
        $g_k \gets $ \PIU($\boldsymbol{\theta}_k,N_p)$\\
        \label{alg:piu_f:run_piu2}
        \lIf{$g_k > g^*$}
        {
            \label{alg:piu_f:best2}
            $g^* \gets g_k, \boldsymbol{\theta^*} \gets \boldsymbol{\theta_k}$ \label{alg:piu_f:best3}
        }
    }
    $\mu,\Sigma \gets update\_gaussian(g_{1 \sim b},\boldsymbol{\theta_{1 \sim b}})$\\
    \label{alg:piu_f:up_g}
}
\Return $g^*,\boldsymbol{\theta^*}$\label{alg:piu_f:rt_umo}

\SetKwProg{Fn}{Function}{:}{}
\Fn{\PIU($\boldsymbol{\theta}$,$Np$)}{
    \label{alg:piu_f:1}
    Construct update model $\pi_{\boldsymbol{\theta}}$\\
    \label{alg:piu_f:2}

    \For{$j \gets 1$ \KwTo $N_p$}
    {
        \label{alg:piu_f:4}
        \leIf{$j = 1$}
        {
            \label{alg:piu_f:init1}
            $\boldsymbol{\omega} \gets \mathbf{1}$\\
            \label{alg:piu_f:3}
        }
        {
            \label{alg:piu_f:init2}
            $\boldsymbol{\omega} \gets \pi_{\boldsymbol{\theta}}(\boldsymbol{\omega}, U_{E_g})$
            \label{alg:piu_f:7}
        }
        Construct guidance graph $G_g(V_g,E_g,\boldsymbol{\omega})$ \\
        \label{alg:piu_f:avg1}
        \For{$q \gets 1$ \KwTo $N_{e\_piu} $}
        {
            \label{alg:piu_f:5}
            $U_{E_g}^{(q)}, f^{(q)} \gets simulate(G_g)$\\
        }
        \label{alg:piu_f:avg2}
        $U_{E_g} \gets \frac{1}{N_{e\_piu}}\sum_{q=1}^{N_{e\_piu}} U_{E_g}^{(q)}$\\
        \label{alg:piu_f:avg4}
        $f \gets \frac{1}{N_{e\_piu}}\sum_{q=1}^{N_{e\_piu}} f^{(q)}$\\
        \label{alg:piu_f:6}
    }
    \Return $f$
    \label{alg:piu_f:8}
}

\end{algorithm}

\noindent \textbf{Update Model Optimization.}
To train the update model, we run the PIU algorithm for $N_p$ iterations with update model $\pi_{\boldsymbol{\theta}}$ given parameters $\boldsymbol{\theta}$.
We search for optimal parameters $\boldsymbol{\theta^*} = \argmax_{\boldsymbol{\theta} \in \boldsymbol{\Theta}} \texttt{PIU}(\boldsymbol{\theta},N_p)$ using CMA-ES.
\Cref{alg:piu_f:10,alg:piu_f:for_umo,alg:piu_f:sample,alg:piu_f:run_piu1,alg:piu_f:run_piu2,alg:piu_f:up_g,alg:piu_f:best2,alg:piu_f:best3,alg:piu_f:rt_umo} of \Cref{alg:cma-es-piu} show the pseudocode. Starting with a given initial multivariate Gaussian distribution (\Cref{alg:piu_f:10}), the algorithm samples $b$ parameter vectors (\Cref{alg:piu_f:sample}) and runs PIU with them (\Cref{alg:piu_f:run_piu1,alg:piu_f:run_piu2}). Based on the returned throughput values, it keeps track of the best update model (\Cref{alg:piu_f:best2,alg:piu_f:best3}) and updates the Gaussian distribution (\Cref{alg:piu_f:up_g}), starting a new iteration. The optimization ends after running the above process for $I$ iterations (\Cref{alg:piu_f:for_umo}). Finally, the algorithm returns the best update model and the corresponding throughput (\Cref{alg:piu_f:rt_umo}).

\noindent \textbf{On the Advantage of PIU.}
Compared to directly using CMA-ES, the advantage of optimizing the update model and using PIU to generate guidance graph is two-folds. First, optimizing the update model reduces the dimension of search space.
Although solving GGO directly with CMA-ES is versatile and applicable to various lifelong MAPF algorithms and maps, its effectiveness diminishes in high-dimensional search spaces, making it challenging to use CMA-ES to search for edge weights directly for large maps.
Specifically, CMA-ES employs a full-rank $n \times n$ covariance matrix to model its Gaussian distribution in an $n$-dimensional space, leading to quadratic increases in both time and space complexity~\citep{Varelas2018large_cma_es}. In the case of GGO, the number of edge weights of a guidance graph increases at least linearly with the number of vertices, while our update model maintains a consistent number of parameters regardless of the size of the guidance graph, offering a more scalable solution than directly applying CMA-ES.
Second, the optimized update model is not specific to the map it is optimized on. Different maps with similar layouts could potentially have similar high-throughput guidance graphs that can be generated by the same update model. The guidance graph optimized by CMA-ES, on the other hand, consists of edge weights for a specific map.

\section{Experimental Evaluation} \label{sec:exp}

In this section, we compare guidance graphs optimized by CMA-ES and PIU with various baselines and assess the capability of PIU to generate high-throughput guidance graphs for maps of larger sizes with similar layouts.

\subsection{Experiment Setup} \label{sec:exp_setup}

\begin{table}[!t]
    \centering
    \resizebox{\linewidth}{!}{
    \begin{tabular}{cccrrrrc}
    \toprule
    Setup & MAPF & Map & $|E_g|$ & $|E_{wait}|$ & $|E_{move}|$ & $N_a$ & GGO \\
    \midrule
    1 & \multirow{7}{*}[-0.5em]{PIBT} & \randomSmall     & 3,359 & 819 & 2,540     & \multirow{3}{*}{400}    & \multirow{7}{*}[-0.5em]{CMA-ES \& PIU} \\
    2 &                               & \warehouselargeW & 4,074 & 948 & 3,126     &                        &       \\
    3 &                               & \mazeSmall       & 3,484 & 790 & 2,694     &                        &       \\
    \cmidrule(lr){3-7}
    4 &                               & \emptyMid        & 11,328 & 2,304 & 9,024  & 1,000                  &       \\
    \cmidrule(lr){3-7}
    5 &                               & \roomLarge       & 14,340 & 3,232 & 11,108 & \multirow{2}{*}{1,500} &       \\
    6 &                               & \randomLarge     & 13,568 & 3,270 & 10,298 &                        &       \\
    \cmidrule(lr){3-7}
    7 &                               & \denSmall        & 11,227 & 2,445 & 8,782  & 1,200                  &       \\
    \midrule
    8 & RHCR                          & \warehouselargeW & 4,074 & 948 & 3,126     & 220                    & CMA-ES \\
    \midrule
    9 & DPP                           & \warehouseSmallR & 1,478 & 320 & 1,158     & 88                     & CMA-ES \\
    \midrule
    10 & PIBT                  & \warehouselargeW & 4,074 & 948 & 3,126     & 150   & CMA-ES \& PIU \\
    \bottomrule
    \end{tabular}
    }
    \caption{
    Summary of the experiment setup. $N_a$ is the number of agents. 
    $|E_g|$, $|E_{wait}|$, and $|E_{move}|$ are the number of wait edges, movement edges, and all edges in the guidance graph, respectively. 
    Setups 1 to 9 compare our optimized guidance graphs with the baselines.    
    Setup 10 compares PIBT with GGO against RHCR without GGO when there are fewer agents.
    }
    \label{tab:exp-setup}
\end{table}

\begin{table}[!t]
    \centering
    \resizebox{0.95\linewidth}{!}{
        \begin{tabular}{ccrrr}
\toprule
Setup & MAPF + GGO &  SR & Throughput & CPU Runtime (s) \\
\midrule
\multirow{6}{*}{1} & \textbf{PIBT + CMA-ES} & $\textbf{100\%}$ & $\textbf{7.78} \pm \textbf{0.02}$ & $1.31 \pm 0.01$\\
  & PIBT + PIU & $\textbf{100\%}$ & $7.46 \pm 0.02$ & $1.29 \pm 0.02$\\
  & PIBT + Crisscross & $\textbf{100\%}$ & $6.84 \pm 0.02$ & $1.24 \pm 0.02$\\
  & PIBT + HM Cost & $\textbf{100\%}$ & $5.98 \pm 0.02$ & $\textbf{1.17} \pm \textbf{0.02}$\\
  & PIBT + Traffic Flow & $\textbf{100\%}$ & $7.43 \pm 0.02$ & $1.19 \pm 0.02$\\
  & PIBT + Unweighted & $\textbf{100\%}$ & $5.52 \pm 0.01$ & $1.20 \pm 0.02$\\
\midrule
\multirow{6}{*}{2} & \textbf{PIBT + CMA-ES} & $\textbf{100\%}$ & $\textbf{7.64} \pm \textbf{0.01}$ & $1.27 \pm 0.01$\\
  & PIBT + PIU & $\textbf{100\%}$ & $7.28 \pm 0.01$ & $1.22 \pm 0.01$\\
  & PIBT + Crisscross & $\textbf{100\%}$ & $6.65 \pm 0.01$ & $\textbf{1.21} \pm \textbf{0.01}$\\
  & PIBT + HM Cost & $\textbf{100\%}$ & $5.63 \pm 0.01$ & $1.24 \pm 0.01$\\
  & PIBT + Traffic Flow & $\textbf{100\%}$ & $5.84 \pm 0.01$ & $1.23 \pm 0.01$\\
  & PIBT + Unweighted & $\textbf{100\%}$ & $5.22 \pm 0.01$ & $1.25 \pm 0.01$\\
\midrule
\multirow{6}{*}{3} & PIBT + CMA-ES & $\textbf{100\%}$ & $1.40 \pm 0.03$ & $0.60 \pm 0.01$\\
  & \textbf{PIBT + PIU} & $\textbf{100\%}$ & $\textbf{1.47} \pm \textbf{0.02}$ & $0.60 \pm 0.01$\\
  & PIBT + Crisscross & $\textbf{100\%}$ & $1.18 \pm 0.03$ & $0.59 \pm 0.01$\\
  & PIBT + HM Cost & $\textbf{100\%}$ & $1.16 \pm 0.02$ & $0.65 \pm 0.01$\\
  & PIBT + Traffic Flow & $\textbf{100\%}$ & $0.95 \pm 0.02$ & $0.63 \pm 0.01$\\
  & PIBT + Unweighted & $\textbf{100\%}$ & $1.09 \pm 0.02$ & $\textbf{0.58} \pm \textbf{0.01}$\\
\midrule
\multirow{6}{*}{4} & PIBT + CMA-ES & $\textbf{100\%}$ & $24.04 \pm 0.01$ & $2.86 \pm 0.05$\\
  & \textbf{PIBT + PIU} & $\textbf{100\%}$ & $\textbf{25.98} \pm \textbf{0.01}$ & $\textbf{2.28} \pm \textbf{0.03}$\\
  & PIBT + Crisscross & $\textbf{100\%}$ & $23.84 \pm 0.02$ & $2.82 \pm 0.04$\\
  & PIBT + HM Cost & $\textbf{100\%}$ & $20.90 \pm 0.02$ & $2.79 \pm 0.05$\\
  & PIBT + Traffic Flow & $\textbf{100\%}$ & $19.90 \pm 0.02$ & $2.73 \pm 0.04$\\
  & PIBT + Unweighted & $\textbf{100\%}$ & $19.48 \pm 0.03$ & $2.78 \pm 0.04$\\
\midrule
\multirow{6}{*}{5} & PIBT + CMA-ES & $\textbf{100\%}$ & $3.12 \pm 0.01$ & $4.91 \pm 0.06$\\
  & \textbf{PIBT + PIU} & $\textbf{100\%}$ & $\textbf{3.13} \pm \textbf{0.01}$ & $4.59 \pm 0.05$\\
  & PIBT + Crisscross & $\textbf{100\%}$ & $2.75 \pm 0.01$ & $4.67 \pm 0.07$\\
  & PIBT + HM Cost & $\textbf{100\%}$ & $2.41 \pm 0.01$ & $4.71 \pm 0.05$\\
  & PIBT + Traffic Flow & $\textbf{100\%}$ & $2.87 \pm 0.01$ & $\textbf{4.55} \pm \textbf{0.05}$\\
  & PIBT + Unweighted & $\textbf{100\%}$ & $2.51 \pm 0.01$ & $4.95 \pm 0.06$\\
\midrule
\multirow{6}{*}{6} & \textbf{PIBT + CMA-ES} & $\textbf{100\%}$ & $\textbf{9.00} \pm \textbf{0.07}$ & $3.52 \pm 0.06$\\
  & PIBT + PIU & $\textbf{100\%}$ & $8.43 \pm 0.11$ & $4.25 \pm 0.07$\\
  & PIBT + Crisscross & $\textbf{100\%}$ & $7.31 \pm 0.09$ & $3.55 \pm 0.05$\\
  & PIBT + HM Cost & $\textbf{100\%}$ & $6.45 \pm 0.10$ & $3.75 \pm 0.06$\\
  & PIBT + Traffic Flow & $\textbf{100\%}$ & $6.00 \pm 0.12$ & $\textbf{3.39} \pm \textbf{0.05}$\\
  & PIBT + Unweighted & $\textbf{100\%}$ & $6.01 \pm 0.09$ & $3.45 \pm 0.05$\\
\midrule
\multirow{6}{*}{7} & \textbf{PIBT + CMA-ES} & $\textbf{100\%}$ & $\textbf{4.98} \pm \textbf{0.01}$ & $2.71 \pm 0.04$\\
  & PIBT + PIU & $\textbf{100\%}$ & $4.87 \pm 0.01$ & $\textbf{2.61} \pm \textbf{0.03}$\\
  & PIBT + Crisscross & $\textbf{100\%}$ & $4.16 \pm 0.01$ & $3.27 \pm 0.05$\\
  & PIBT + HM Cost & $\textbf{100\%}$ & $3.99 \pm 0.02$ & $2.90 \pm 0.06$\\
  & PIBT + Traffic Flow & $\textbf{100\%}$ & $3.06 \pm 0.01$ & $2.73 \pm 0.06$\\
  & PIBT + Unweighted & $\textbf{100\%}$ & $3.05 \pm 0.01$ & $3.17 \pm 0.05$\\
\midrule
\multirow{5}{*}{8} & \textbf{RHCR + CMA-ES} & $\textbf{100\%}$ & $\textbf{6.58} \pm \textbf{0.04}$ & $\textbf{91.24} \pm \textbf{15.66}$\\
  & RHCR + Crisscross & $\textbf{100\%}$ & $5.59 \pm 0.20$ & $3771.71 \pm 659.20$\\
  & RHCR + HM Cost & $0\%$ & N/A & N/A\\
  & RHCR + Traffic Flow & $32\%$ & $3.57 \pm 0.20$ & $324.12 \pm 63.91$\\
  & RHCR + Unweighted & $0\%$ & N/A & N/A\\
\midrule
\multirow{5}{*}{9} & \textbf{DPP + CMA-ES} & $\textbf{100\%}$ & $\textbf{5.17} \pm \textbf{0.00}$ & $28.75 \pm 0.53$\\
  & DPP + Crisscross & $\textbf{100\%}$ & $4.76 \pm 0.01$ & $\textbf{17.36} \pm \textbf{0.31}$\\
  & DPP + HM Cost & $\textbf{100\%}$ & $4.32 \pm 0.01$ & $29.94 \pm 1.39$\\
  & DPP + Traffic Flow & $\textbf{100\%}$ & $4.07 \pm 0.00$ & $578.43 \pm 53.61$\\
  & DPP + Unweighted & $\textbf{100\%}$ & $4.34 \pm 0.01$ & $20.30 \pm 0.49$\\
\midrule
\multirow{4}{*}{10} & PIBT + CMA-ES & $\textbf{100\%}$ & $4.74 \pm 0.01$ & $0.58 \pm 0.00$\\
  & PIBT + PIU & $\textbf{100\%}$ & $4.77 \pm 0.01$ & $0.58 \pm 0.01$\\
  & PIBT + Unweighted & $\textbf{100\%}$ & $3.75 \pm 0.00$ & $\textbf{0.50} \pm \textbf{0.00}$\\
  & \textbf{RHCR + Unweighted} & $\textbf{100\%}$ & $\textbf{4.95} \pm \textbf{0.00}$ & $87.37 \pm 0.72$\\
\bottomrule
        \end{tabular}
    }
    \caption{Success rates (\textit{SR}), throughput, and CPU runtimes of the simulations on different guidance graphs.
   For RHCR and DPP, the success rate is the percentage of simulations that end without congestion. For PIBT, it is the percentage of simulations that end without timeout. %
   We measure the throughput and CPU runtime over only successful simulations.}
    \label{tab:numerical-result}
\end{table}

\begin{figure*}[!t]
    \centering
    \includegraphics[width=0.95\textwidth]{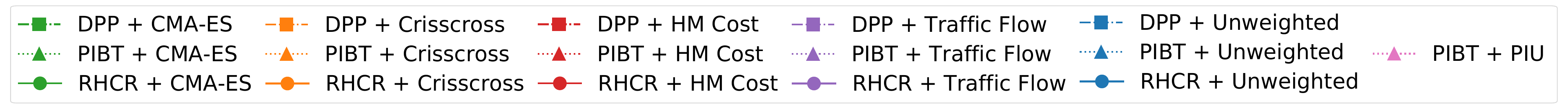}\\
     \begin{subfigure}{0.241\textwidth}
        \centering
        \includegraphics[width=1\textwidth]{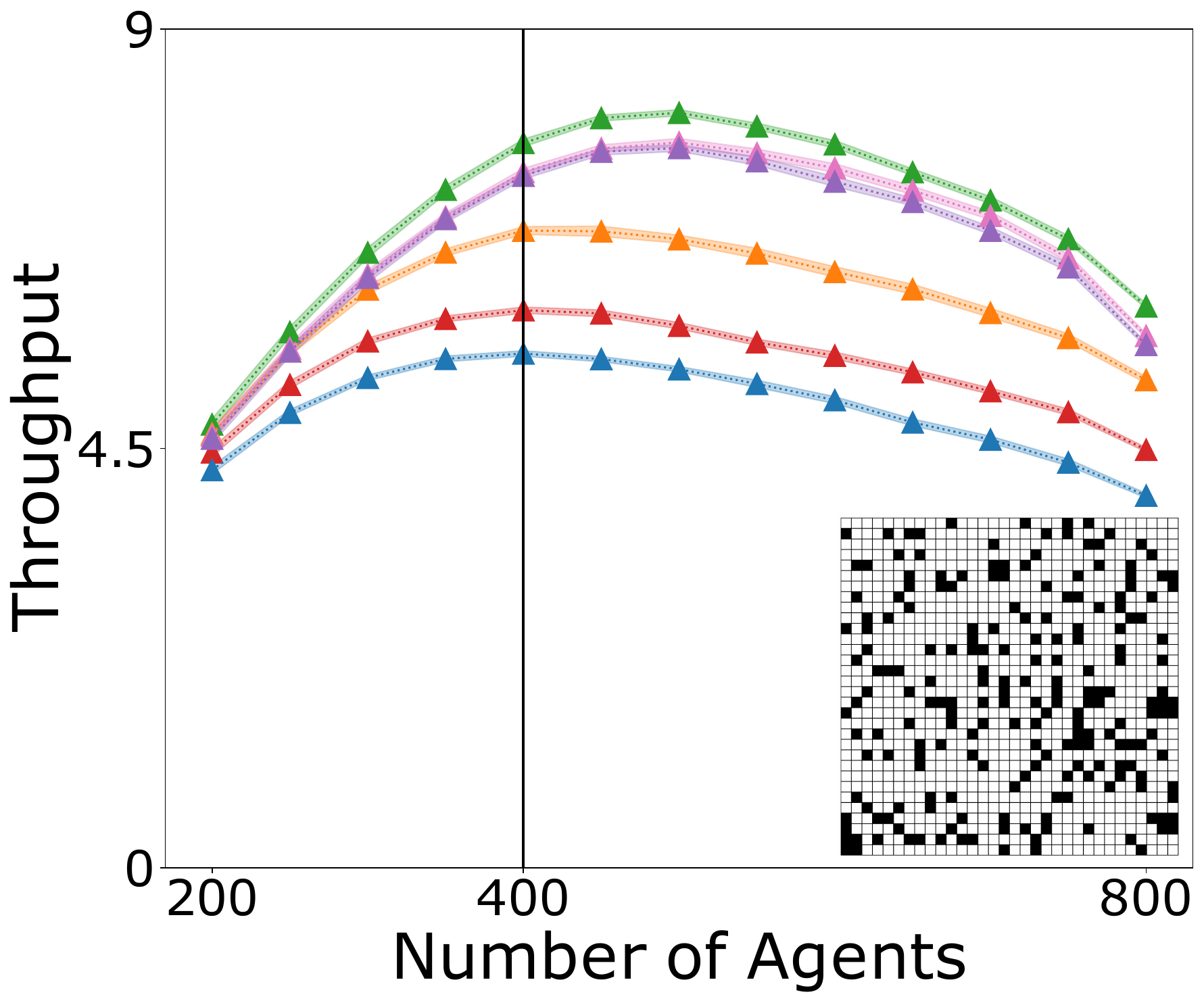}
        \caption{Setup 1: \randomSmall}
        \label{fig:random-32-32}
    \end{subfigure}%
    \hfill
    \begin{subfigure}{0.241\textwidth}
        \centering
        \includegraphics[width=1\textwidth]{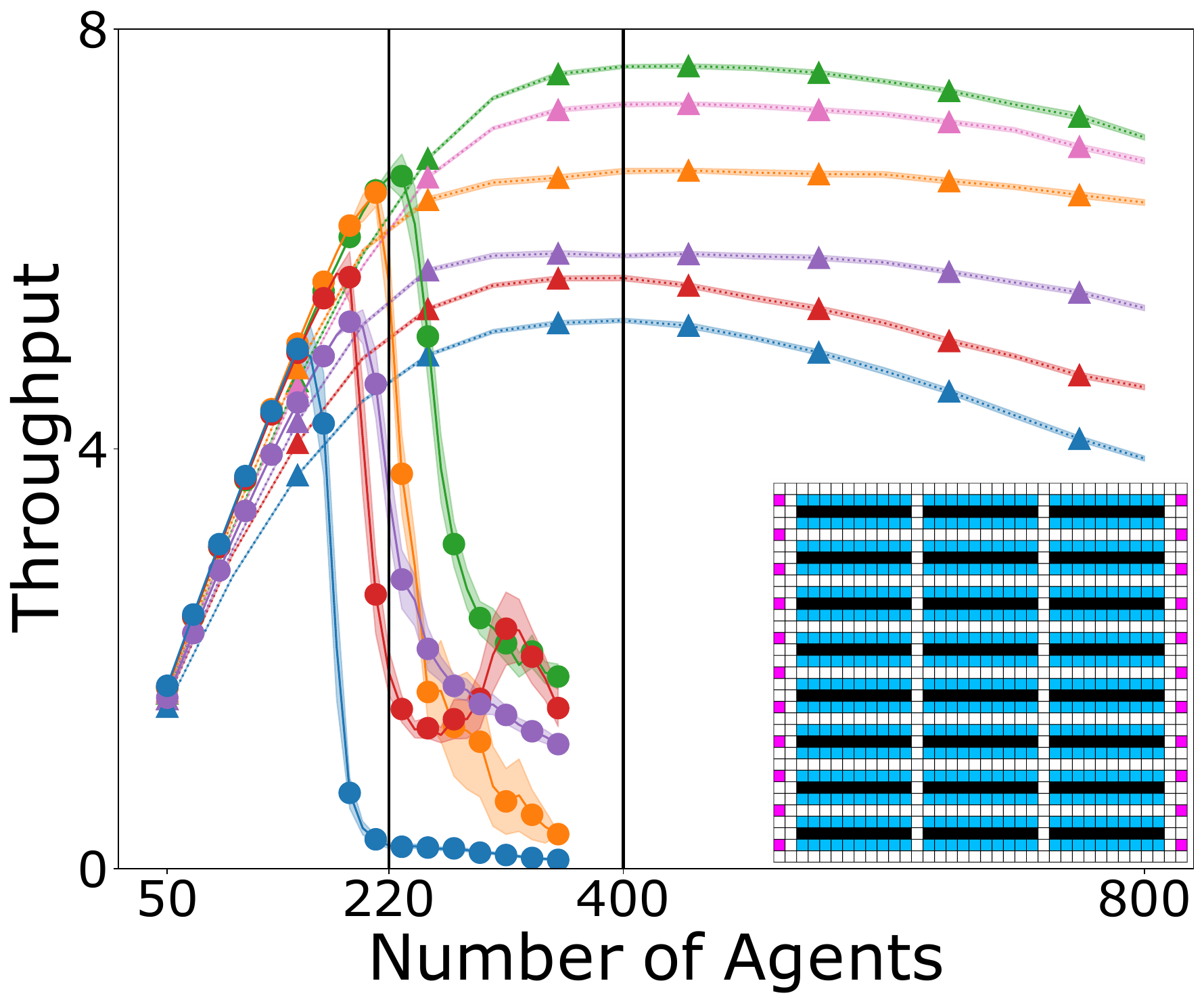}
        \caption{Setup 2 \& 8: \warehouselargeW}
        \label{fig:warehouse-large-w}
    \end{subfigure}
    \begin{subfigure}{0.241\textwidth}
        \centering
        \includegraphics[width=1\textwidth]{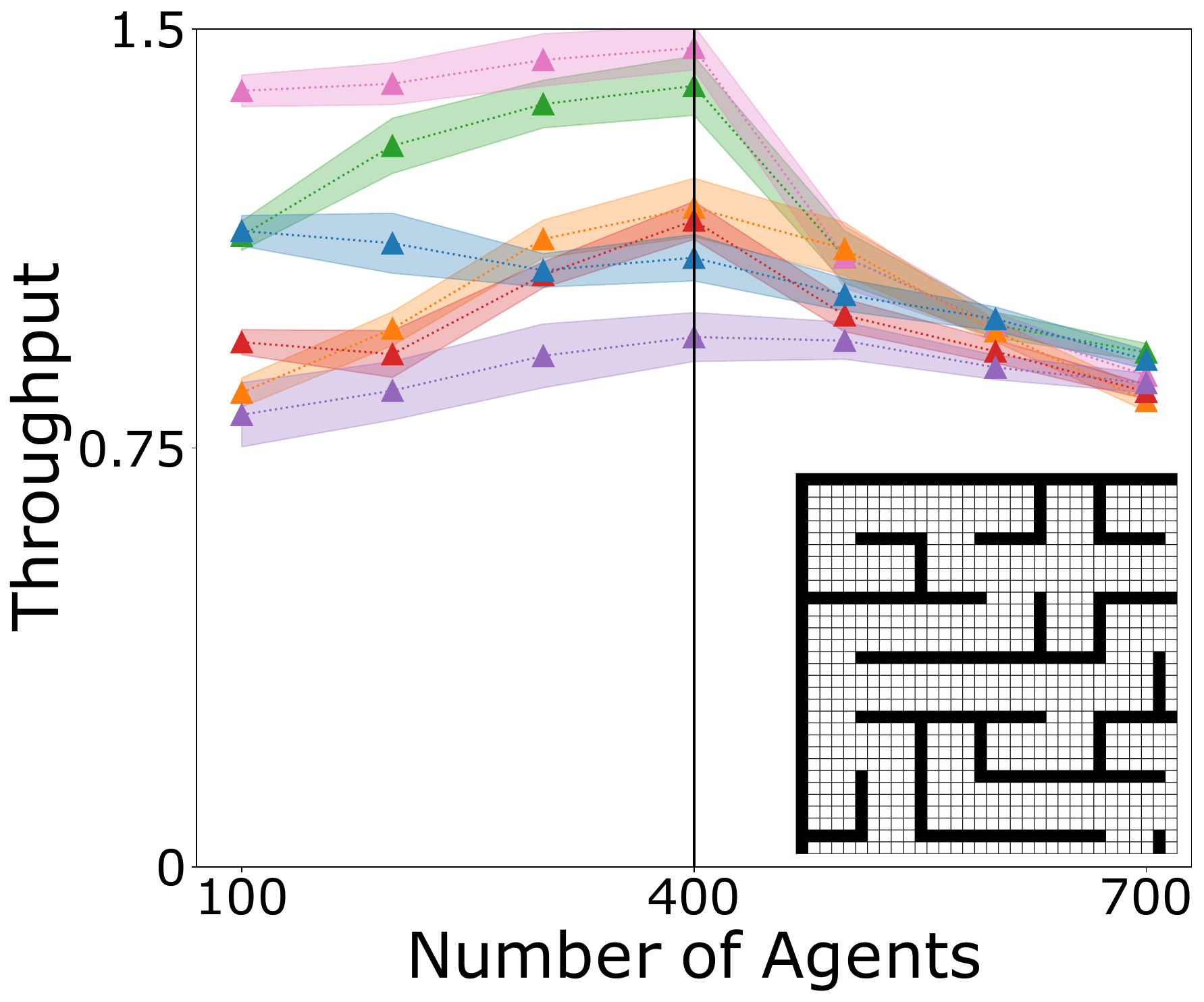}
        \caption{Setup 3: \mazeSmall}
        \label{fig:maze-32-32-4}
    \end{subfigure}%
    \hfill
    \begin{subfigure}{0.241\textwidth}
        \centering
        \includegraphics[width=1\textwidth]{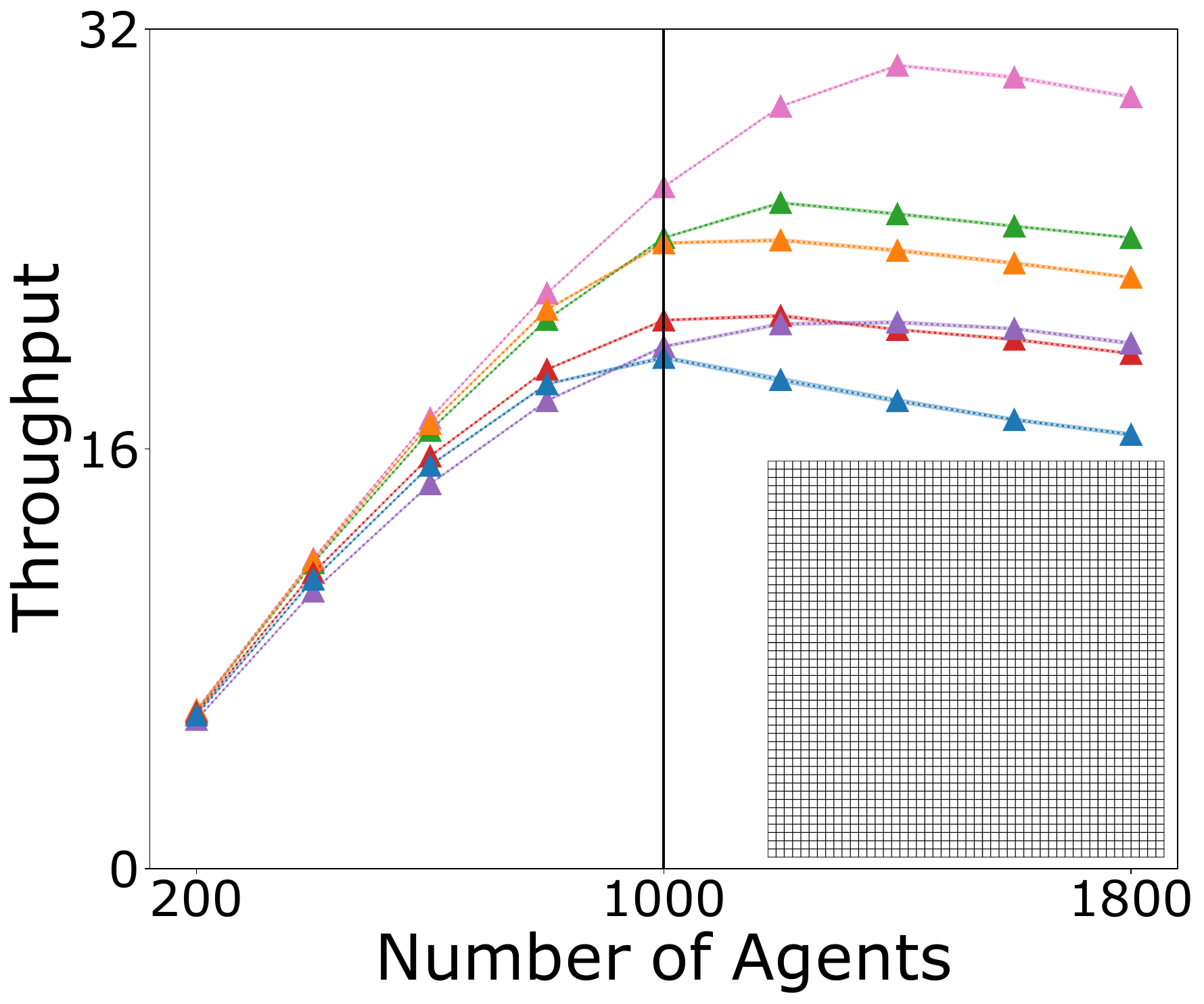}
        \caption{Setup 4: \emptyMid}
        \label{fig:empty-48-48}
    \end{subfigure}\\
    \hfill
    \begin{subfigure}{0.241\textwidth}
        \centering
        \includegraphics[width=1\textwidth]{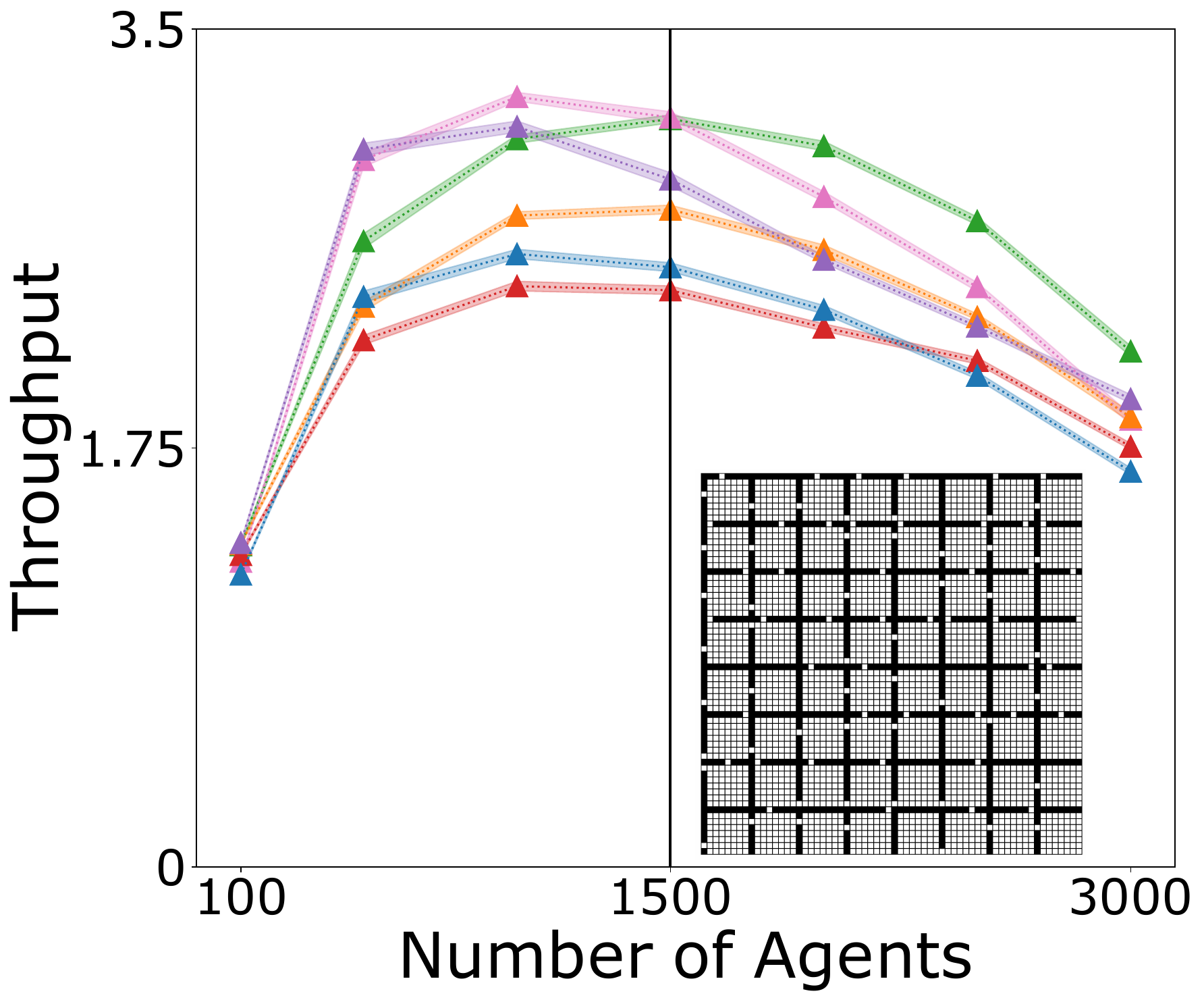}
        \caption{Setup 5: \roomLarge}
        \label{fig:room-64-64}
    \end{subfigure}
    \hfill
    \begin{subfigure}{0.241\textwidth}
        \centering
        \includegraphics[width=1\textwidth]{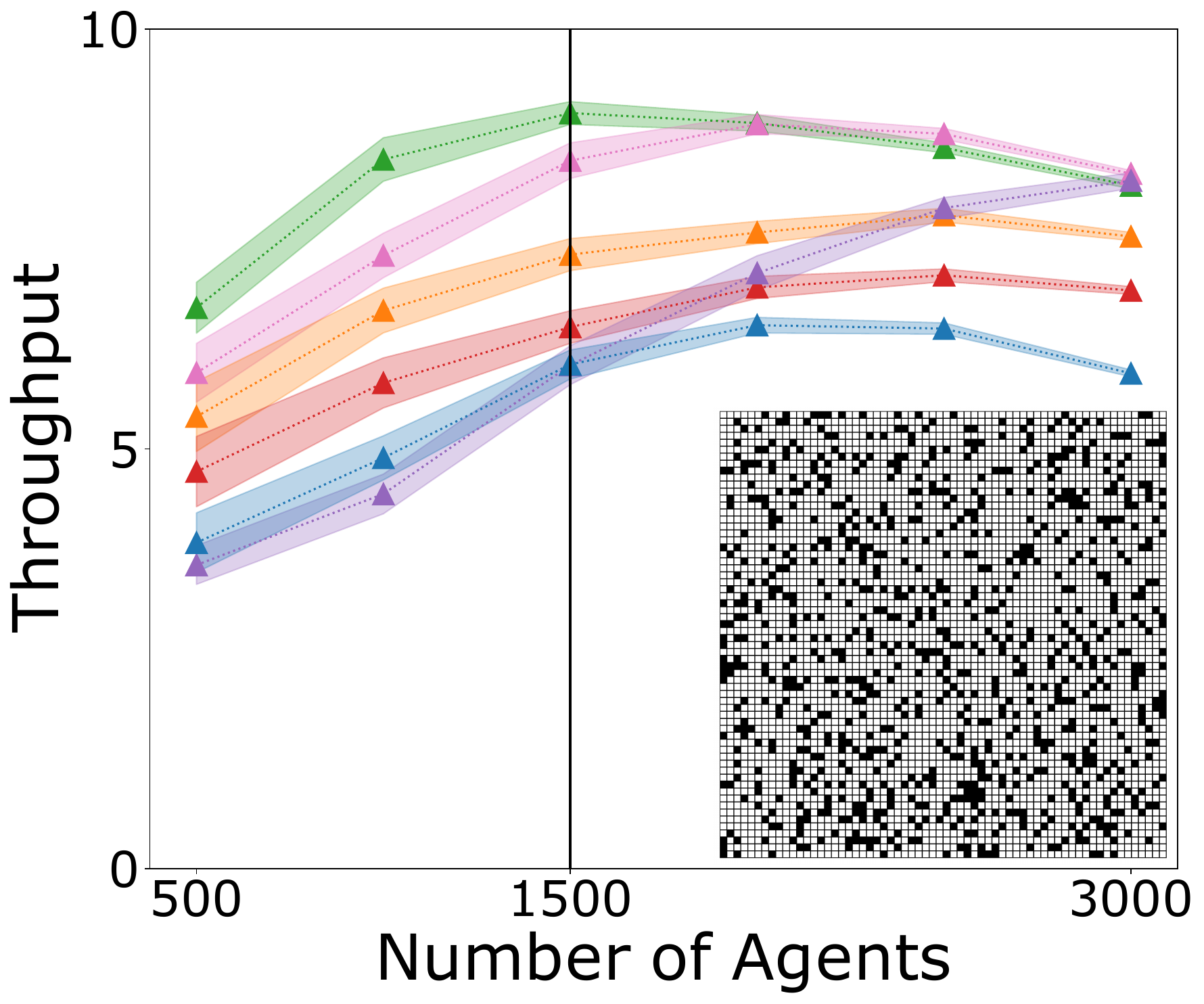}
        \caption{Setup 6: \randomLarge}
        \label{fig:random-64-64-20}
    \end{subfigure}%
    \hfill
    \begin{subfigure}{0.241\textwidth}
        \centering
        \includegraphics[width=1\textwidth]{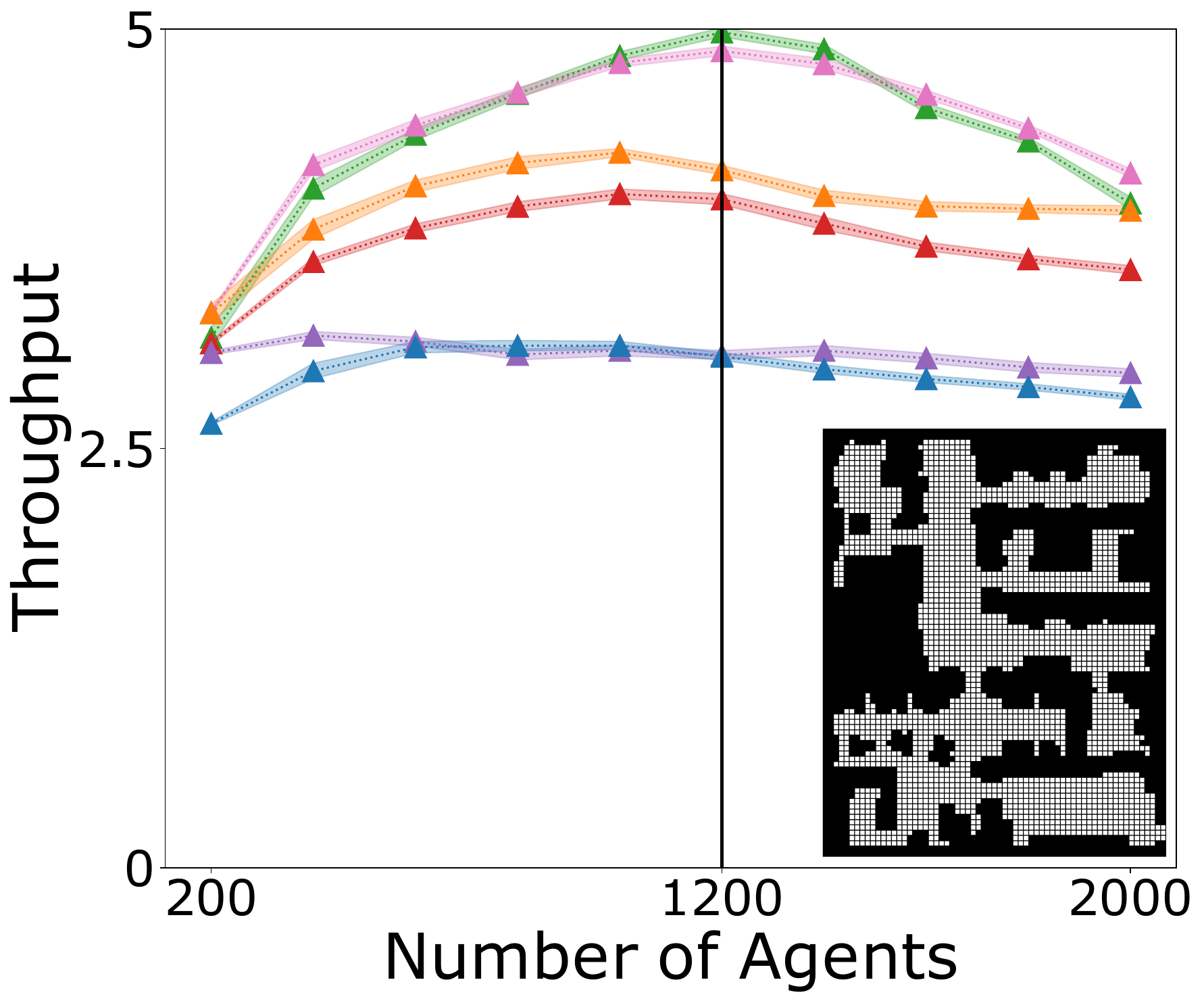}
        \caption{Setup 7: \denSmall}
        \label{fig:den312d}
    \end{subfigure}%
    \hfill
     \begin{subfigure}{0.241\textwidth}
        \centering
        \includegraphics[width=1\textwidth]{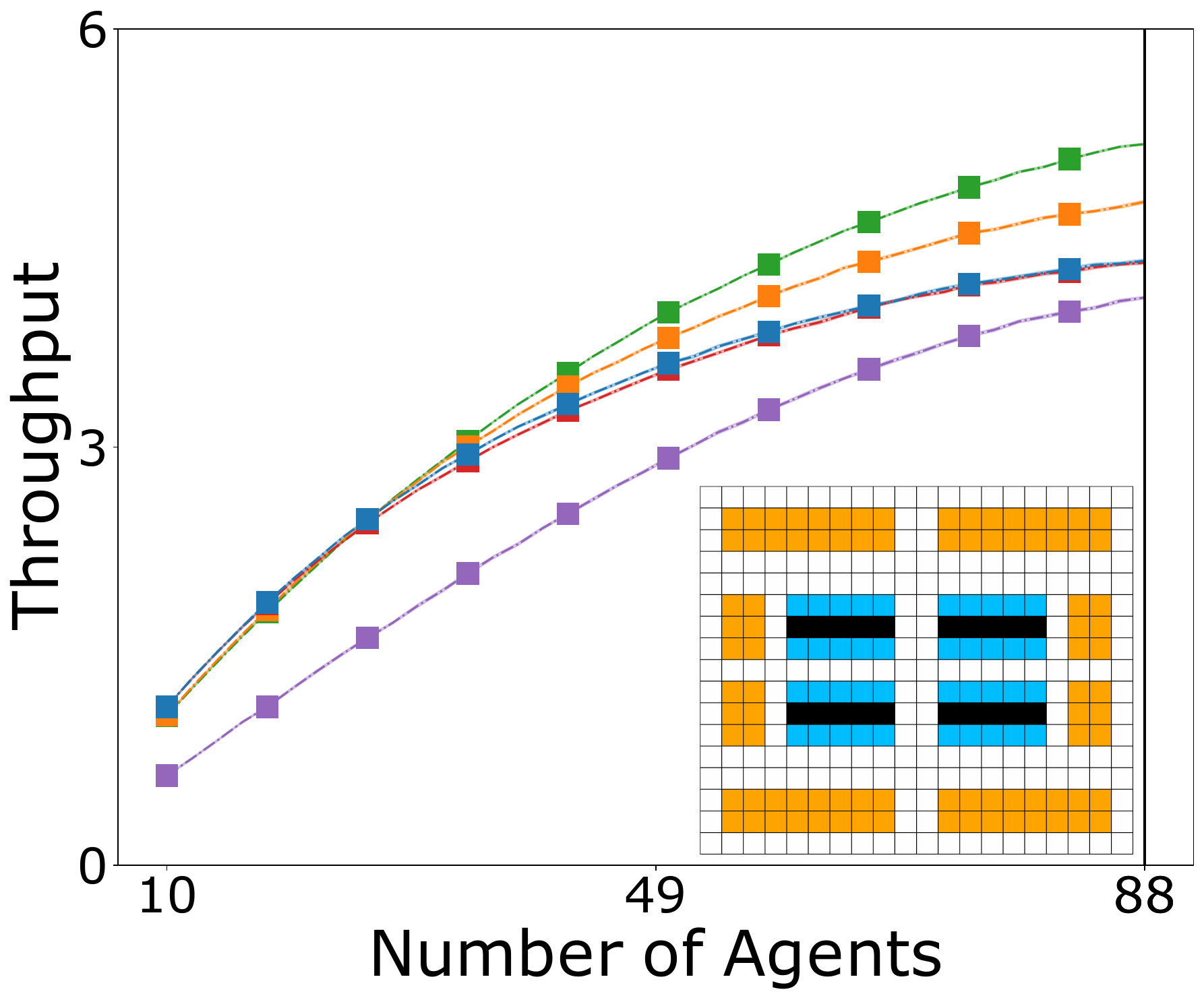}
        \caption{Setup 9: \warehouseSmallR}
        \label{fig:warehouse-small-r}
    \end{subfigure}%
    \caption{Throughput with different numbers of agents. The guidance graphs are optimized with $N_a$ agents, which is indicated by the black vertical lines. In (b), the black vertical lines at 220 and 400 agents indicate $N_a$ for setups 8 and 2, respectively.}
    \label{fig:major-result}
\end{figure*}

\noindent \textbf{General Setups.}
\Cref{tab:exp-setup} outlines our experimental setup. 
Column 2 shows the lifelong MAPF algorithms. Following the recommendations of previous works~\citep{Li2020LifelongMP,zhangLayout23}, we use PBS~\citep{MaAAAI19} and  SIPP~\citep{PhillipsICRA11} as the MAPF solver and the single-agent solver, respectively, in both RHCR and DPP and use $h=5$ and $w=10$ in RHCR.

Column 3 outlines the maps, all being 4-neighbor girds, including two warehouse maps (\warehouselargeW and \warehouseSmallR) from previous works~\citep{Li2020LifelongMP,zhangLayout23} and six maps (\randomSmall, \mazeSmall, \emptyMid, \roomLarge, \randomLarge, and \denSmall) from the MAPF benchmark~\citep{SternSoCS19}. We choose multiple maps for PIBT to demonstrate that both CMA-ES and PIU work for different maps. 
We show the maps at the corners of \Cref{fig:major-result}.
In all maps, black tiles are obstacles, and non-black tiles are traversable. In warehouse maps, orange tiles are home locations, blue tiles are endpoints, and purple tiles are workstations.
In \warehouseSmallR, agents start from orange tiles and move constantly between blue tiles. In \warehouselargeW, agents start from non-black tiles and move between blue and purple tiles alternatively. In all other maps, agents start from and move between white tiles. The agents' goals are assigned uniform randomly.

Columns 4 to 6 show the number of edges in the guidance graphs of the corresponding maps. In setups 1 to 7, we optimize all $|E_g| = |E_{wait}| + |E_{move}|$ edges.
In setups 8 and 9, however, SIPP cannot handle different wait costs at different vertices.
Therefore, we optimize the wait costs of all vertices as one variable, resulting in $|E_{move}| + 1 = 3,127$ and $1,159$ variables to be optimized in the guidance graphs in setups 8 and 9, respectively.
Column 7 is the number of agents used in lifelong MAPF simulations, with a larger $N_a$ for PIBT compared to RHCR and DPP to demonstrate that both CMA-ES and PIU work for congested scenarios.

Column 8 shows the GGO algorithms we run for each setup. We apply CMA-ES across all setups to demonstrate its versatility. However, due to computational constraints, we focus on using PIU primarily with PIBT. While both PIU and CMA-ES conduct the same number of simulations, there is a notable difference in their execution. In CMA-ES, all $N_{e\_cma}$ simulations in each guidance graph evaluation can be parallelized. In contrast, PIU runs the simulations sequentially, resulting in slower runtime.

We choose the hyperparameters of CMA-ES and PIU such that they run the same number of simulations to ensure a fair comparison. In particular, we set batch size $b = 100$ and the number of iterations $I = 100$ for both CMA-ES and PIU, resulting in a total of $b \times I = 10$k objective function evaluations for both algorithms. In each iteration, we select the top $N_{best} = 50$ solutions to update the Gaussian distribution. 
For CMA-ES, each evaluation runs $N_{e\_cma} = 5$ simulations, resulting in $50$k simulations. For PIU, each evaluation runs the PIU algorithm for $N_p=5$ iterations and each iteration runs $N_{e\_piu} = 1$ simulation, resulting in $50$k simulations, identical to CMA-ES. We run each simulation for 1,000 timesteps. For RHCR and DPP, we stop the simulation early in case of congestion, which happens if more than half of the agents wait at their current location.

\noindent \textbf{Update Model.} Given our use of grid maps in the experiments, we use a Convolutional Neural Network (CNN) as our update model, which can generate guidance graphs for maps of arbitrary sizes. The CNN has 3 convolutional layers of kernel sizes 3, 1, 1, respectively. Each layer is followed by a ReLU activation and a batch normalization layer. The update model has 4,231 parameters. For a map of dimension $h \times w$, we represent the edge weights $\boldsymbol{\omega}$ of the guidance graph as a tensor of size $h \times w \times 5$, where the first four channels are the movement costs and the last channel is the wait costs.

\noindent \textbf{Baselines.} As mentioned in \Cref{sec:guidance-review}, we have 4 baseline guidance graphs, namely (1) \textit{Unweighted}, (2) \textit{Crisscross}~\citep{lironPhDthesis}, (3) \textit{HM Cost}~\citep{liron_highway16}, and (4) \textit{Traffic Flow}~\citep{ChenAAAI24}. 
We discuss the methods of generating the baseline guidance graphs in \Cref{appen:baseline}.

\noindent \textbf{Evaluation.}
To evaluate PIU, we use the optimized update model to run PIU with $N_{e\_piu} = 1$ to generate the guidance graph. In \Cref{appen:piu-val}, we show that the choice of $N_{e\_piu}$ does not have significant impact on the throughput of the generated guidance graphs.
When we evaluate a guidance graph from CMA-ES, PIU, or baselines with a given number of agents, we run 50 simulations, each for 1,000 timesteps, and report the results with both means and standard errors.

\noindent \textbf{Implementation.} We implement the update model in PyTorch~\citep{Paszke2019PyTorchAI}, CMA-ES in Pyribs~\citep{pyribs}, and Traffic Flow and HM Cost guidance graph generation in Python. We implement the lifelong MAPF algorithms in C++ based on openly available implementation from previous works~\citep{Li2020LifelongMP,okumura2019priority}.

\noindent \textbf{Compute Resource.} We run our experiments on two machines: (1) a local machine with a 64-core AMD Ryzen Threadripper 3990X CPU, 192 GB of RAM, and an Nvidia
RTX 3090Ti GPU, and (2) a high-performing cluster with numerous 64-core AMD EPYC 7742 CPUs, each with 256 GB of RAM. We measure all CPU runtime on machine (1).

\subsection{Results}

\noindent \textbf{GGO vs Baselines.}
We first compare our optimized guidance graphs with the baseline guidance graphs. 
For each guidance graph, we run 50 simulations and report the numerical results in \Cref{tab:numerical-result} in the format of $x \pm y$, where $x$ is the average and $y$ is the standard error. Both CMA-ES and PIU outperform all baseline guidance graphs in all setups in terms of throughput. Specifically, CMA-ES outperforms all baseline guidance graphs in all setups, showing the versatility of the algorithm. 
For the baseline methods, the human-designed crisscross guidance performs quite well in setups 2, 3, 4, 6, 7, 8, and 9, outperforming all other baselines. Traffic Flow is more competitive in setups 1 and 5.
When comparing the throughput of CMA-ES and PIU, no clear winner emerges: CMA-ES wins in setups 1, 2, 6, and 7, PIU wins in 3 and 4, and they perform similarly in setup 5. %
\Cref{appen:gg-viz} visualizes all the optimized guidance graphs.

We also report CPU runtimes in \Cref{tab:numerical-result}, although runtimes are not the optimization objectives of our GGO algorithms. 
PIBT runs very fast and finishes all 1,000-timestep simulations in 5 seconds, implying each timestep takes less than 0.005 seconds. Therefore, the runtime difference between different methods is negligible in practice. RHCR and DPP are significantly slower than PIBT. For RHCR, CMA-ES leads to the best runtime. For DPP, it runs PBS to solve a one-shot MAPF instance for all agents in the first timestep and then, in each timestep, replans only for agents that have just reached their goals. We conjecture that the slower runtime of DPP with CMA-ES than baselines comes from the slower runtime in the first timestep. This is because the optimized guidance graph could encourage the agents to take longer paths in order to avoid congestion.

To further understand the performance of the optimized guidance graphs, we vary the number of agents and plot the throughput in \Cref{fig:major-result}. The trends are similar in all maps, with CMA-ES and PIU generally outperforming all baselines, except that Traffic Flow matches PIU with fewer agents In \randomSmall and \roomLarge. However, Traffic Flow is less competitive in all other maps, indicating that the performance of Traffic Flow depends on the map structures.

\begin{figure}[!t]
    \centering
    \includegraphics[width=0.44\textwidth]{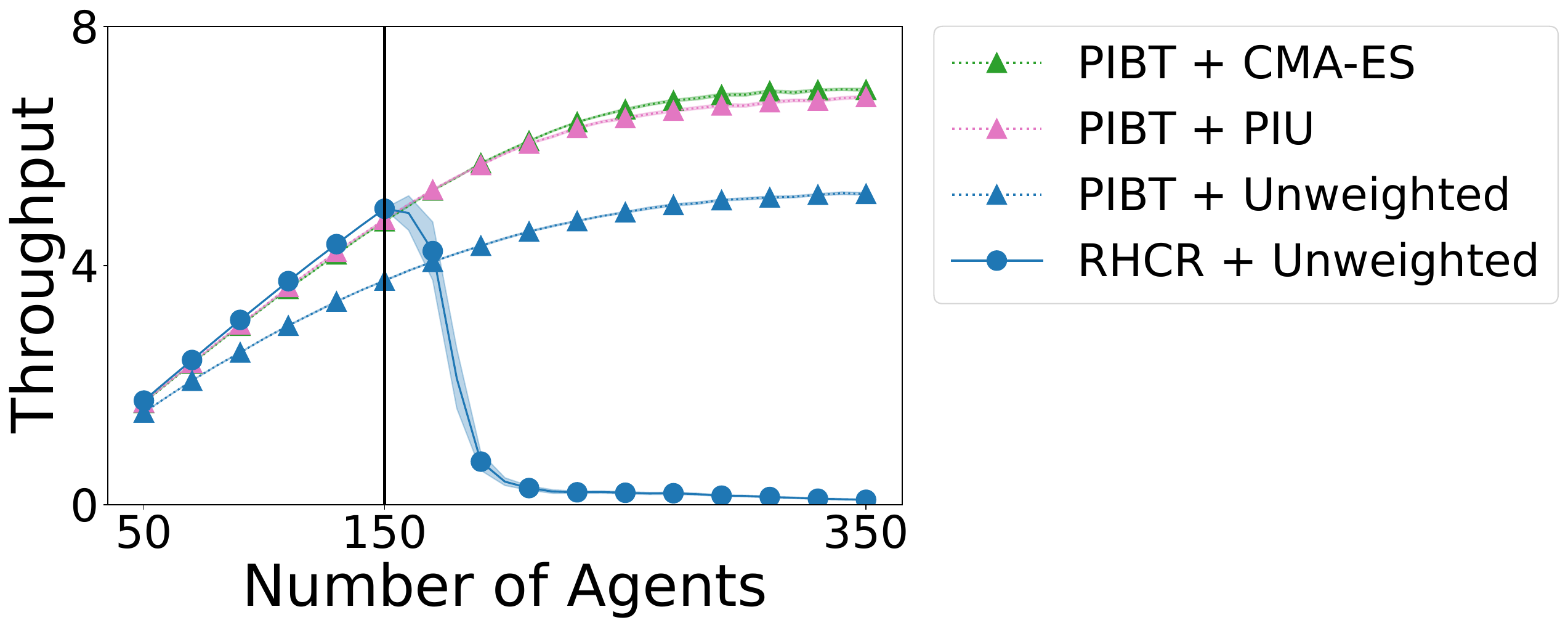}
    \caption{Setup 10: An optimized guidance graph enables PIBT to have competitive throughput with RHCR despite the advantage of RHCR with fewer agents.}
    \label{fig:rhcr_pibt_gap_150}
\end{figure}

\noindent \textbf{RHCR vs PIBT+GGO.}
Given the considerable runtime advantage of PIBT over RHCR (and DPP), we conduct an additional experiment, as detailed in setup 10 of \Cref{tab:exp-setup}, to explore whether PIBT with an optimized guidance graph can achieve higher throughput than RHCR without guidance graphs. %
We choose $N_a = 150$ because it is the largest number with which RHCR without guidance graphs can maintain a 100\% success rate.
Setup 10 in \Cref{tab:numerical-result} shows the numerical results. %
While RHCR still has the highest throughput, both GGO methods significantly reduce the throughput gap between PIBT and RHCR, from 24.2\% to less than 4.2\%.
Therefore, with the help of our optimized guidance graph, we can enable a greedy, distributed, yet extremely fast rule-based MAPF algorithm (PIBT) to achieve performance comparable to a centralized, computationally heavy, search-based MAPF algorithm (RHCR). %
\Cref{fig:rhcr_pibt_gap_150} further compares PIBT and RHCR with various numbers of agents. The throughput of RHCR quickly drops after 150 agents, while that of PIBT maintains an increasing trend with more agents.

\begin{figure}
    \centering
    \includegraphics[width=0.48\textwidth]{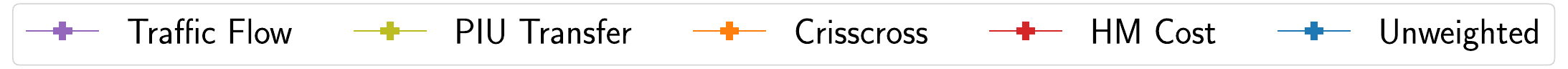}
    \includegraphics[width=0.48\textwidth]{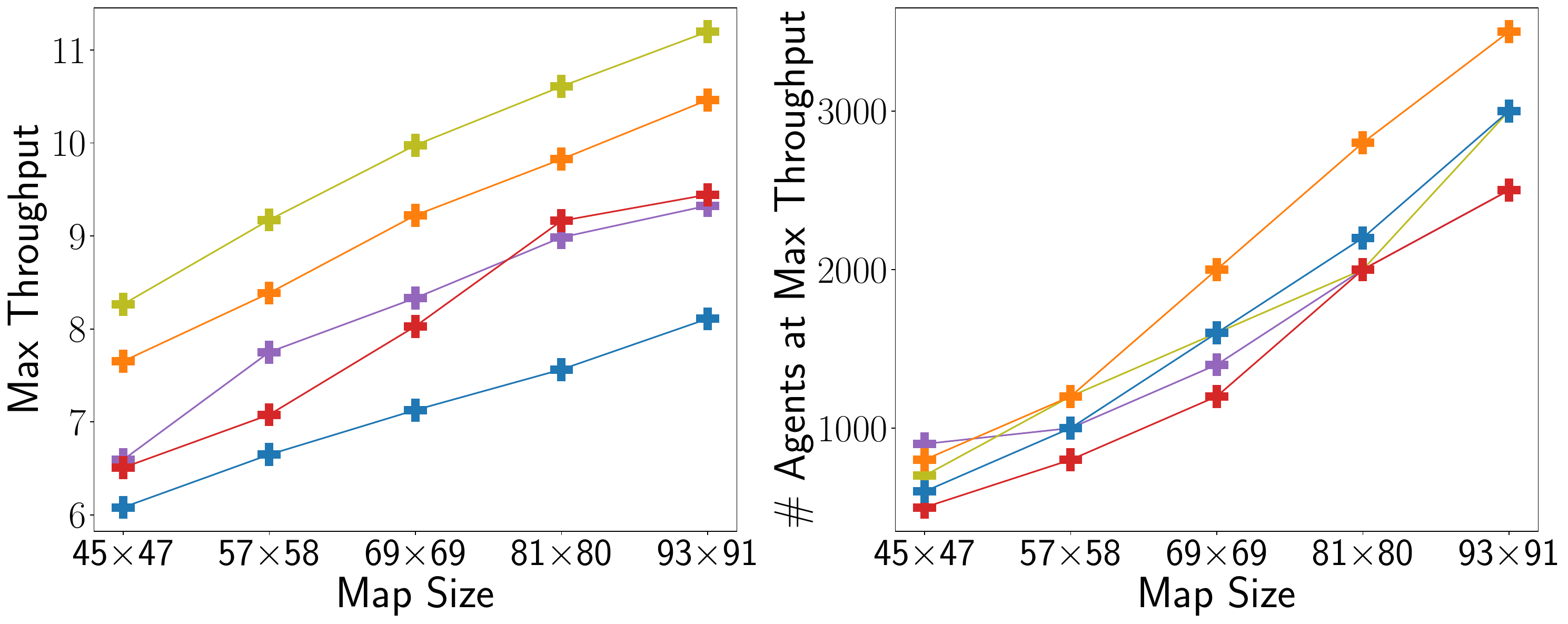}
    \caption{Max throughput and the number of agents at this maximum. PIU Transfer refers to using the update model optimized in setup 2 to generate guidance graphs.}
    \label{fig:scalability}
\end{figure}

\noindent \textbf{Update Model Transferability.}
We attempt to transfer the update model optimized with setup 2 to larger warehouse maps with similar layouts.  
We mimic the layout pattern of \warehouselargeW and design larger maps with sizes up to 93 $\times$ 91 by repeating blocks of 10 shelves and endpoints and placing workstations on the left and right borders.
\Cref{fig:warehouse-scale-map} in \Cref{appen:thr-large-warehouse} plots the resulting maps. 
We use the optimized update model from setup 2 to generate guidance graphs for these maps with an increasing number of agents, ranging from around 10\% to 90\% of the non-black tiles in the maps. We then run 50 simulations in each of the generated guidance graphs with PIBT. We plot the maximum throughput achieved in each map and the corresponding number of agents in \Cref{fig:scalability}, comparing PIU Transfer with baseline guidance graphs. We observe that PIU Transfer dominates all baselines with all sizes in terms of throughput. Notably, while crisscross has the second highest throughput across different map sizes, PIU Transfer can achieve higher throughput with an equal or smaller number of agents. 
\Cref{appen:thr-large-warehouse} shows the throughput with different numbers of agents in larger warehouse maps.

\section{Conclusion} 

We define guidance graphs and GGO to maximize the throughput of lifelong MAPF, reviewing previous works on guidance in MAPF and highlighting the generality of our guidance graph. We propose CMA-ES and PIU that optimize guidance graphs across different algorithms and maps. We also show that the update model can generate guidance graphs for larger maps with similar patterns.

Our work is limited in many ways, yielding numerous future directions. 
First, both CMA-ES and PIU are computationally expensive, requiring a large number of lifelong MAPF simulations, taking 1.2 hours (setup 1) to 55 hours (setup 8) to run on machine (2) in \Cref{sec:exp_setup}.
Future works can focus on reducing the computational requirements of both methods. 
Second, our optimized guidance graphs improve throughput but such improvement lacks explainability.
Future works can focus on either generating more explainable guidance graphs or analyzing the explainability of our optimized guidance graphs.
Third, although our guidance graph can be used with online update mechanisms introduced in previous works~\citep{ChenAAAI24,Yu2023}, we limit our experiment settings without such mechanisms. Integrating these mechanisms with our GGO algorithms could further enhance MAPF guidance utility.

\section*{Acknowledgments}

This work used Bridge-$2$ at Pittsburgh Supercomputing Center (PSC) through allocation CIS$220115$ from the Advanced Cyberinfrastructure Coordination Ecosystem: Services \& Support (ACCESS) program, which is supported by National Science Foundation grants \#$2138259$, \#$2138286$, \#$2138307$, \#$2137603$, and \#$2138296$.

\bibliographystyle{named}
\bibliography{ijcai24}

\clearpage

\appendix

\section{Baseline Guidance Graphs} \label{appen:baseline}

In this section, we discuss the baseline guidance graphs that we compare with in \Cref{sec:exp}. As mentioned in \Cref{sec:guidance-review}, we have 4 baseline guidance graphs, namely (1) \textit{Unweighted}, (2) \textit{Crisscross}~\citep{lironPhDthesis}, (3) \textit{HM Cost}~\citep{liron_highway16}, and (4) \textit{Traffic Flow}~\citep{ChenAAAI24}.

\begin{algorithm}[!t] 
\caption{Traffic Flow Guidance Graph Generation}\label{alg:traffic-flow}\LinesNumbered
\SetKwInput{KwInput}{Input}             %
\SetKwInput{KwOutput}{Output}           %
\SetKwProg{Fn}{Function}{:}{}

\KwInput
{
$G_g(V_g, E_g, \boldsymbol{\omega})$, where $\boldsymbol{\omega} = 1$\newline
$N_{base}$: number of iterations \newline
$S,T \subset V_g$: potential start and goal locations of the agents \newline
$plan\_path$: function to find single agent path given start, goal, and guidance graph.
}

Initialize vertex usage $U_{V_g}(v) \gets 0, \forall v \in V_g$ \\
\label{alg:tf:init-v-use}
Initialize edge usage $U_{E_g}(u,v) \gets 0, \forall (u,v) \in E_g$ \\
\label{alg:tf:init-e-use}
\For{$i \gets 1$ \KwTo $N_{base}$}{
\label{alg:tf:singlepp1}
    Sample $s_i \in_{R} S,g_i \in_{R} T$ s.t. $s_i \neq g_i$ \\ 
    \label{alg:tf:singlepp2}
    $P_i \gets plan\_path(s_i, g_i, G_g)$ \\
    \label{alg:tf:singlepp3}
    \For{Vertex $v \in P_i$}{
    \label{alg:tf:singlepp4}
        $U_{V_g}(v) \gets U_{V_g}(v) + 1$ \\
        \label{alg:tf:singlepp5}
    }
    \For{Edge $(u,v) \in P_i$}{
    \label{alg:tf:singlepp6}
        $U_{E_g}(u,v) \gets U_{E_g}(u,v) + 1$ \\
        \label{alg:tf:singlepp7}
    }
    \For{Edge $(u,v) \in E_g$}{
    \label{alg:tf:tf-for}
        $p(v) \gets  \ceil{\frac{U_{V_g}(v) - 1}{2}}$\\
        \label{alg:tf:tf-p}
        $c(u,v) \gets U_{E_g}(u,v) \times U_{E_g}(v,u)$\\
        \label{alg:tf:tf-c}
        $\boldsymbol{\omega}(u,v) \gets 1 + c(u,v) + p(v)$\\
        \label{alg:tf:tf-w}
    }
}
$\boldsymbol{\omega}_{TF} \gets \boldsymbol{\omega}$\\
\label{alg:tf:tf-set-w}
Return $G_{TF}(V_g, E_g, \boldsymbol{\omega}_{TF})$\\
\label{alg:tf:tf-rt}
\end{algorithm}

\subsection{Unweighted and Crisscross}

Both unweighted and crisscross guidance graphs are human-designed. We define unweighted guidance graph as follows:

\begin{definition}[Unweighted Guidance Graph]
    We define a guidance graph $G(V_g,E_g,\boldsymbol{\omega})$ where $\boldsymbol{\omega} = 1$ as an unweighted guidance graph.
\end{definition}

Our crisscross guidance graph follows the definition of crisscross highways~\citep{lironPhDthesis}. In particular:

\begin{definition}[Crisscross Guidance Graph]
    Given an unweighted guidance graph $G(V_g,E_g,\boldsymbol{\omega})$ for a 4-neighbor grid-based map in which agents can move up, down, left, or right at each vertex, we select a subset of edges $E_c \subset E_g$ such that: 
    \begin{enumerate}
        \item in the even rows, all edges pointing right are chosen, 
        \item in the odd rows, all edges pointing left are chosen, 
        \item in the even columns, all edges pointing up are chosen, 
        \item in the odd columns, all edges pointing down are chosen.
    \end{enumerate}
     We let the edge weights of edges in $E_c$ be 0.5 and all other edges $E_g \setminus E_c $ be 1, promoting the agents to use edges in $E_c$.
\end{definition}

\subsection{Traffic Flow and HM Cost}

\begin{algorithm}[!t] 
\caption{HM Cost Guidance Graph Generation}\label{alg:hm-cost}\LinesNumbered
\SetKwInput{KwInput}{Input}             %
\SetKwInput{KwOutput}{Output}           %
\SetKwProg{Fn}{Function}{:}{}

\KwInput
{
$G_g(V_g, E_g, \boldsymbol{\omega})$, where $\boldsymbol{\omega} = 1$\newline
$N_{base}$: number of iterations \newline
$S,T \subset V_g$: potential start and goal locations of the agents \newline
$plan\_path$: function to find single agent path given start, goal, and guidance graph.\newline
$\alpha,\beta,\gamma \in \mathbb{R}$: hyperparameters of computing follow preference $p$, interference cost $t$, and saturation cost $t$, respectively.
}

Initialize vertex usage $U_{V_g}(v) \gets 0, \forall v \in V_g$ \\
\label{alg:hm:init-v-use}
Initialize edge usage $U_{E_g}(u,v) \gets 0, \forall (u,v) \in E_g$ \\
\label{alg:hm:init-e-use}
\For{$i \gets 1$ \KwTo $N_{base}$}{
\label{alg:hm:singlepp1}
    Sample $s_i \in_{R} S,g_i \in_{R} T$ s.t. $s_i \neq g_i$ \\ 
    \label{alg:hm:singlepp2}
    $P_i \gets plan\_path(s_i, g_i, G_g)$ \\
    \label{alg:hm:singlepp3}
    \For{Vertex $v \in P_i$}{
    \label{alg:hm:singlepp4}
        $U_{V_g}(v) \gets U_{V_g}(v) + 1$ \\
        \label{alg:hm:singlepp5}
    }
    \For{Edge $(u,v) \in P_i$}{
    \label{alg:hm:singlepp6}
        $U_{E_g}(u,v) \gets U_{E_g}(u,v) + 1$ \\
        \label{alg:hm:singlepp7}
    }
    \For{Edge $(u,v) \in E_g$}{
    \label{alg:hm:for}
        $p(u,v) \gets \alpha \times \frac{U_{E_g}(u,v)}{N_{base}}$\\
        \label{alg:hm:p}
        $t(u,v) \gets \beta \times \frac{U_{E_g}(v,u)}{N_{base}}$ \\
        \label{alg:hm:t}
        $s(u,v) \gets \gamma^{\frac{U_{E_g}(u,v) + U_{E_g}(v,u)}{2N_{base}}}$ \\
        \label{alg:hm:s}
        $c(u,v) \gets 1 - p(u,v) + t(u,v) + s(u,v)$\\
        \label{alg:hm:c}
        $\boldsymbol{\omega}(u,v) \gets c(u,v)$\\
        \label{alg:hm:w}
    }
}

Set $E_{HM} \subset E_g$ with lowest $c(u,v)$ s.t. $|E_{HM}| = \frac{1}{7} |E_g|$ and $u \neq v$\\
\label{alg:hm:e-em}
Set $E_{HM}' \gets $ randomly sample $\frac{1}{5}$ of edges from $E_{HM}$ \\
\label{alg:hm:e-em-p}
Initialize edge weights $\boldsymbol{\omega}_{HM}$ \\
\label{alg:hm:e-em-w}
\For{Edge $(u,v) \in E_g$}{
\label{alg:hm:hwy-for}
    \eIf{$(u,v) \in E_{HM}'$}{
    \label{alg:hm:hwy-if}
        $\boldsymbol{\omega}_{HM}(u,v) \gets 0.5$\\
        \label{alg:hm:hwy-0.5}
    }{
    \label{alg:hm:hwy-else}
        $\boldsymbol{\omega}_{HM}(u,v) \gets 1$\\
        \label{alg:hm:hwy-1}
    }
}
Return $G_{HM}(V_g, E_g, \boldsymbol{\omega}_{HM})$\\
\label{alg:hm:rt}
\end{algorithm}

The previous work on traffic flow guidance~\citep{ChenAAAI24} is developed for lifelong MAPF with an online update mechanism. The work on HM Cost guidance~\citep{liron_highway16} is developed for one-shot MAPF. Therefore, we adapt both methods to construct guidance graphs for lifelong MAPF. 

\noindent \textbf{Traffic Flow.} \Cref{alg:traffic-flow} describes the adapted Traffic Flow guidance graph generation procedure. On a high level, we iteratively plan single-agent paths based on the current guidance graph and update the graph based on the congestion of these paths. Starting with an unweighted guidance graph, we sample a pair of start and goal locations (\Cref{alg:tf:singlepp2}) and search for a path $P_i$ that minimizes the sum of its edge weights on the current guidance graph (\Cref{alg:tf:singlepp3}). Then, we increment the usages of vertices and edges on $P_i$ by 1 (\Cref{alg:tf:singlepp4,alg:tf:singlepp5,alg:tf:singlepp6,alg:tf:singlepp7}). Afterward, we follow the previous work~\citep{ChenAAAI24} to compute the vertex congestion averaged over vertex usage for each vertex (\Cref{alg:tf:tf-p}) and the contraflow congestion for each edge (\Cref{alg:tf:tf-c}) and then update the edge weights of the guidance graph by summing the vertex congestion, the contraflow congestion, and 1 (\Cref{alg:tf:tf-w}), where the 1 indicates the zero congestion cost. The updated edge weights inflate the cost of frequently used edges so that the following single-agent paths are encouraged to avoid these edges.
Traffic Flow repeats the procedure for $N_{base}$ iterations and returns the edge weights from the last iteration as the guidance graph (\Cref{alg:tf:tf-set-w,alg:tf:tf-rt}). Since the Traffic Flow algorithm does not consider wait costs, we set wait costs of all vertices to be 1.

\begin{figure}[!t]
    \centering
    \begin{subfigure}{0.24\textwidth}
        \centering
        \includegraphics[width=1\textwidth]{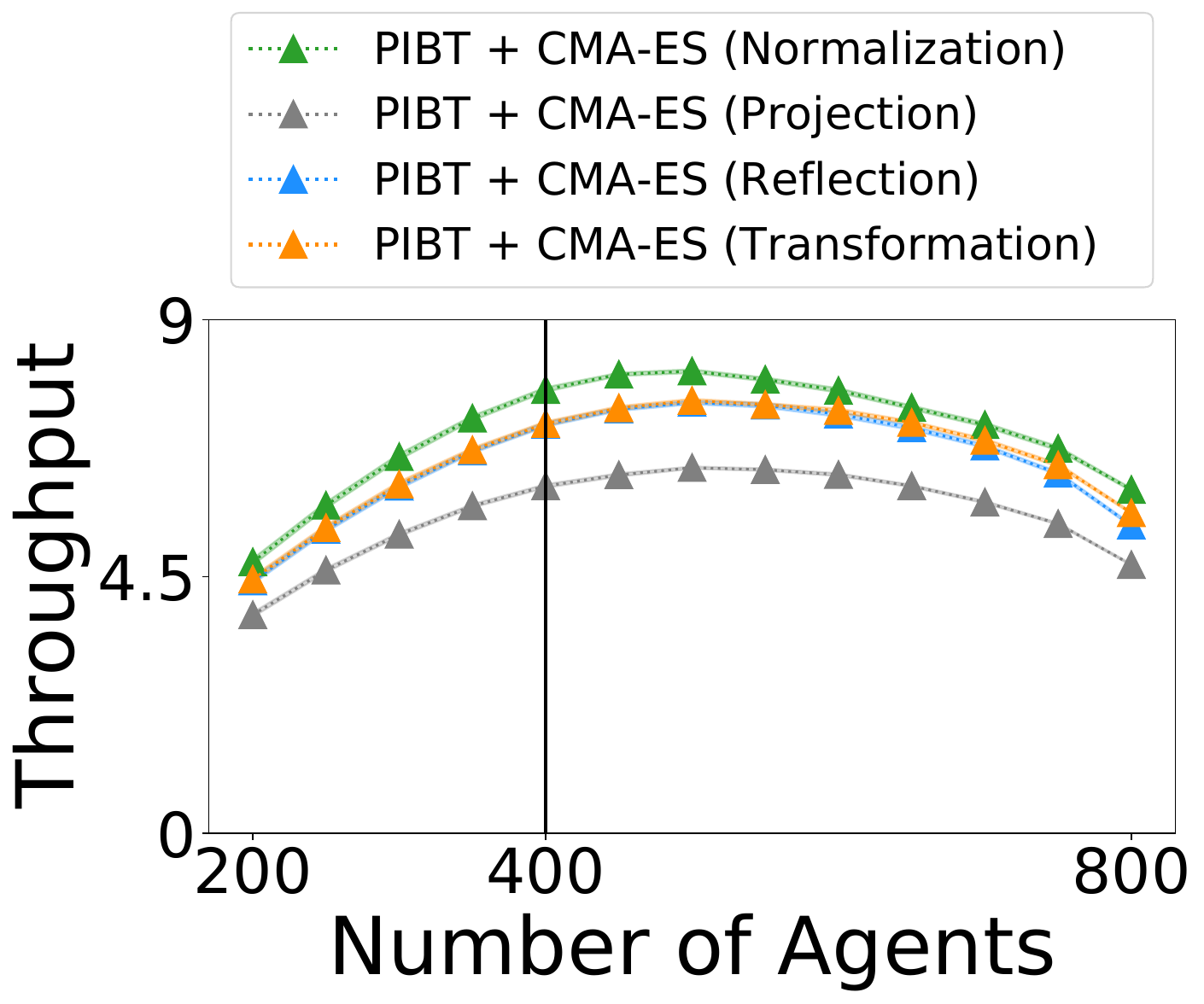}
        \caption{Setup 1: \randomSmall}
        \label{fig:random-32-32-bounds-handle}
    \end{subfigure}%
    \hfill
    \begin{subfigure}{0.24\textwidth}
        \centering
        \includegraphics[width=1\textwidth]{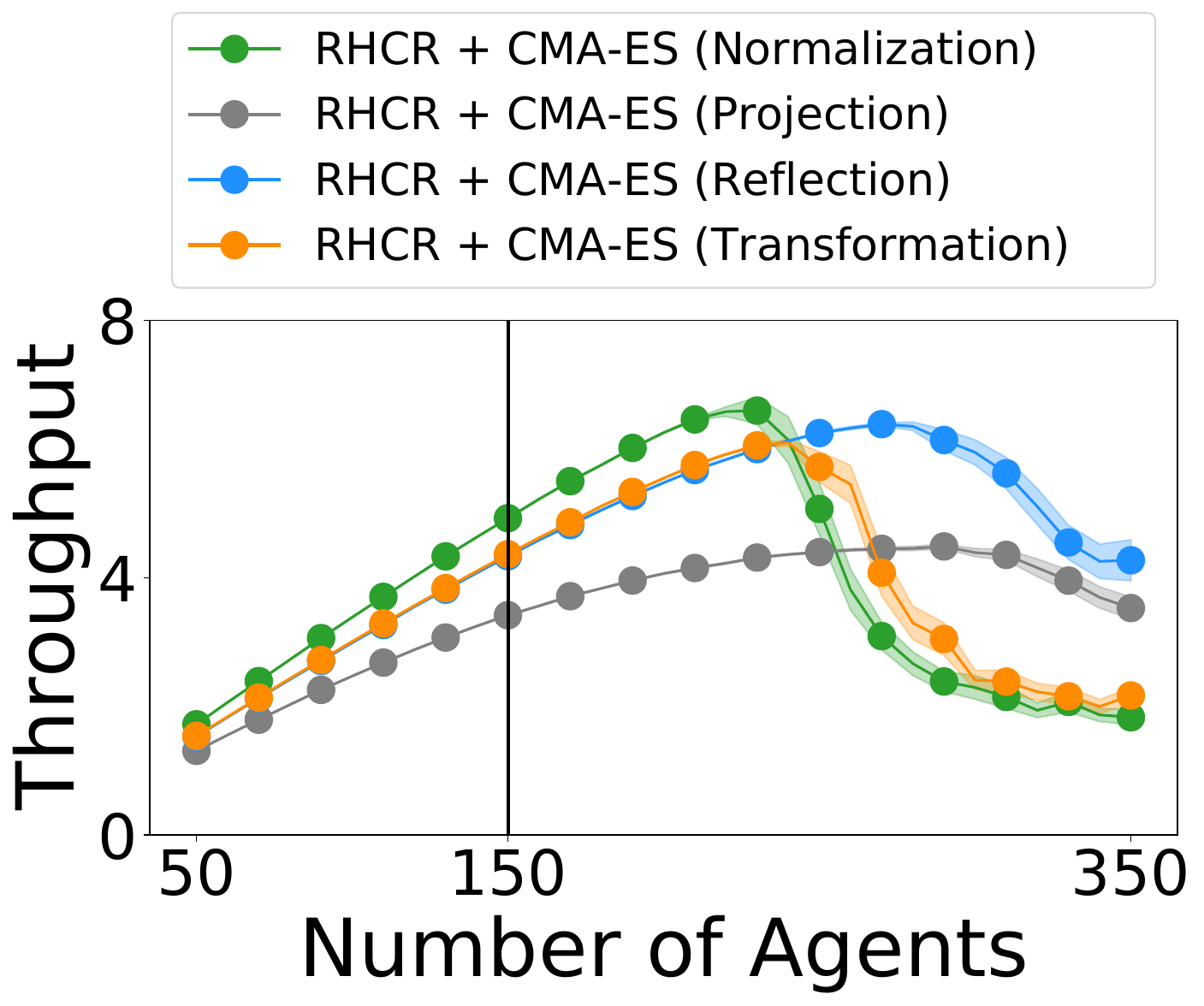}
        \caption{Setup 8: \warehouselargeW}
        \label{fig:warehouse-large-bounds-handle}
    \end{subfigure}%
    \caption{Comparison of different bounds handling methods in CMA-ES in setup 1 and 4. The black vertical lines indicates the number of agents $N_a$ used to optimize the guidance graph.}
    \label{fig:bounds-handle}
\end{figure}

\noindent \textbf{HM Cost.} \Cref{alg:hm-cost} describes the adapted HM Cost guidance graph generation procedure. Similar to Traffic Flow, we first sample start and goal locations (\Cref{alg:hm:singlepp2}) and run single-agent path planning on the current guidance graph (\Cref{alg:hm:singlepp3}), updating vertex and edge usage (\Cref{alg:hm:singlepp4,alg:hm:singlepp5,alg:hm:singlepp6,alg:hm:singlepp7}). Then we follow the previous work~\citep{liron_highway16} to compute (1) the follow preference $p$ (\Cref{alg:hm:p}), which encourages the agents to traverse through previously used edges, (2) the interference cost $t$, which discourages the agents to traverse through edges in the opposite directions of previously used edges, and (3) the saturation cost $s$. The HM cost $c$ is a linear combination of the above three variables (\Cref{alg:hm:c}), with smaller HM costs indicating edges that are used more frequently. We then set the HM cost as the edge weights of the guidance graph and repeat the iteration for $N_{base}$ times (\Cref{alg:hm:w}).

Nevertheless, after running $N_{base}$ single agent path planning, HM Cost includes additional procedures to generate the final guidance graph from the computed HM cost. Following previous work~\citep{liron_highway16}, we select the top $\frac{1}{7}$ of the edges with the smallest HM cost and then randomly sample $\frac{1}{5}$ of them as the set of highway edges (\Cref{alg:hm:e-em,alg:hm:e-em-p}). We then set the weights of the highway edges as 0.5 and non-highway edges as 1, forming a guidance graph (\Cref{alg:hm:e-em-w,alg:hm:hwy-for,alg:hm:hwy-if,alg:hm:hwy-0.5,alg:hm:hwy-else,alg:hm:hwy-1}). This is equivalent to setting $c=2$ following the definition of highway in \Cref{sec:guidance-review} and previous works~\citep{Cohen2015FeasibilitySU,liron_highway16,li2023study} (i.e. the weights of the highway edges are 1 and non-highway edges 2).
In HM Cost, all selected edges are not self-edges. Therefore, the wait costs of all vertices are 1, same as the non-highway edges.

\subsection{Hyperparameters for Baseline Methods}

In both Traffic Flow and HM Cost, we use $N_{base} = 10,000$ to generate the guidance graphs for all maps. In HM Cost, we follow the previous work~\citep{liron_highway16} to use $\alpha = 0.5$, $\beta = 1.2$, and $\gamma = 1.3$.

\section{Additional Experiments} \label{appen:add_exp}

We include the following additional experiments: 
(1) for bounds handling of CMA-ES, we show ablation experiments on normalization comparing with other representative bounds handling methods presented in \citep{BIEDRZYCKI_cma-es_bounds2020},
(2) for PIU, we test guidance graph generation with different values of $N_{e\_piu}$ to test if running more simulations in PIU can improve the generated guidance graph,
and (3) the throughput with different number of agents in large warehouse maps shown in \Cref{fig:warehouse-scale-map}.

\subsection{Bounds Handling in CMA-ES} \label{appen:bounds-handle}

A previous work~\citep{BIEDRZYCKI_cma-es_bounds2020} compares a number of bounds handling approaches for CMA-ES. Their comparison demonstrates that resampling, reflection, projection, and transformation are among the most popular and empirically best choices of bounds handling methods.
Therefore, we first briefly present these methods and compare them with our proposed bounds handling, namely min-max normalization. For simplicity, we use normalization to refer to min-max normalization in the following texts.

\subsubsection{Bounds-Handling Methods}

\begin{figure}[!t]
    \centering
    \includegraphics[width=0.45\textwidth]{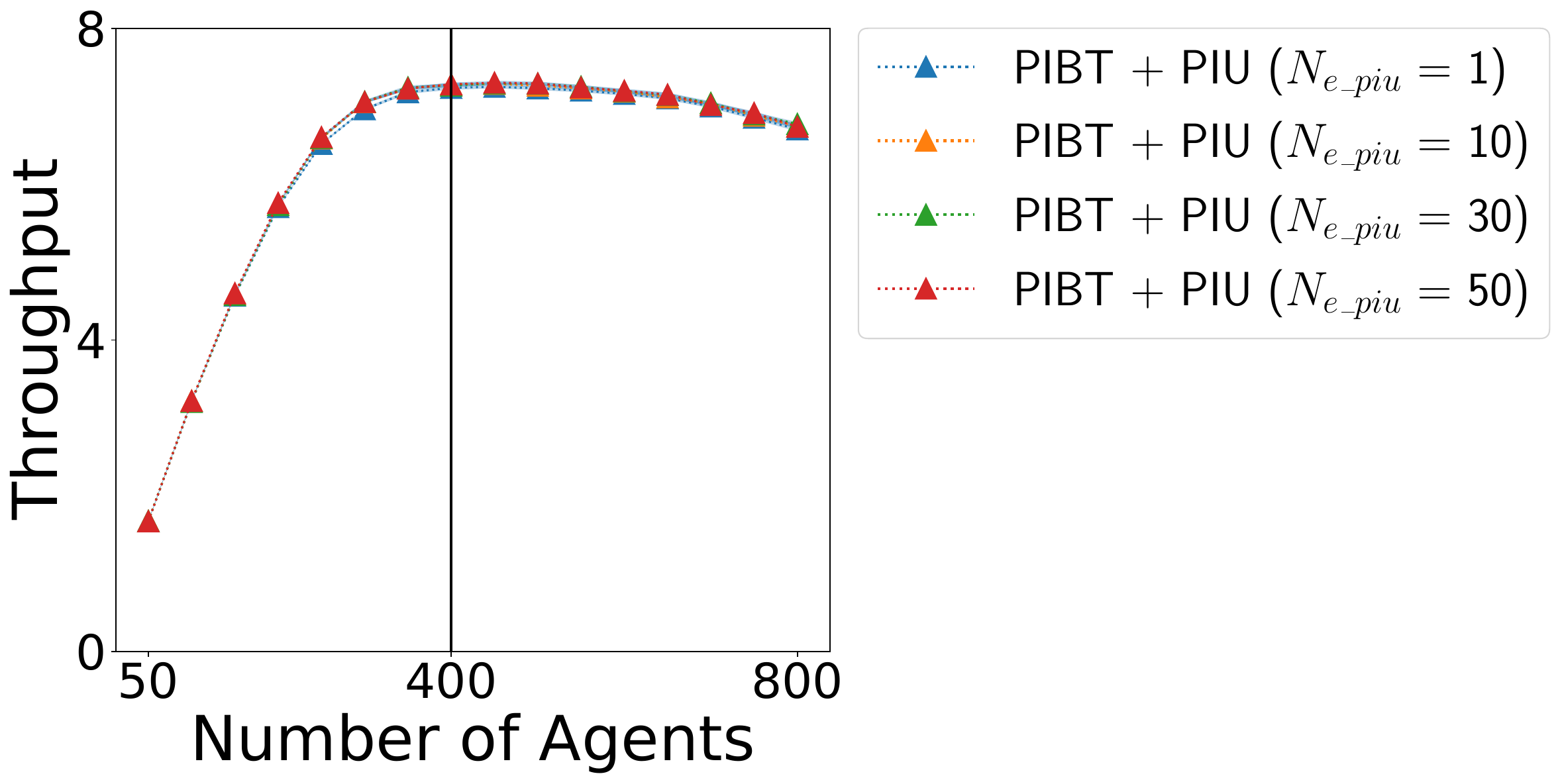}
    \caption{Effect of different values of $N_{e\_piu}$. The black vertical line indicates the number of agents used to optimize the update model.}
    \label{fig:N-e-piu_ablation}
\end{figure}

\begin{figure*}[!ht]
    \centering

    \begin{subfigure}[t]{0.19\textwidth}
        \centering
        \includegraphics[width=1\textwidth]{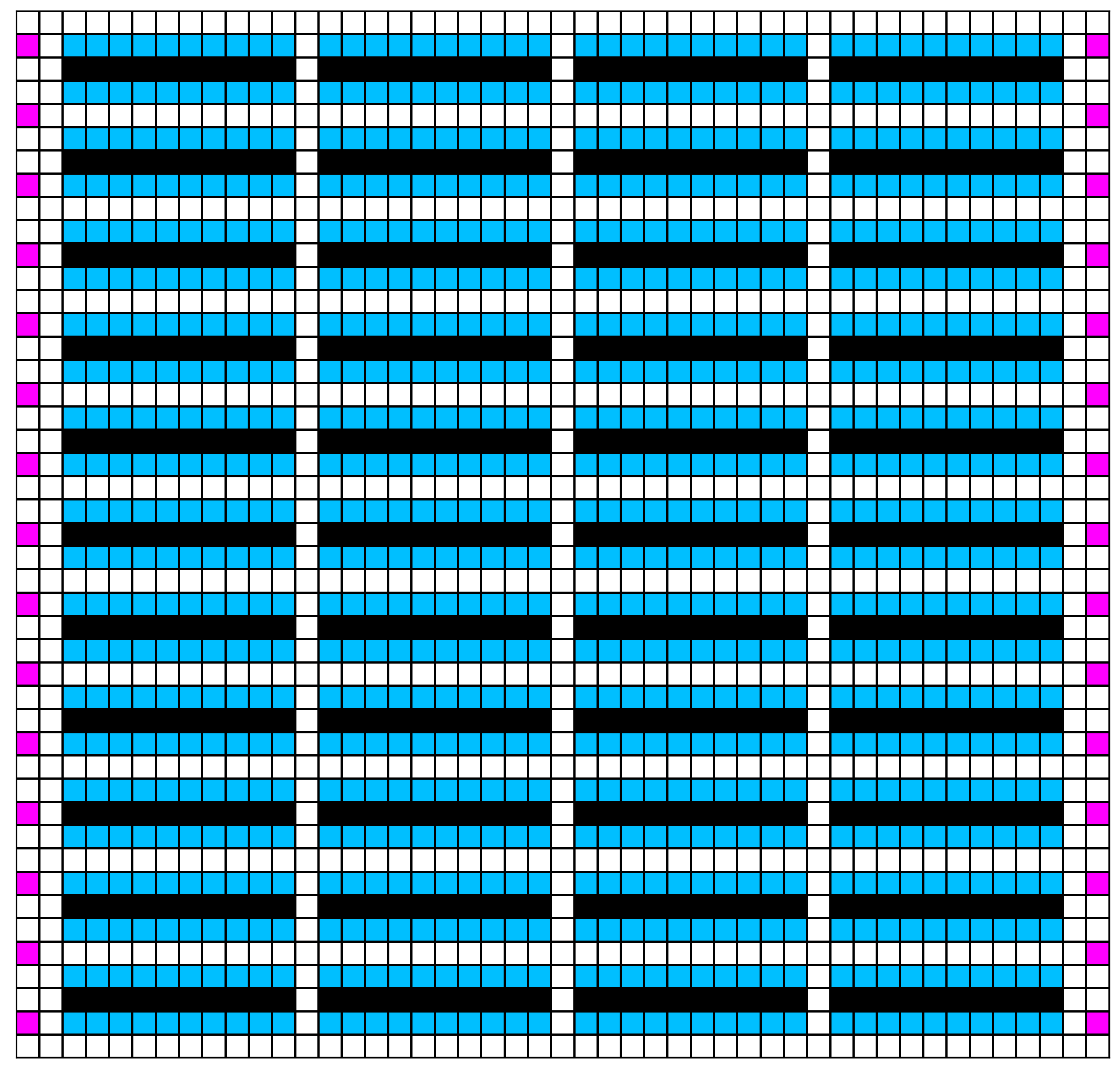}
        \caption{45 $\times$ 47}
    \end{subfigure}%
    \hfill
    \begin{subfigure}[t]{0.19\textwidth}
        \centering
        \includegraphics[width=1\textwidth]{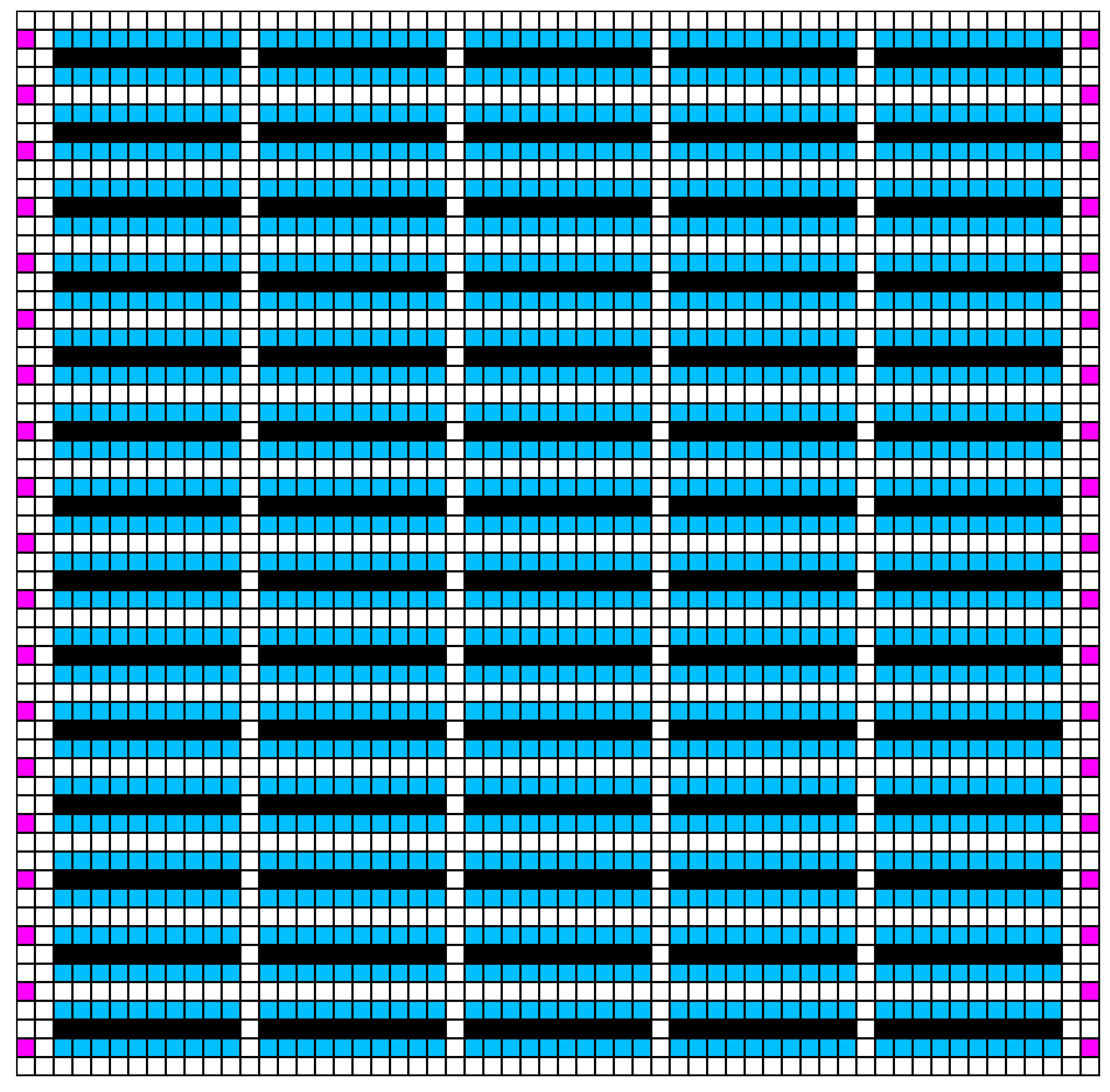}
        \caption{57 $\times$ 58}
    \end{subfigure}
    \hfill
    \begin{subfigure}[t]{0.19\textwidth}
        \centering
        \includegraphics[width=1\textwidth]{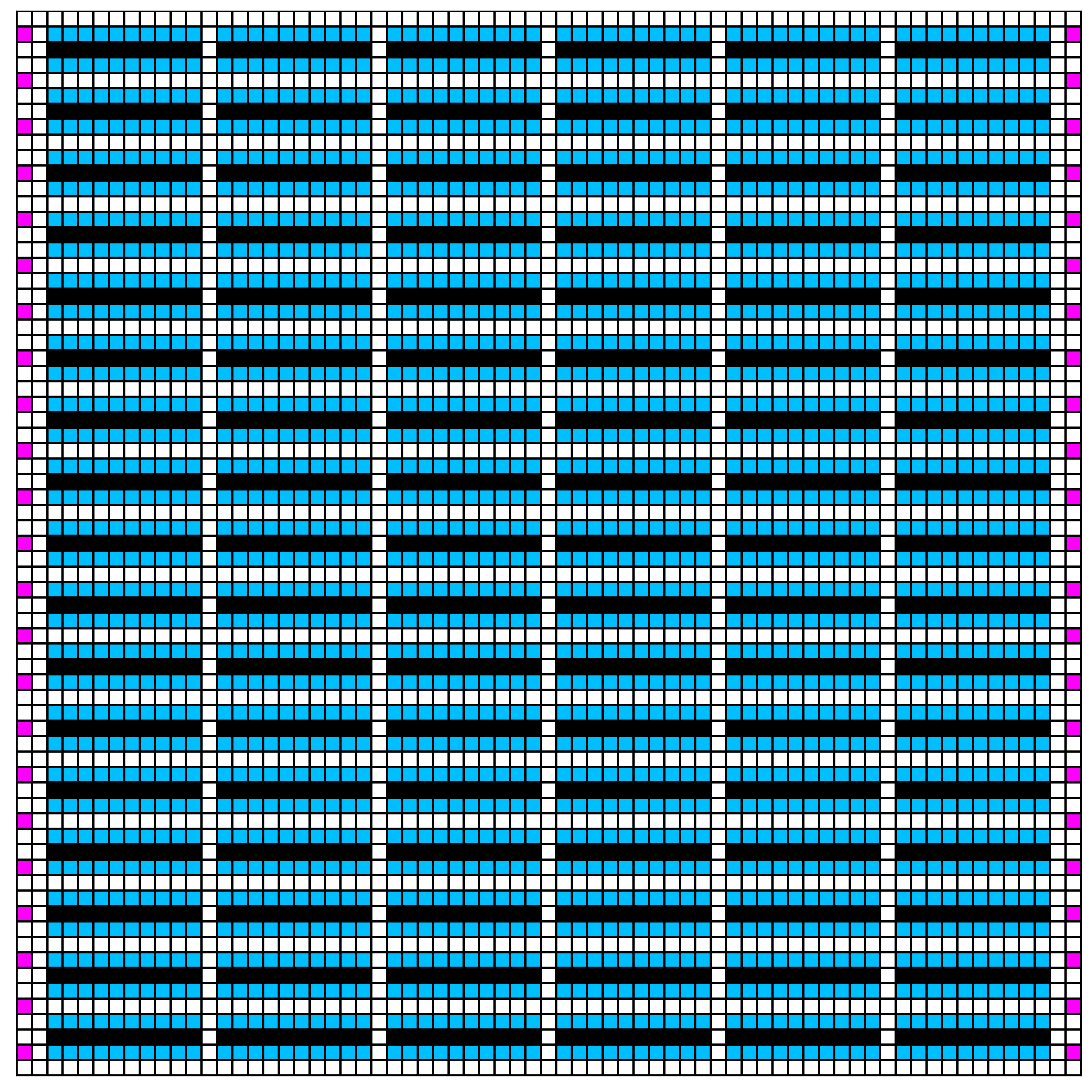}
        \caption{69 $\times$ 69}
    \end{subfigure}%
    \hfill
    \begin{subfigure}[t]{0.19\textwidth}
        \centering
        \includegraphics[width=1\textwidth]{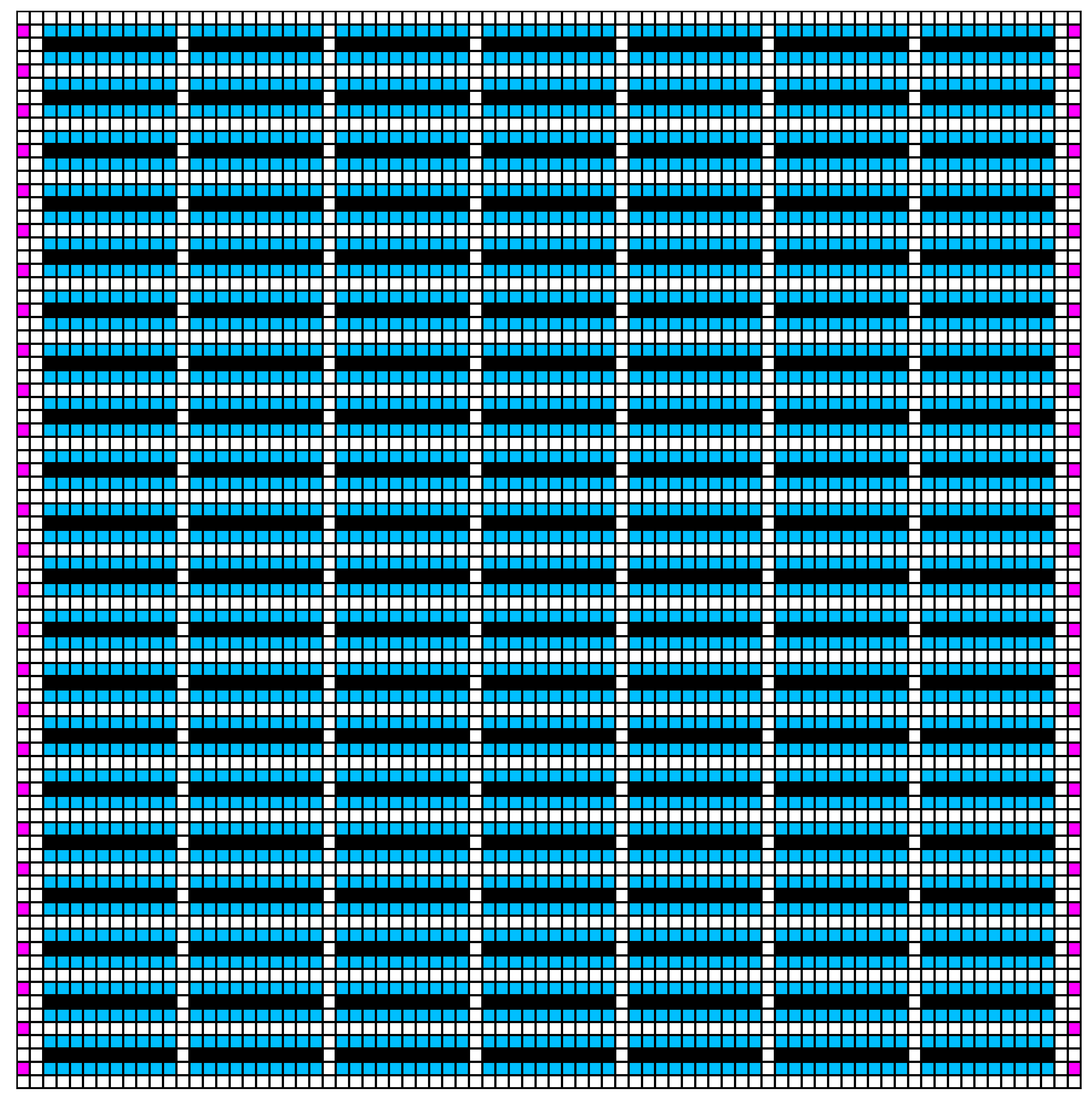}
        \caption{81 $\times$ 80}
    \end{subfigure}%
    \hfill
    \begin{subfigure}[t]{0.19\textwidth}
        \centering
        \includegraphics[width=1\textwidth]{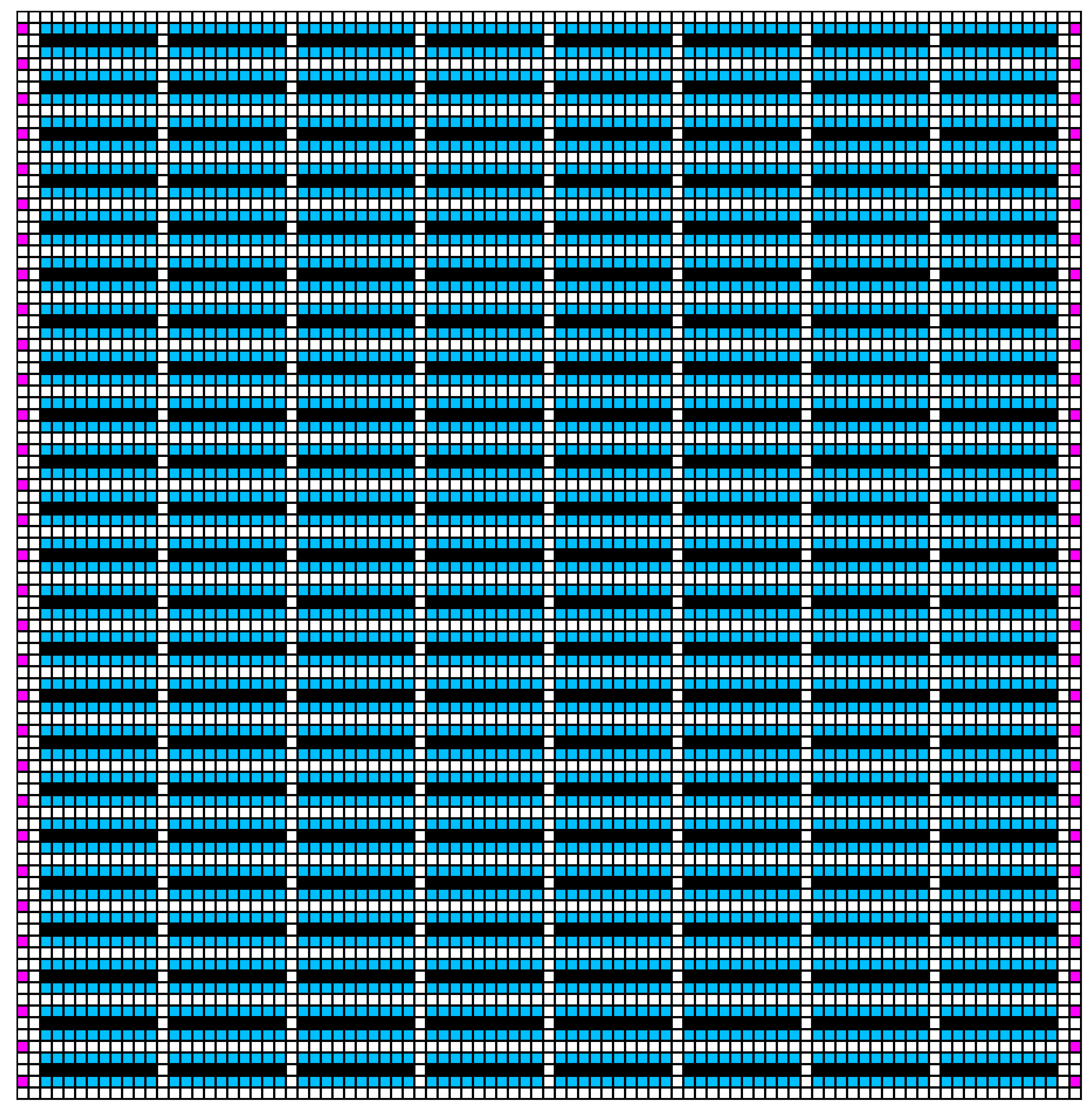}
        \caption{93 $\times$ 91}
    \end{subfigure}%
    \caption{
    Warehouse maps used for testing the transferability of the update model optimized with setup 2 in \Cref{sec:exp}.
    }
    \label{fig:warehouse-scale-map}
\end{figure*}

\begin{figure*}[!ht]
    \centering
    \includegraphics[width=1\textwidth]{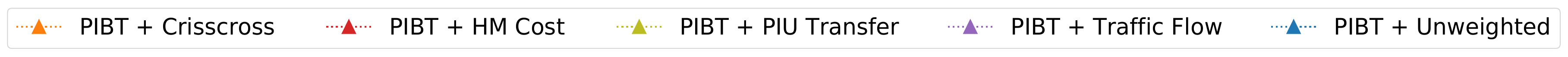}
    \begin{subfigure}[t]{0.19\textwidth}
        \centering
        \includegraphics[width=1\textwidth]{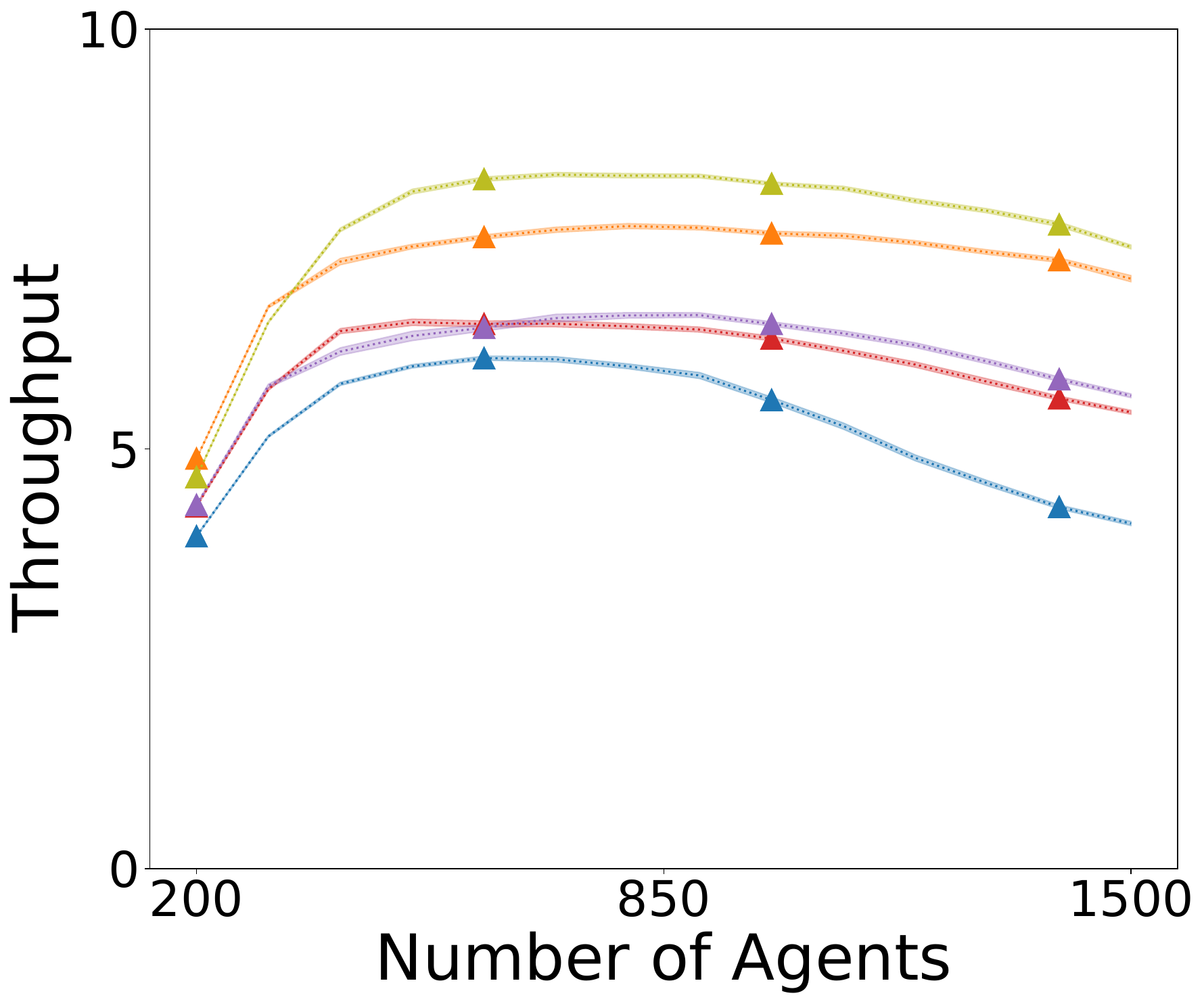}
        \caption{45 $\times$ 47}
    \end{subfigure}%
    \hfill
    \begin{subfigure}[t]{0.19\textwidth}
        \centering
        \includegraphics[width=1\textwidth]{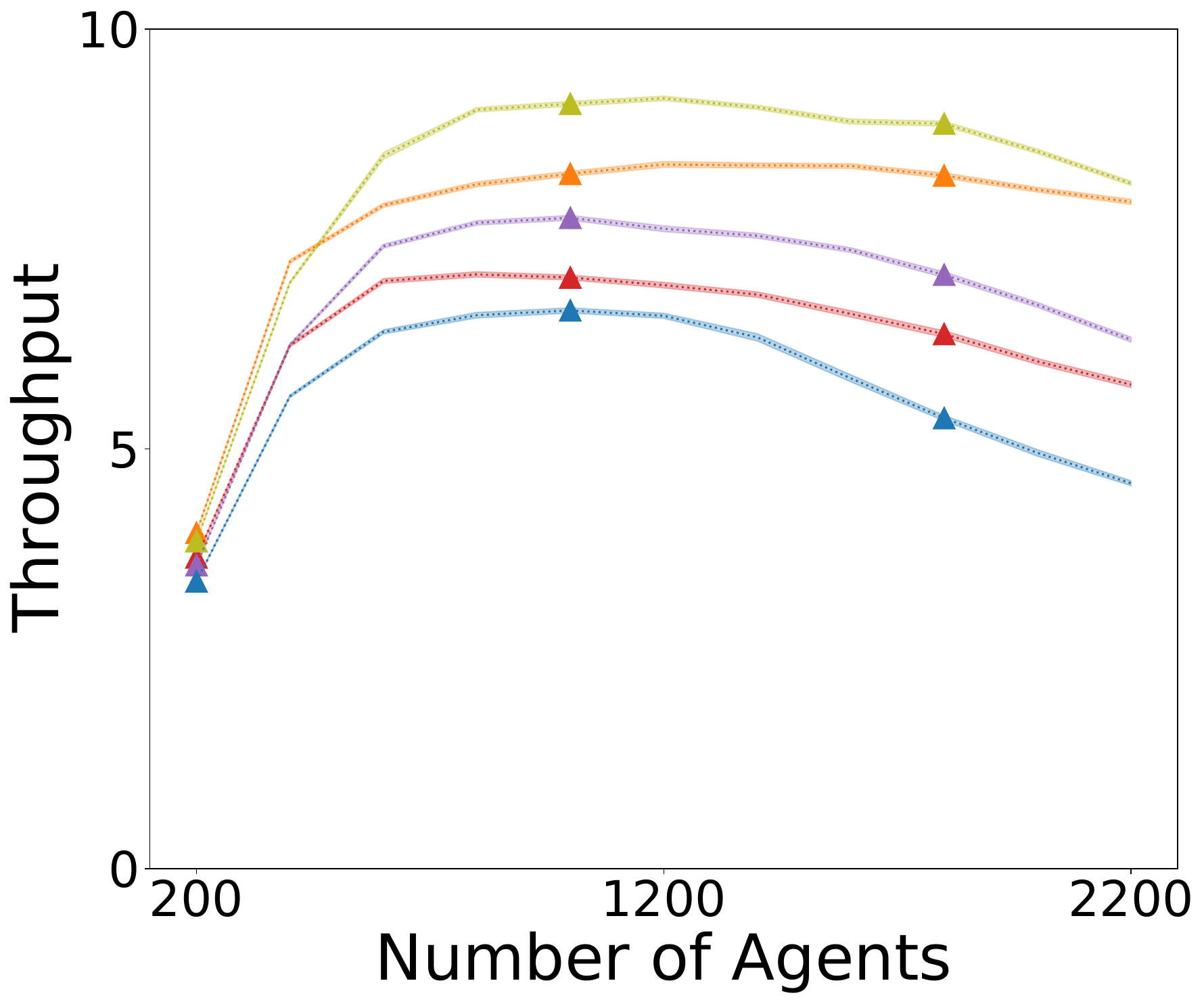}
        \caption{57 $\times$ 58}
    \end{subfigure}
    \hfill
    \begin{subfigure}[t]{0.19\textwidth}
        \centering
        \includegraphics[width=1\textwidth]{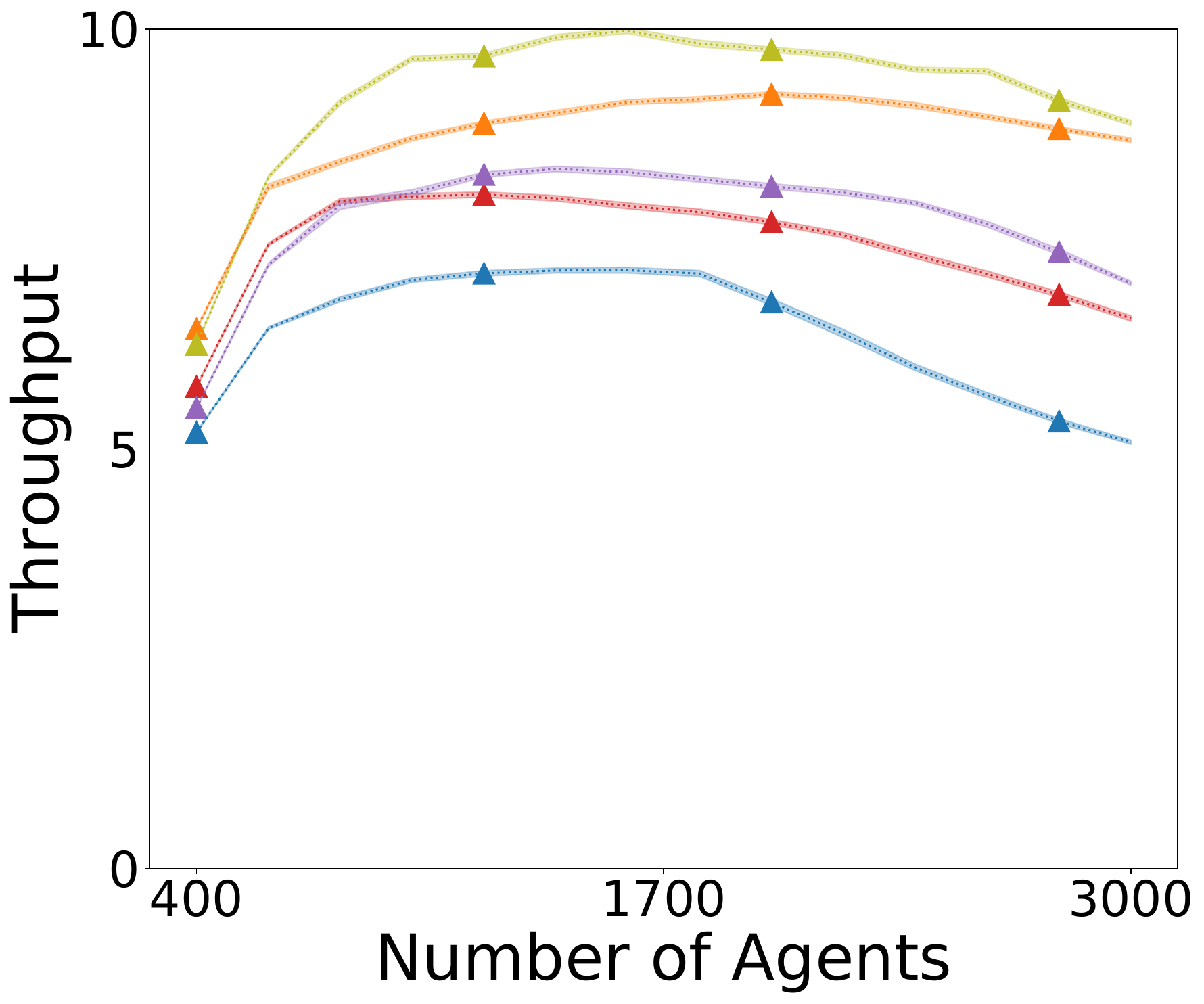}
        \caption{69 $\times$ 69}
    \end{subfigure}%
    \hfill
    \begin{subfigure}[t]{0.19\textwidth}
        \centering
        \includegraphics[width=1\textwidth]{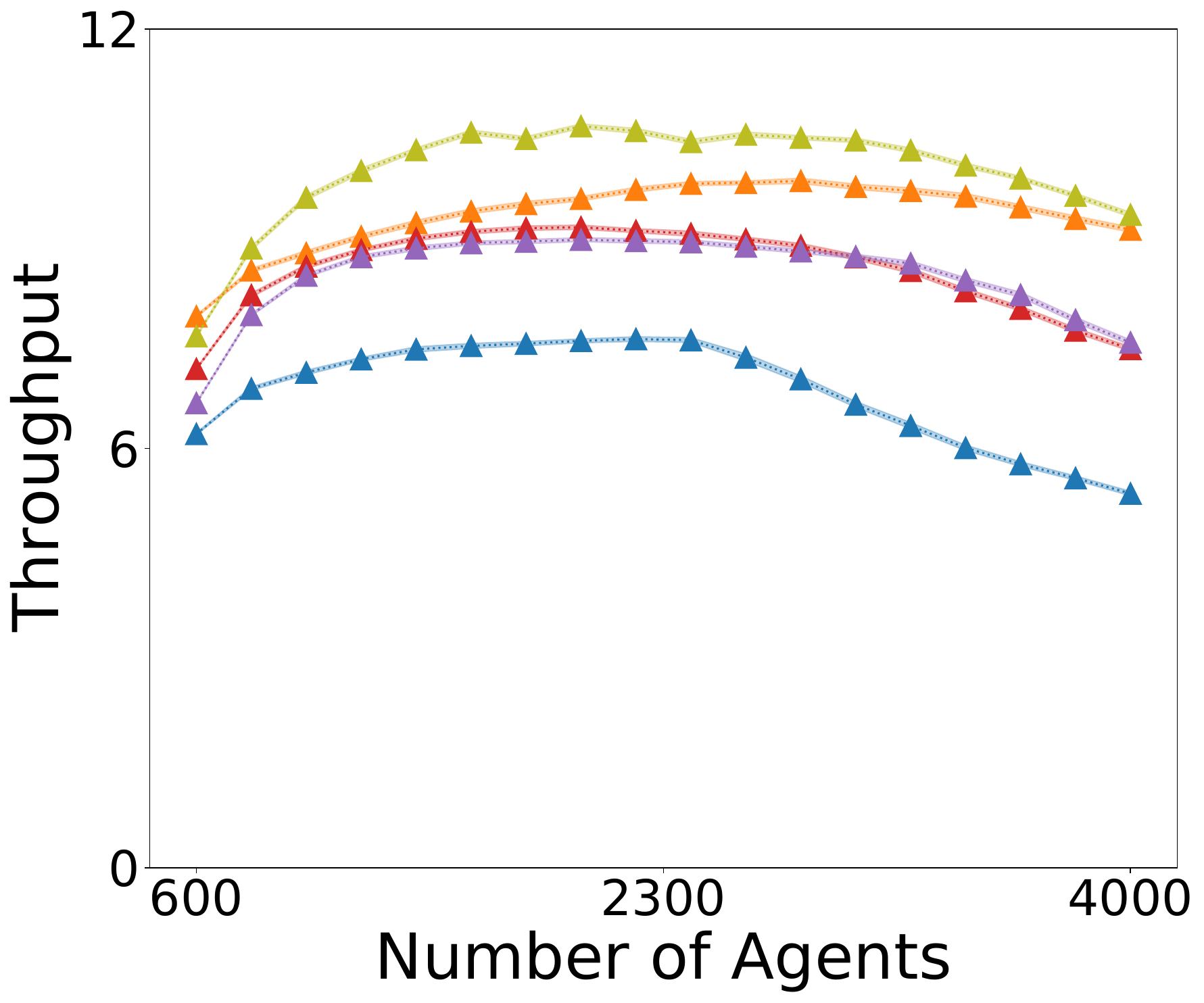}
        \caption{81 $\times$ 80}
    \end{subfigure}%
    \hfill
    \begin{subfigure}[t]{0.19\textwidth}
        \centering
        \includegraphics[width=1\textwidth]{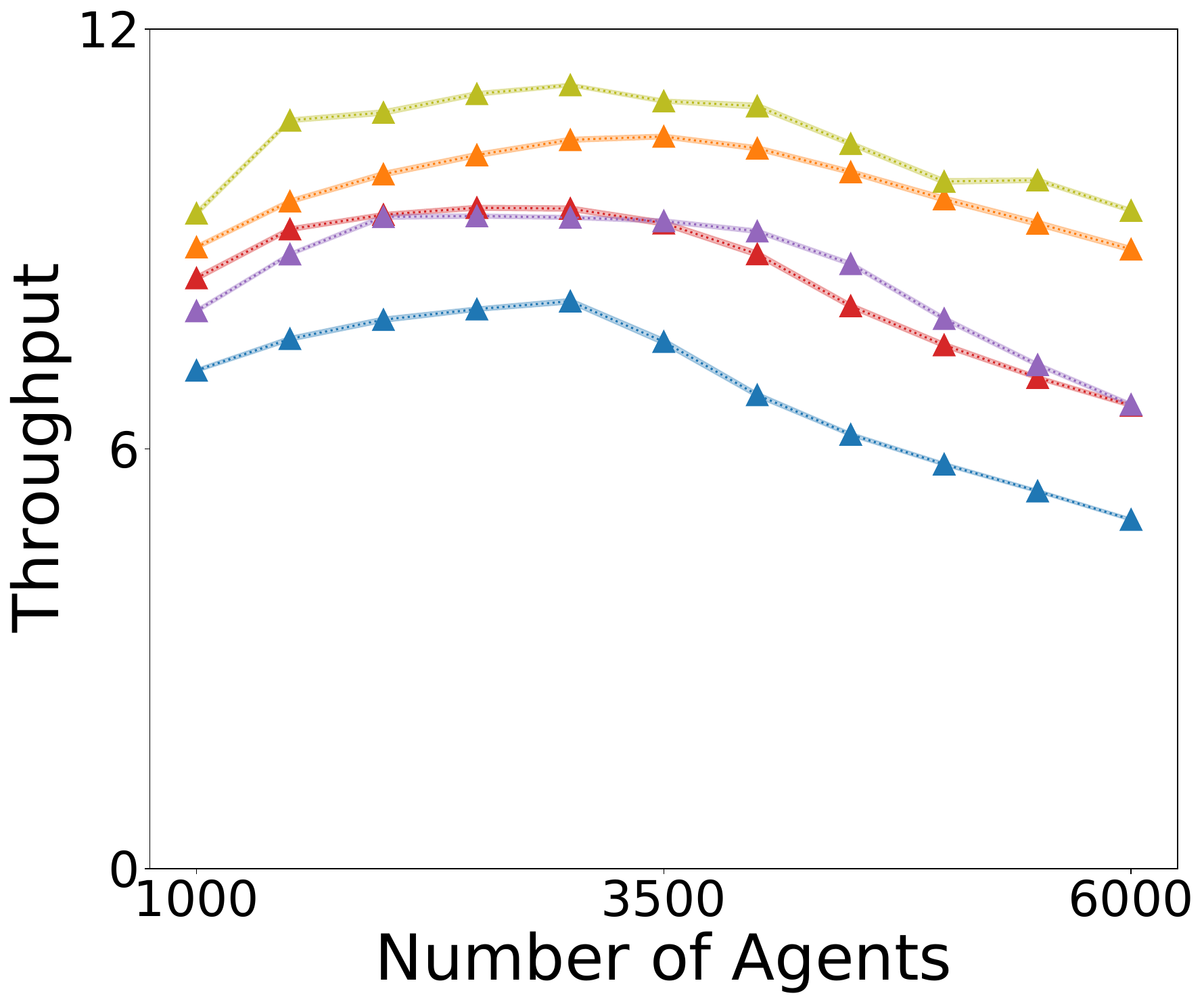}
        \caption{93 $\times$ 91}
    \end{subfigure}%
    \caption{
    Throughput with different numbers of agents in larger warehouse maps.
    }
    \label{fig:warehouse-scale-result}
\end{figure*}

\begin{figure*}[!t]
    \centering
    \begin{subfigure}[b]{0.83\textwidth}
        \centering
        \includegraphics[width=1\textwidth]{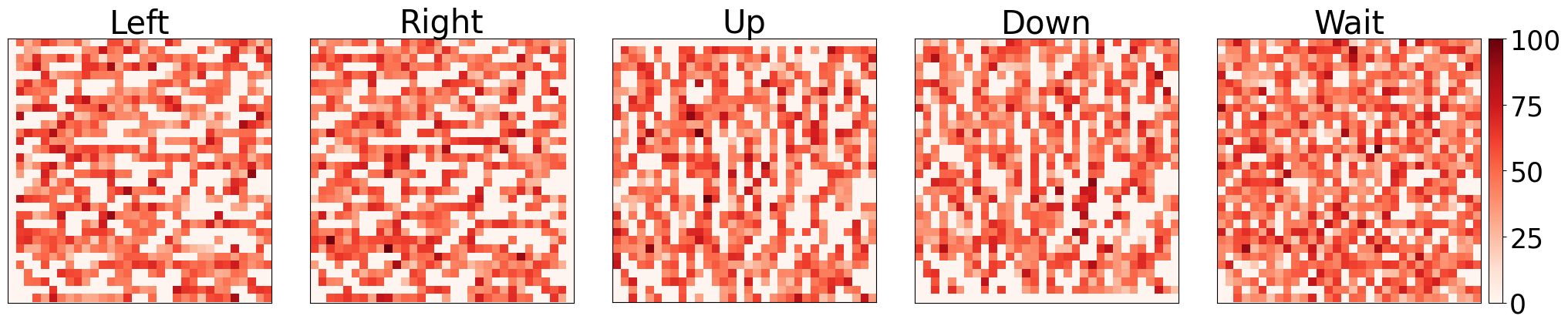}
        \caption{Setup 1 (\randomSmall): PIBT + CMA-ES}
        \label{fig:random-32-32-cma-es-gg}
    \end{subfigure}\\
    \begin{subfigure}[b]{0.83\textwidth}
        \centering
        \includegraphics[width=1\textwidth]{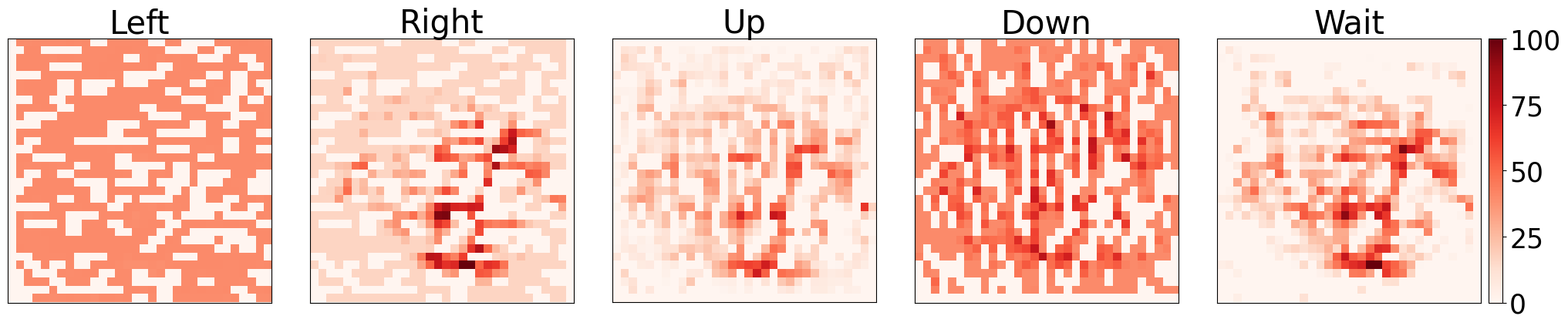}
        \caption{Setup 1 (\randomSmall): PIBT + PIU}
        \label{fig:random-32-32-piu-gg}
    \end{subfigure}\\
    \begin{subfigure}[b]{0.83\textwidth}
        \centering
        \includegraphics[width=1\textwidth]{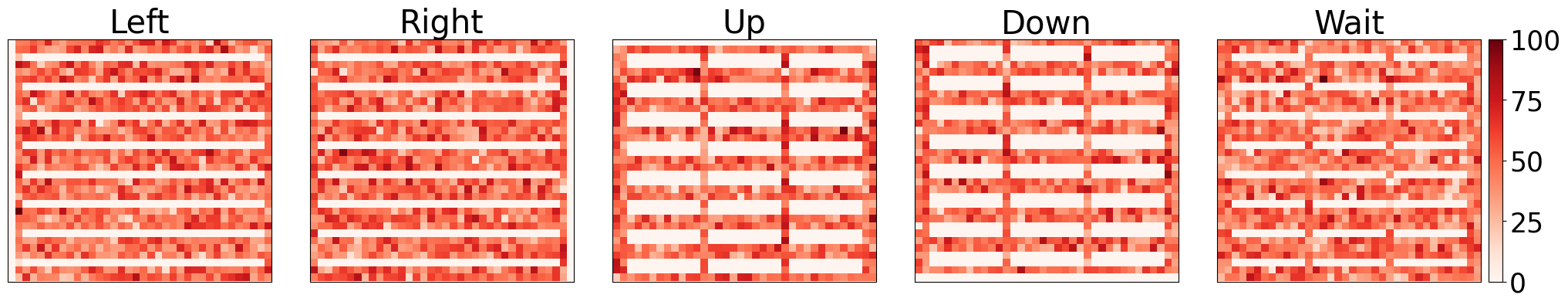}
        \caption{Setup 2 (\warehouselargeW): PIBT + CMA-ES}
        \label{fig:warehouse-large-cma-es-gg}
    \end{subfigure}\\
    \begin{subfigure}[b]{0.83\textwidth}
        \centering
        \includegraphics[width=1\textwidth]{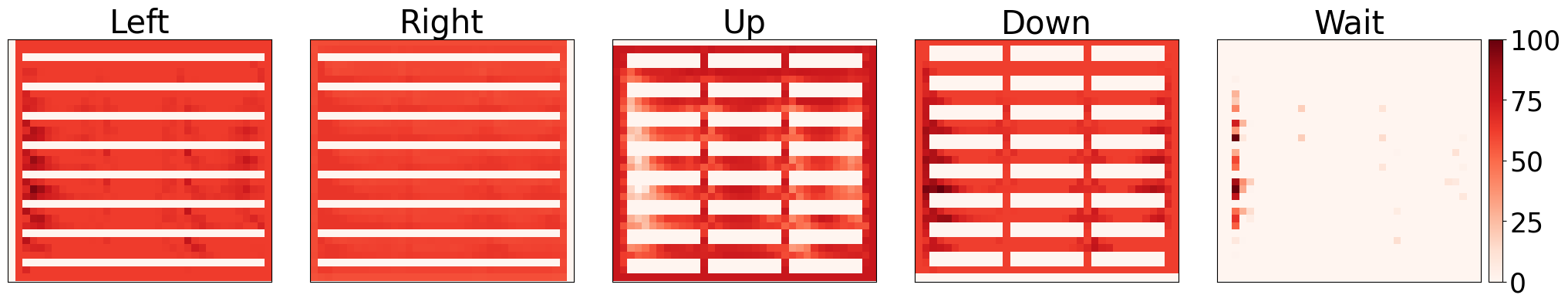}
        \caption{Setup 2 (\warehouselargeW): PIBT + PIU}
        \label{fig:warehouse-large-piu-gg}
    \end{subfigure}\\
    \caption{Optimized guidance graphs of setups 1 and 2.}
    \label{fig:opt-gg12}
\end{figure*}

Assume that we optimize for $\boldsymbol{\omega'} \in \mathbb{R}^n$ such that $\boldsymbol{l} \leq \boldsymbol{\omega'} \leq \boldsymbol{u}$ with $\boldsymbol{l},\boldsymbol{u} \in \mathbb{R}^n$. Then, given a randomly sampled solution $\boldsymbol{\omega} \in \mathbb{R}^n$, the bounds handling methods seek to generate a valid solution $\boldsymbol{\omega'}$ from $\boldsymbol{\omega}$.

\noindent \textbf{Resampling}: The resampling method keeps resampling from the Gaussian distribution until all variables $\boldsymbol{\omega_i} \in \boldsymbol{\omega}$ are within the given bounds.

Notably, the resampling method does not stop until all variables are within the bounds. The probability of sampling a solution from a Gaussian distribution such that all variables are within a given bound depends heavily on the parameters of the distribution and the dimensionality of the search space. Our experiment setups specified in \Cref{tab:exp-setup} have at least 1159 parameters, making resampling inapplicable.

\noindent \textbf{Projection}: The projection method projects out-of-bounds solutions to the lower or upper bounds.

\begin{equation}
    \boldsymbol{\omega'}_i=
    \begin{cases}
        \boldsymbol{\omega_i} & \boldsymbol{l_i} \leq \boldsymbol{\omega_i} \leq \boldsymbol{u_i}\\
        \boldsymbol{l_i} & \boldsymbol{\omega_i} < \boldsymbol{l_i}\\
        \boldsymbol{u_i} & \boldsymbol{\omega_i} > \boldsymbol{u_i}\\
    \end{cases}
\end{equation}

\noindent \textbf{Reflection}: The reflection method reflects the out-of-bounds solutions to within the bounds such that:

\begin{equation}
    \boldsymbol{\omega'}_i=
    \begin{cases}
        \boldsymbol{\omega_i} & \boldsymbol{l_i} \leq \boldsymbol{\omega_i} \leq \boldsymbol{u_i}\\
        2\boldsymbol{l_i} - \boldsymbol{\omega}_i & \boldsymbol{\omega_i} < \boldsymbol{l_i}\\
        2\boldsymbol{u_i} - \boldsymbol{\omega}_i & \boldsymbol{\omega_i} > \boldsymbol{u_i}\\
    \end{cases}
    \label{cma-es-reflection}
\end{equation}

\noindent \textbf{Transformation}: The transformation method maps out-of-bounds solutions to within the bounds such that:

\begin{equation}
    a^l_i = min(\frac{u_i - l_i}{2}, \frac{1 + |l_i|}{20})
\end{equation}

\begin{equation}
    a^u_i = min(\frac{u_i - l_i}{2}, \frac{1 + |u_i|}{20})
\end{equation}

\begin{equation}
    \boldsymbol{\omega'}_i=
    \begin{cases}
        \boldsymbol{\omega_i} & \boldsymbol{l_i} + a^l_i \leq \boldsymbol{\omega_i} \leq \boldsymbol{u_i} - a^u_i \\
        \boldsymbol{l_i} + \frac{(\boldsymbol{\omega}_i - (\boldsymbol{l_i} - a^l_i))^2}{4a^l_i} & \boldsymbol{l_i} - a^l_i \leq \boldsymbol{\omega_i} < \boldsymbol{l_i} + a^l_i\\
        \boldsymbol{u_i} - \frac{(\boldsymbol{\omega}_i - (\boldsymbol{u_i} + a^u_i))^2}{4a^u_i} & \boldsymbol{u_i} - a^u_i < \boldsymbol{\omega_i} \leq \boldsymbol{u_i} + a^u_i\\
    \end{cases}
    \label{cma-es-transformation}
\end{equation}

Intuitively, the transformation does not change the sampled solution if it is within the bound $[\boldsymbol{l_i} + a^l_i, \boldsymbol{u_i} - a^u_i ]$. If the solution falls into $[\boldsymbol{l_i} - a^l_i, \boldsymbol{l_i} + a^l_i)$ or $(\boldsymbol{u_i} - a^u_i, \boldsymbol{u_i} + a^u_i]$, quadratic transformations are applied to map the solution to within the bounds. If the solution is smaller than $\boldsymbol{l_i} - a^l_i$ or larger than $\boldsymbol{u_i} + a^u_i$, the transformation method first uses \Cref{cma-es-reflection} to reflect the solution using $\boldsymbol{l_i} - a^l_i$ or $\boldsymbol{u_i} + a^u_i$ as the bounds. Then \Cref{cma-es-transformation} is applied to further transform the value if necessary. We use the transformation method implemented in Pycma~\citep{nikolaus_hansen_pycma} to run the experiments.

\subsubsection{Empirical Comparison}

We compare normalization with projection, reflection, and transformation on setup 1 and setup 4. Similar to \Cref{sec:exp}, we run lifelong MAPF simulations with varying numbers of agents, each with 50 simulations. \Cref{fig:bounds-handle} shows the throughput. For both setups, normalization empirically achieves the highest throughput with $N_a$ agents. While scaling to more agents, normalization consistently has better throughput than other bounds handling methods in setup 1 with PIBT. While the throughput of RHCR drops more rapidly with normalization after $N_a$ agents, this can be compensated by increasing $N_a$ during the optimization of the guidance graph.

The advantage of normalization comes from the utilization of the guidance graph in lifelong MAPF. As discussed in \Cref{sec:approach}, the absolute magnitude of the edge weights has less or no impact on the MAPF solutions compared to the relative magnitude. Therefore, normalizing these edge weights simplifies the optimization problem, shifting the focus to the \textit{shape} of the Gaussian distribution modeling the edge weights, rather than their precise numerical values. This simplification enables normalization to outperform other bounds-handling methods for CMA-ES.

\subsection{On the Value of \texorpdfstring{$N_{e\_piu}$}{N\_e\_piu}} \label{appen:piu-val}

We use the update model optimized with setup 2 to generate guidance graphs with $N_{e\_piu} \in \{1, 10, 30, 50\}$. We then evaluate the guidance graphs by running lifelong MAPF simulations with various number of agents, each with 50 simulations. We show the result in \Cref{fig:N-e-piu_ablation}. We find no significant difference in throughput with different values of $N_{e\_piu}$.

\subsection{Throughput of Larger Warehouse Maps} \label{appen:thr-large-warehouse}

\Cref{fig:warehouse-scale-map} shows the maps used for the transferability experiment.
\Cref{fig:warehouse-scale-result} shows throughput with different number of agents in large warehouse maps shown in \Cref{fig:warehouse-scale-map}. In general, we observe that PIU Transfer dominates all baselines.

\section{Optimized Guidance Graphs} \label{appen:gg-viz}

\begin{figure*}
    \centering
    \begin{subfigure}[b]{0.83\textwidth}
        \centering
        \includegraphics[width=1\textwidth]{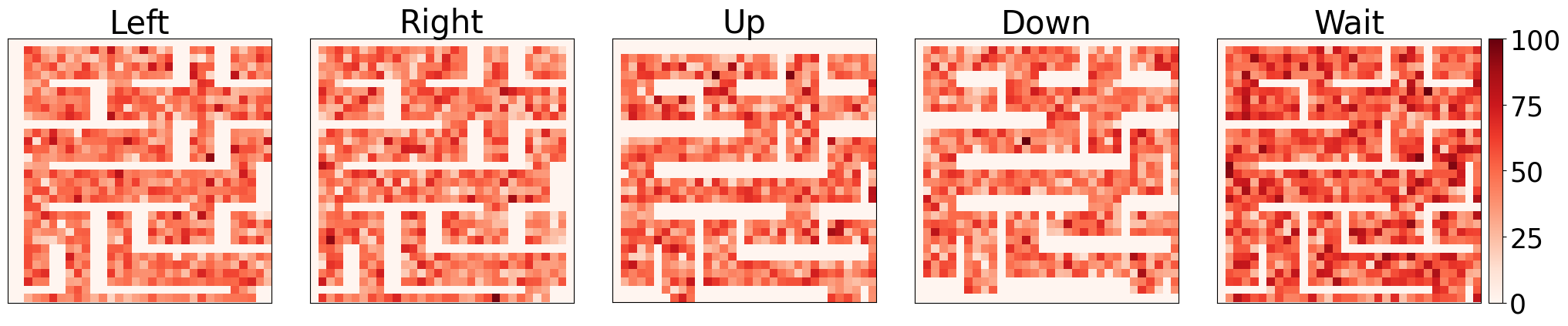}
        \caption{Setup 3 (\mazeSmall): PIBT + CMA-ES}
        \label{fig:maze-32-32-4-cma-es-gg}
    \end{subfigure}\\
    \begin{subfigure}[b]{0.83\textwidth}
        \centering
        \includegraphics[width=1\textwidth]{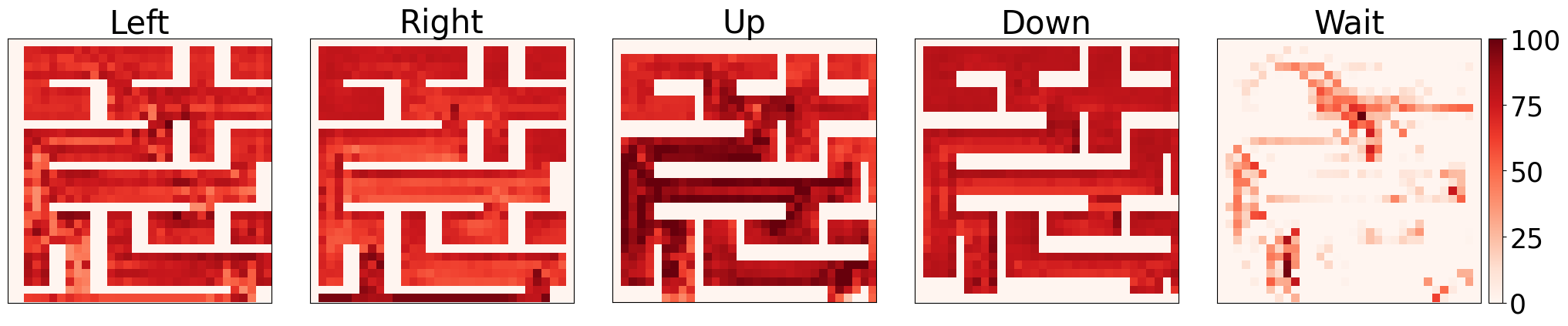}
        \caption{Setup 3 (\mazeSmall): PIBT + PIU}
        \label{fig:maze-32-32-4-piu-gg}
    \end{subfigure}\\
    \begin{subfigure}[b]{0.83\textwidth}
        \centering
        \includegraphics[width=1\textwidth]{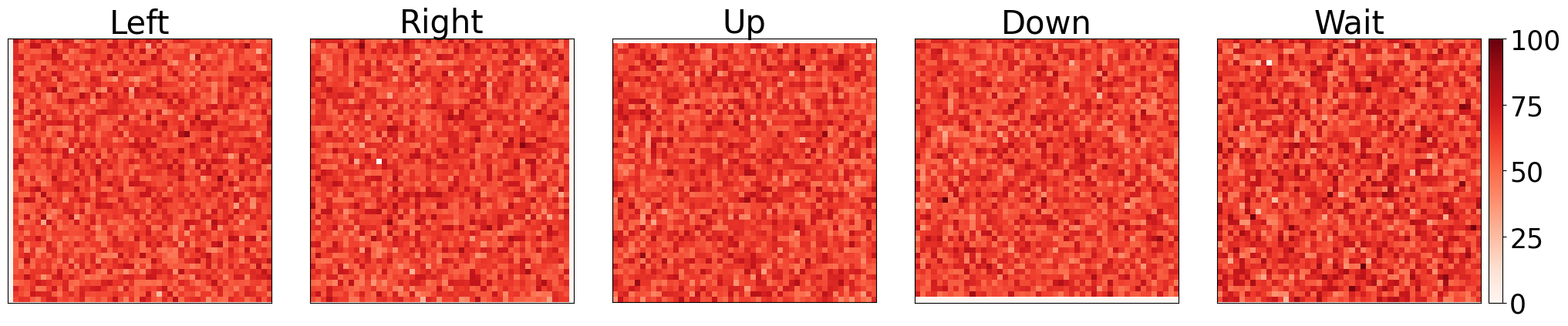}
        \caption{Setup 4 (\emptyMid): PIBT + CMA-ES}
        \label{fig:empty-48-48-cma-es-gg}
    \end{subfigure}\\
    \begin{subfigure}[b]{0.83\textwidth}
        \centering
        \includegraphics[width=1\textwidth]{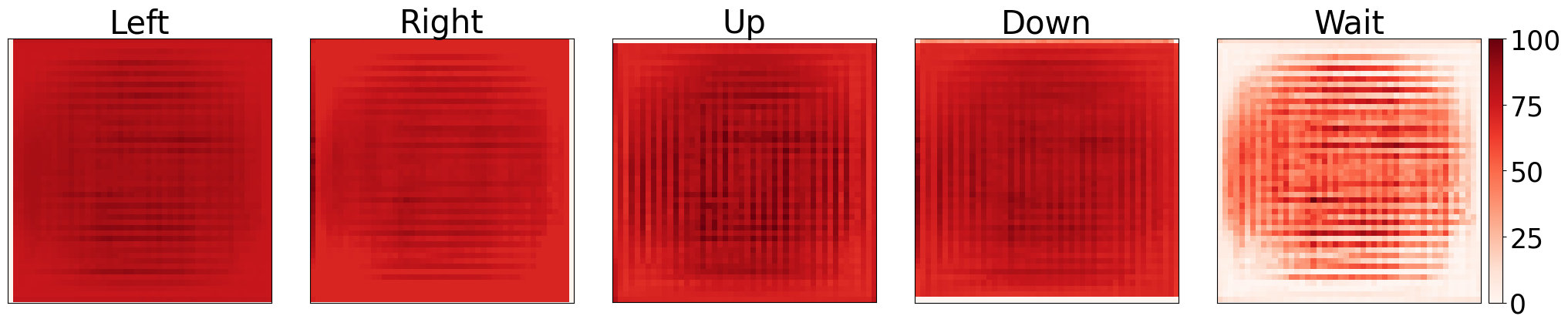}
        \caption{Setup 4 (\emptyMid): PIBT + PIU}
        \label{fig:empty-48-48-piu-gg}
    \end{subfigure}\\
    \begin{subfigure}[b]{0.83\textwidth}
        \centering
        \includegraphics[width=1\textwidth]{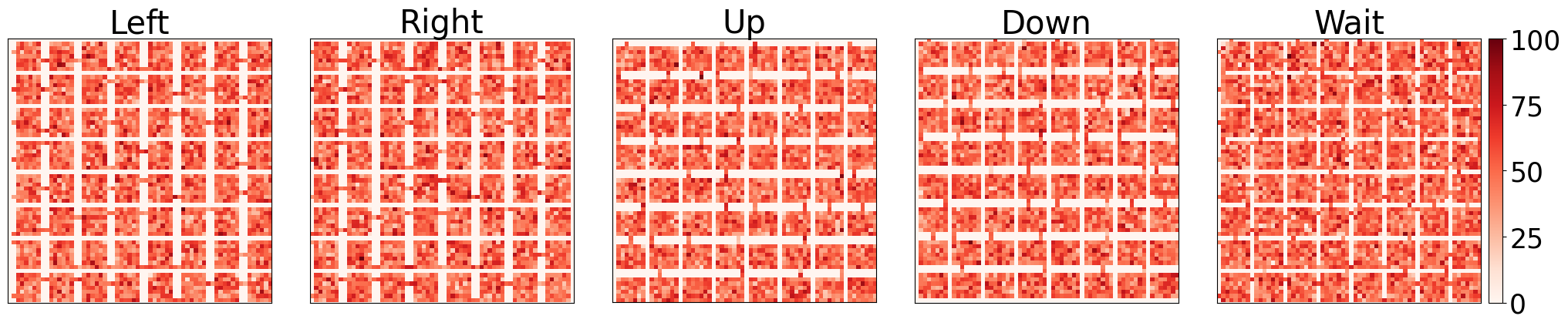}
        \caption{Setup 5 (\roomLarge): PIBT + CMA-ES}
        \label{fig:room-64-64-8-cma-es-gg}
    \end{subfigure}\\
    \begin{subfigure}[b]{0.83\textwidth}
        \centering
        \includegraphics[width=1\textwidth]{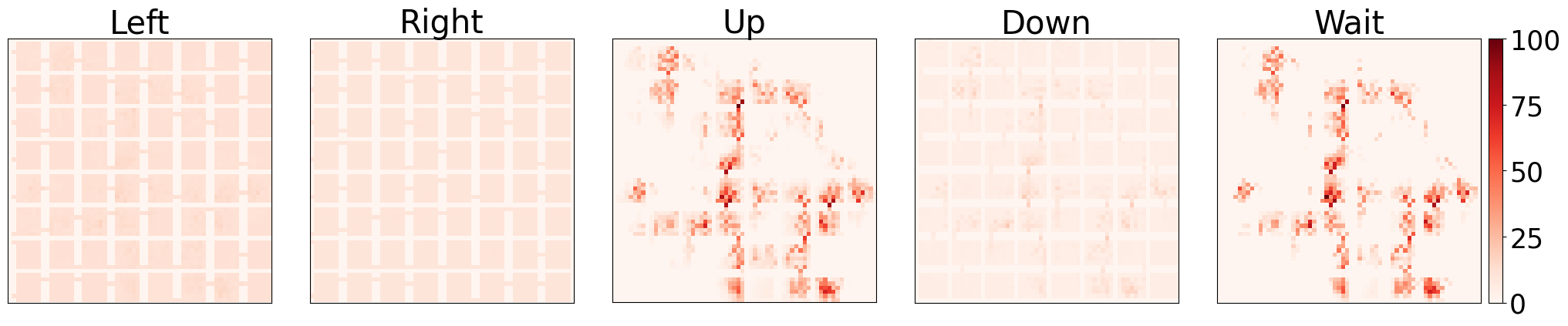}
        \caption{Setup 5 (\roomLarge): PIBT + PIU}
        \label{fig:room-64-64-8-piu-gg}
    \end{subfigure}\\
    \caption{Optimized guidance graphs of setups 3, 4, and 5.}
    \label{fig:opt-gg-345}
\end{figure*}

\begin{figure*}
    \centering
    \begin{subfigure}[b]{0.83\textwidth}
        \centering
        \includegraphics[width=1\textwidth]{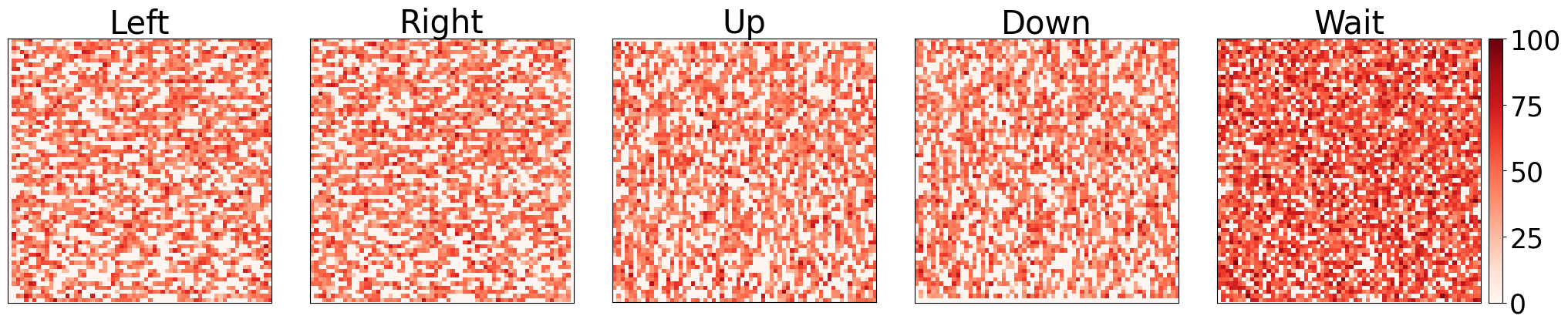}
        \caption{Setup 6 (\randomLarge): PIBT + CMA-ES}
        \label{fig:random-64-64-20-cma-es-gg}
    \end{subfigure}\\
    \begin{subfigure}[b]{0.83\textwidth}
        \centering
        \includegraphics[width=1\textwidth]{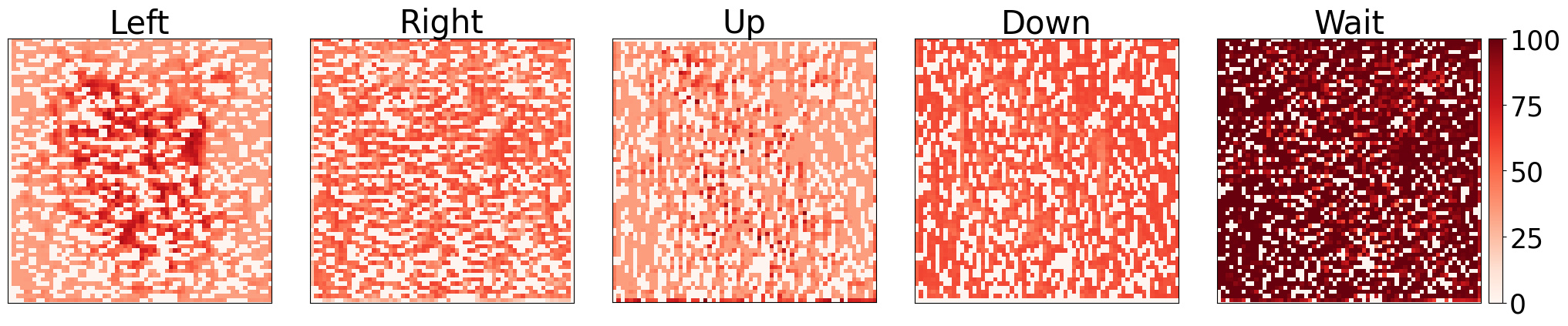}
        \caption{Setup 6 (\randomLarge): PIBT + PIU}
        \label{fig:random-64-64-20-piu-gg}
    \end{subfigure}\\
    \begin{subfigure}[b]{0.83\textwidth}
        \centering
        \includegraphics[width=1\textwidth]{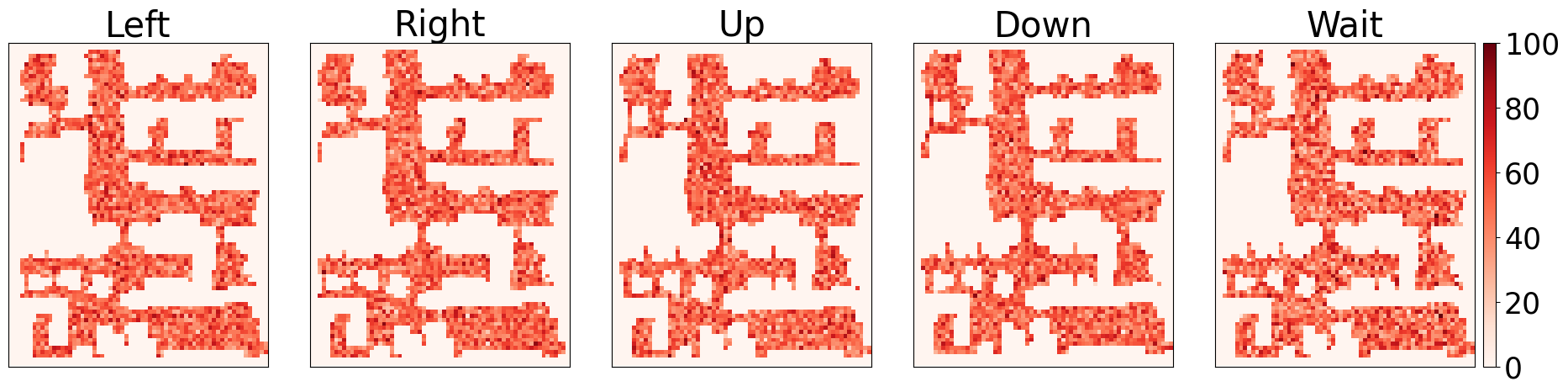}
        \caption{Setup 7 (\denSmall): PIBT + CMA-ES}
        \label{fig:den312d-cma-es-gg}
    \end{subfigure}\\
    \begin{subfigure}[b]{0.83\textwidth}
        \centering
        \includegraphics[width=1\textwidth]{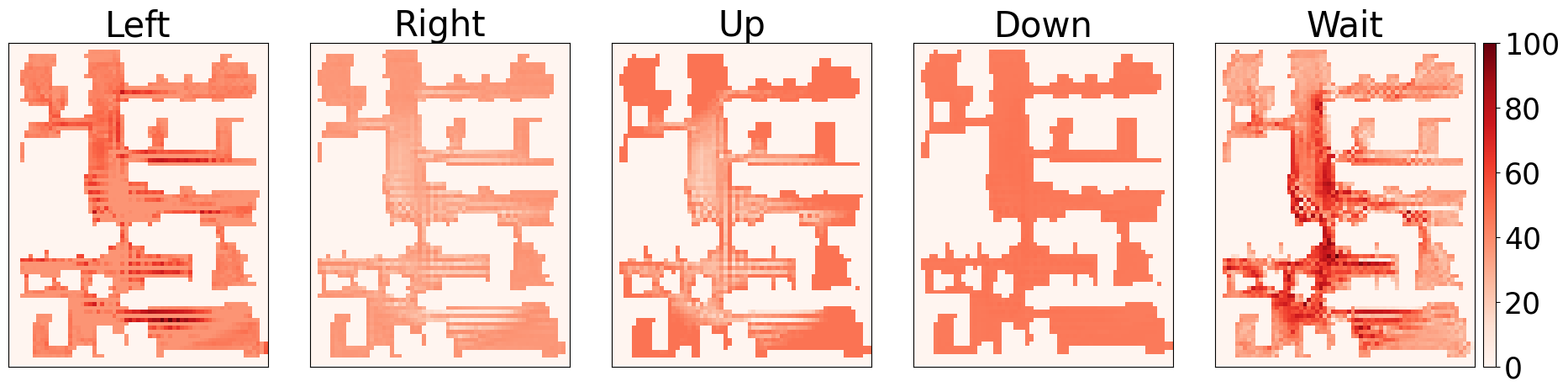}
        \caption{Setup 7 (\denSmall): PIBT + PIU}
        \label{fig:den312d-piu-gg}
    \end{subfigure}\\
    \begin{subfigure}[b]{0.83\textwidth}
        \centering
        \includegraphics[width=1\textwidth]{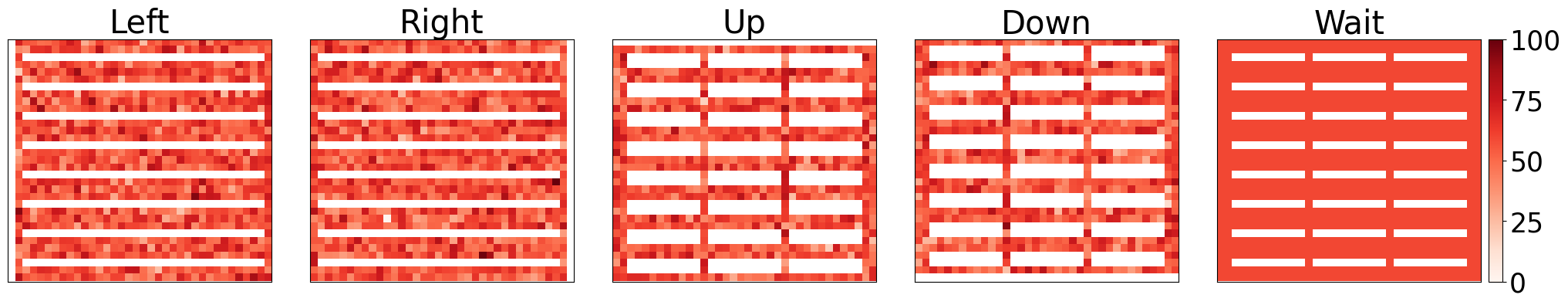}
        \caption{Setup 8 (\warehouselargeW): RHCR + CMA-ES}
        \label{fig:warehouse-large-rhcr-cma-es-gg}
    \end{subfigure}\\
    \caption{Optimized guidance graphs of setups 6, 7, and 8.}
    \label{fig:opt-gg-678}
\end{figure*}

\begin{figure*}
    \centering
    \begin{subfigure}[b]{0.83\textwidth}
        \centering
        \includegraphics[width=1\textwidth]{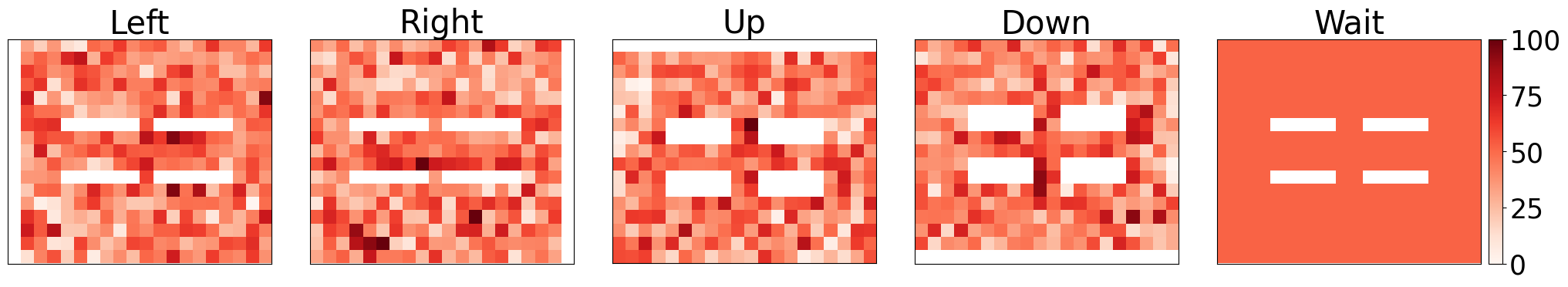}
        \caption{Setup 9 (\warehouseSmallR): DPP + CMA-ES}
        \label{fig:warehouse-large-rhcr-piu-gg}
    \end{subfigure}\\
    \begin{subfigure}[b]{0.83\textwidth}
        \centering
        \includegraphics[width=1\textwidth]{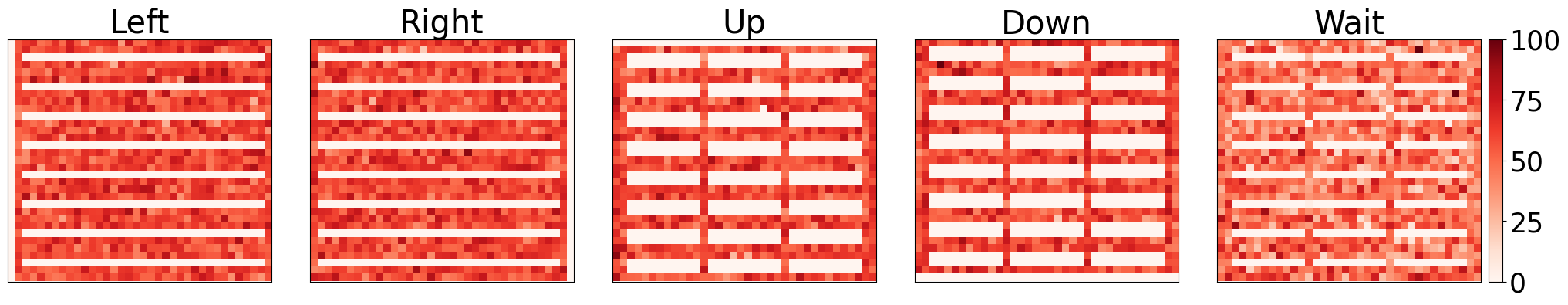}
        \caption{Setup 10 (\warehouselargeW): PIBT + CMA-ES (150 agents)}
        \label{fig:warehouse-large-pibt-150-cma-es-gg}
    \end{subfigure}\\
    \begin{subfigure}[b]{0.83\textwidth}
        \centering
        \includegraphics[width=1\textwidth]{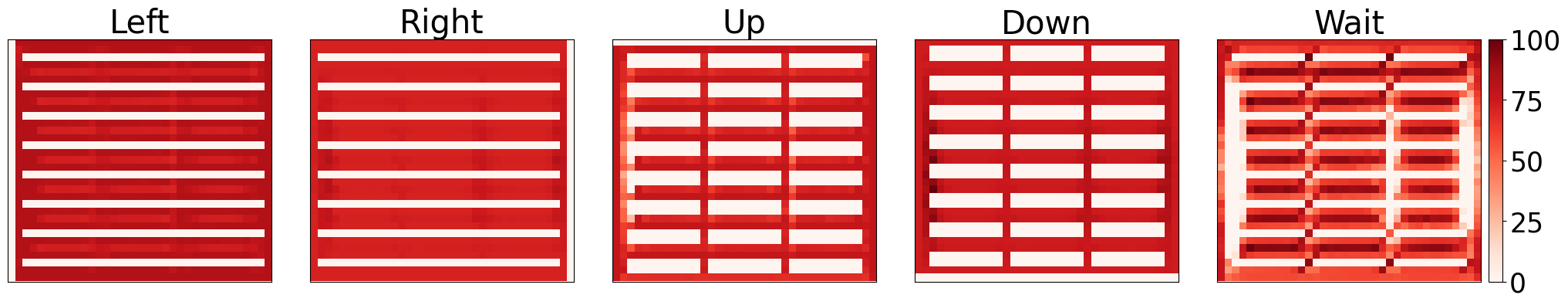}
        \caption{Setup 10 (\warehouselargeW): PIBT + PIU (150 agents)}
        \label{fig:warehouse-large-pibt-150-piu-gg}
    \end{subfigure}\\
    \caption{Optimized guidance graphs of setups 9 and 10.}
    \label{fig:opt-gg-910}
\end{figure*}

We show the optimized guidance graphs with CMA-ES and PIU in \Cref{fig:opt-gg12,fig:opt-gg-345,fig:opt-gg-678,fig:opt-gg-910}. In general, the guidance graphs optimized by CMA-ES are hardly explainable. Those optimized by PIU possess more patterns. For example, in \emptyMid (\Cref{fig:empty-48-48-piu-gg}), the optimized wait costs are lower than movement costs, promoting agents to wait instead of moving in case of congestion. In \randomLarge (\Cref{fig:random-64-64-20-piu-gg}), on the other hand, the wait costs are generally larger than movement costs. Nevertheless, it is non-trivial to explain why such guidance graphs improve throughput in these maps. We leave generating explainable guidance graphs or explaining optimized guidance graphs to future works.

\end{document}